\documentclass[aps,prb,twocolumn,superscriptaddress,nobibnotes]{revtex4-1}

\usepackage{graphicx}

\newcommand{\upperRomannumeral}[1]{\uppercase\expandafter{\romannumeral#1}}

\usepackage{hyperref}
\pdfoutput=1

\usepackage{amsmath}
\usepackage{bm}
\usepackage{amssymb}
\usepackage{textcomp}
\usepackage{stmaryrd} 
\usepackage{ulem}
\usepackage{gensymb}

\usepackage{import}
\usepackage{color}
\usepackage{verbatim}

\usepackage{float}

\usepackage{sistyle} 

\newcommand{\Vac}{V_{\mathrm {ac}}}
\newcommand{\Vdc}{V_{\mathrm {dc}}}
\newcommand{\V}{\mathrm V}
\newcommand{\nm}{\mathrm {nm}}
\newcommand{\mV}{\mathrm {mV}}
\newcommand{\ed}{\varepsilon_{\mathrm{d}}}
\newcommand{\es}{\varepsilon_{\mathrm{s}}}
\newcommand{\edh}{\varepsilon_{\mathrm{d}}^{\mathrm{h}}}
\newcommand\abs[1]{\left| #1 \right|}

\def\XXint#1#2#3{{\setbox0=\hbox{$#1{#2#3}{\int}$}
     \vcenter{\hbox{$#2#3$}}\kern-.5\wd0}}

\begin{document}

\title{Dynamically-enhanced strain in atomically thin resonators}

\author{Xin Zhang}
\email{zhxsemi@gmail.com}
\affiliation{Universit\'e de Strasbourg, CNRS, Institut de Physique et Chimie des Mat\'eriaux de Strasbourg, UMR 7504, F-67000 Strasbourg, France}
\author{Kevin Makles}
\affiliation{Universit\'e de Strasbourg, CNRS, Institut de Physique et Chimie des Mat\'eriaux de Strasbourg, UMR 7504, F-67000 Strasbourg, France}
\author{L\'eo Colombier}
\affiliation{Universit\'e de Strasbourg, CNRS, Institut de Physique et Chimie des Mat\'eriaux de Strasbourg, UMR 7504, F-67000 Strasbourg, France}
\author{Dominik Metten}
\affiliation{Universit\'e de Strasbourg, CNRS, Institut de Physique et Chimie des Mat\'eriaux de Strasbourg, UMR 7504, F-67000 Strasbourg, France}
\author{Hicham  Majjad}
\affiliation{Universit\'e de Strasbourg, CNRS, Institut de Physique et Chimie des Mat\'eriaux de Strasbourg, UMR 7504, F-67000 Strasbourg, France}
\author{Pierre Verlot}
\affiliation{School of Physics and Astronomy, University of Nottingham, Nottingham, NG7 2RD, United Kingdom}
\affiliation{Institut Universitaire de France, 1 rue Descartes, 75231 Paris cedex 05, France}
\author{St\'ephane Berciaud}
\email{stephane.berciaud@ipcms.unistra.fr}
\affiliation{Universit\'e de Strasbourg, CNRS, Institut de Physique et Chimie des Mat\'eriaux de Strasbourg, UMR 7504, F-67000 Strasbourg, France}
\affiliation{Institut Universitaire de France, 1 rue Descartes, 75231 Paris cedex 05, France}

~

\begin{abstract}
Graphene and related two-dimensional (2D) materials associate remarkable mechanical, electronic, optical and phononic properties. As such, 2D materials are promising for hybrid systems that couple their elementary excitations (excitons, phonons) to their macroscopic mechanical modes. These built-in systems may yield enhanced strain-mediated coupling compared to bulkier architectures, e.g., comprising a single quantum emitter coupled to a nano-mechanical resonator. Here, using micro-Raman spectroscopy on pristine monolayer graphene drums, we demonstrate that the macroscopic flexural vibrations of graphene induce dynamical optical phonon softening. This softening is an unambiguous fingerprint of dynamically-induced tensile strain that reaches values up to $\mathbf{\approx 4 \times 10^{-4}}$ under strong non-linear driving. Such non-linearly enhanced strain exceeds the values predicted for harmonic vibrations with the same root mean square (RMS) amplitude by more than one order of magnitude. Our work holds promise for dynamical strain engineering and dynamical strain-mediated control of light-matter interactions in 2D materials and related heterostructures.
\end{abstract}

\maketitle


\begin{center}
    \textbf{INTRODUCTION}
\end{center}

Since the first demonstration of mechanical resonators made from suspended graphene layers~\cite{Bunch2007}, considerable progress has been made to conceive nano-mechanical systems based on 2D materials~\cite{Castellanos-Gomez2015,Geim2013} with well-characterized performances~\cite{Chen2009,Weber2014,Davidovikj2016,Davidovikj2017,Lee2018}, for applications in mass and force sensing\cite{Weber2016} but also for studies of heat transport~\cite{Barton2012,Morell2019}, non-linear mode coupling~\cite{DeAlba2016,Mathew2016,Guttinger2017} and optomechanical interactions~\cite{Weber2014,Singh2014,Song2014}. These efforts triggered the study of 2D resonators beyond graphene, made for instance from  transition metal dichalcogenide layers~\cite{Castellanos-Gomez2013,Morell2016,Morell2019,Lee2018} and van der Waals heterostructures~\cite{Will2017,Ye2017,Kim2018}. In suspended atomically thin membranes, a moderate out-of-plane stress gives rise to large and swiftly tunable strains, in excess of $1\%$~\cite{Koenig2011,Lloyd2017}, opening numerous possibilities for strain-engineering~\cite{Dai2019}. These assets also position 2D materials as promising systems to achieve enhanced strain-mediated coupling~\cite{Arcizet2011,Teissier2014,Ovartchaiyapong2014,Yeo2014} of macroscopic flexural vibrations to quasiparticles (excitons, phonons) and/or degrees of freedom (spin, valley). Such developments require sensitive probes of dynamical strain. Among the approaches employed to characterise strain in 2D materials, micro-Raman scattering spectroscopy~\cite{Ferrari2013} stands out as a local, contactless and minimally invasive technique that has been extensively exploited in the static regime to quantitatively convert the frequency softening or hardening of the Raman active modes into an amount of tensile or compressive strain, respectively~\cite{Mohiuddin2009,Metten2014,Androulidakis2015,Zhangx2015}. Recently, the interplay between electrostatically-induced strain and doping has been probed in the static regime in suspended graphene monolayers~\cite{Metten2016}. Dynamically-induced strain has been investigated using Raman spectroscopy in bulkier micro electro-mechanical systems~\cite{Pomeroy2008,Xue2007}, including mesoscopic graphite cantilevers~\cite{Reserbat-Plantey2012} but  remains unexplored in resonators made from 2D materials.

In this article, using micro-Raman scattering spectroscopy in resonators made from pristine suspended graphene monolayers, we demonstrate efficient strain-mediated coupling between ``built-in'' quantum degrees of freedom (here the Raman-active optical phonons of graphene) of the 2D resonator, and its macroscopic flexural vibrations. The dynamically-induced strain is quantitatively determined from the frequency of the Raman-active modes and is found to attain anomalously large values, exceeding the levels of strain expected under harmonic vibrations  by more than one order of magnitude. Our work introduces resonators made from graphene and related 2D materials as promising systems  for hybrid opto-electro-mechanics~\cite{Midolo2018} and dynamical strain-mediated control of light-matter interactions.

\begin{figure*}[!ht]
\begin{center}
\includegraphics[width=14.5cm]{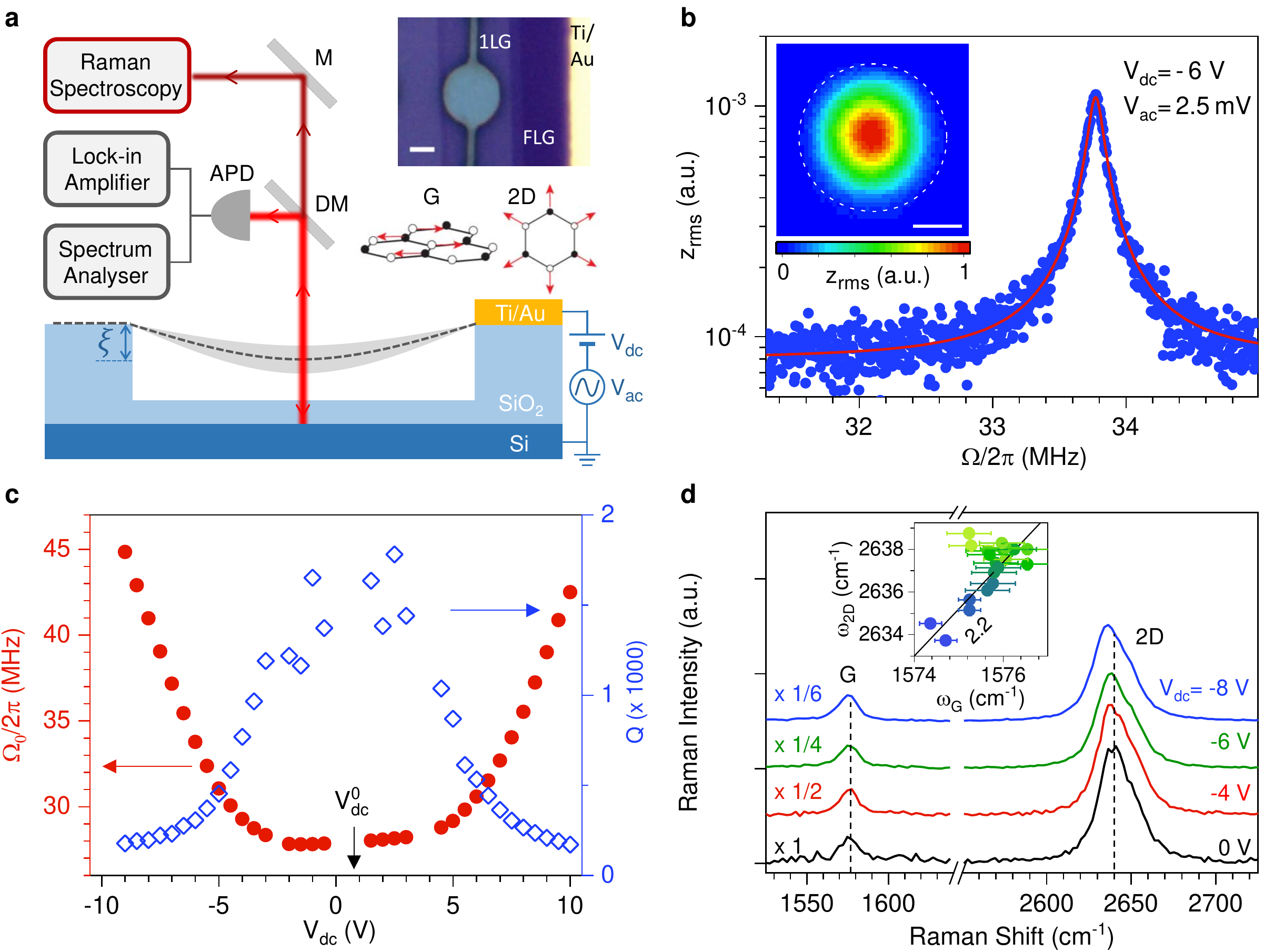}
\caption{{\bf Experimental setup and characterisation of pristine graphene drums.} {\bf a}, Sketch of our experiment combining electrostatic actuation, optical readout of the displacement and micro-Raman spectroscopy of a circular graphene drum (device 1). The graphene layer (with its static displacement $\xi$) is represented by the dark grey dashed line; its flexural motion is sketched with the light grey shade.  M, DM, APD represent a mirror, a dichroic mirror, an avalanche photodiode, respectively. Upper inset: optical image of a suspended graphene monolayer (1L) contacted by a Ti/Au lead (scale bar: 2 $\mu$m). A thicker, few-layer graphene flake (FLG) is also visible. Lower inset: sketch of the atomic displacements contributing to the Raman G mode and 2D mode. {\bf b}, RMS mechanical amplitude $z_{\rm {rms}}$ (blue dots) as a function of the drive frequency $\Omega/2\pi$ at $\Vdc=-6~\mathrm{V}$ and $\Vac=2.5~\mathrm{mV}$. The red line is a fit based on linear response theory (Supplementary Note 6). Inset: map of the resonant mechanical (scale bar: 2 $\mu$m). {\bf c}, Resonance frequency $\Omega_0/2\pi$ and corresponding quality factor $Q$ as a function of $\Vdc$, with $\Vdc^0$ indicating the charge neutrality point in graphene. {\bf d}, Raman spectra measured at the centre of the drum at $\Vdc=0,~-4,~-6,~-8~\mathrm{V}$ and $\Vac=0~\mV$. Inset: correlation between the G- and 2D-mode frequencies ($\omega_{\mathrm{2D}}$ and $\omega_{\mathrm G}$), extracted from Raman spectra measured with $\Vdc$ varying from $-9$~V to 10~V. The light green-to-blue color scale in circles encodes the increase of $\vert \Vdc-\Vdc^0\vert$.  The straight black line with a slope of 2.2 is a guide to the eye corresponding to strain-induced phonon softening. }
\label{Fig1}
\end{center}
\end{figure*}


~

\begin{center}
    \textbf{RESULTS}
\end{center}

\textbf{Measurement Scheme --}
As illustrated in Fig.~\ref{Fig1}a, the system we have developed for probing dynamical strain in the 2D limit is a graphene monolayer, mechanically exfoliated and transferred as is onto a pre-patterned Si/SiO$_2$ substrate. The resulting graphene drum is capacitively driven using a time-dependent gate bias $V_{\mathrm g}(t)=\Vdc+\Vac\cos{\Omega t}$, with $\Vac\ll \Vdc$ and $\Omega/2\pi$ the drive frequency. The DC component of the resulting force ($\propto V_{\mathrm g}^2$, see Methods) enables to control the electrostatic pressure applied to the graphene membrane (and hence its static deflection $\xi$, see Fig.~\ref{Fig1}a), whereas the AC bias leads to a harmonic driving force $(\propto \Vdc\Vac\cos{\Omega t})$. A single laser beam is used to interferometrically measure the frequency-dependent mechanical susceptibility at the drive frequency, akin to Ref.~\onlinecite{Bunch2007} and, at the same time, to record the micro-Raman scattering response of the atomically thin membrane. We have chosen electrostatic rather than photothermal actuation~\cite{Sampathkumar2006} to attain large RMS amplitudes while at the same time avoiding heating and photothermal backaction effects~\cite{Barton2012,Morell2019}, possibly leading to additional damping~\cite{Lee2018}, self-oscillations~\cite{Barton2012}, mechanical instabilities and sample damage. All measurements were performed at room temperature under high vacuum (see Methods and Supplementary Notes  1 to 8). 

~

\textbf{Raman spectroscopy in strained graphene --}
The Raman spectrum of graphene displays two main features: the G mode and the 2D mode, arising from one zone-center (that is, zero momentum) phonon and from a pair of near-zone edge phonons with opposite momenta, respectively (see Fig.~\ref{Fig1}a and Supplementary Note 1)~\cite{Ferrari2013}. Both features are uniquely sensitive to external perturbations. Quantitative methods have been developed to unambiguously separate the share of strain, doping, and possibly heating effects that affect the frequency, full width at half maximum (FWHM) and integrated intensity of a Raman feature~\cite{Pisana2007,Lee2012a,Froehlicher2015,Metten2014,Metten2016} (hereafter denoted $\omega_i$, $\Gamma_i$, $I_i$, respectively, here with $i=\mathrm G,\mathrm{2D}$). Biaxial strain is expected around the centre of circular graphene drums~\cite{Koenig2011} and the large  Gr\"uneisen parameters of graphene ($\gamma_{\rm G}=1.8$ and $\gamma_{2\rm D}=2.4$, with  $\gamma_{i}=\frac{1}{2\omega_i}\frac{\partial \omega_i}{\partial \varepsilon}$ and $\varepsilon$ the level of biaxial strain)~\cite{Metten2014,Androulidakis2015} allow detection of strain levels down to a few $10^{-5}$. The characteristic slope $\frac{\partial \omega_{\mathrm 2D}}{\partial \omega_{\mathrm G}}\approx 2.2$ in graphene under biaxial strain is much larger than in the case of electron or hole doping, where the corresponding slope is significantly smaller than 1~\cite{Lee2012a,Froehlicher2015}. This difference allows a clear disambiguation between strain and doping (see Methods for details).

~

\textbf{Mechanical and Raman characterisation --}
Figure~\ref{Fig1}b presents the main characteristics of a circular graphene drum (device 1) in the linear response regime. A Lorentzian mechanical resonance is observed at $\Omega_0/2\pi\approx33.8~\mathrm {MHz}$ for $\Vdc=-6~\mathrm V$ (Fig.~\ref{Fig1}b and Supplementary Notes 5 and 6). The mechanical mode profile shows radial symmetry (inset in Fig.~\ref{Fig1}b) as expected for the fundamental flexural resonance of a circular drum~\cite{Davidovikj2016}. The mechanical resonance frequency is widely gate-tunable: it increases by $\sim 70~\%$ as $\abs{\Vdc}$ is ramped up to 10~V and displays a symmetric, ``U-shaped'' behavior with respect to a near-zero DC bias $\Vdc^0= 0.75~\mathrm V$, at which graphene only undergoes a built-in tension.  These two features are characteristic of a low built-in tension~\cite{Chen2009,Barton2012,Lee2018,Singh2010} that we estimate to be $T_0=\left(4\pm0.4\right)\times 10^{-2}~\mathrm {Nm}^{-1}$, corresponding to a built-in static strain $\varepsilon_{\rm s}^0=T_0\left(1-\nu\right)/E_{\rm{1LG}}\approx \left(1.0\pm0.1\right)\times10^{-4}$, where $E_{\rm{1LG}}=340~\rm{N\,m^{-1}}$, $\nu=0.16$ are the Young modulus and Poisson ratio of pristine monolayer graphene~\cite{Lee2008} (Supplementary Note 6). The quality factor $Q$ is high, in excess of 1500 near charge neutrality. As $\left|\Vdc\right|$ increases, $Q$ drops down to $\sim 200$ due to electrostatic damping\cite{Lee2018}.\\
Figure~\ref{Fig1}d shows that the Raman response of suspended graphene is tunable by application of a DC gate bias, as extensively discussed in Ref.~\onlinecite{Metten2016}. Once $\Vdc$ is large enough to overcome $\varepsilon_{\rm s}^0$, the membrane starts to bend downwards and the downshifts of the G- and 2D-mode features measured at the centre of the drum are chiefly due to biaxial strain ($\partial \omega_{\mathrm 2D}/\partial \omega_{\mathrm G}\approx 2.2$, see inset in Fig.~\ref{Fig1}d) with negligible contribution from electrostatic doping~\cite{Metten2016} (see Methods for details).
At $\Vdc=-9~\mathrm V$, the $ 4\pm0.5~\rm{ cm^{-1}}$ 2D-mode downshift relative to its value near $V_{\rm{dc}}^0$ yields a gate-induced static strain $\varepsilon_{\rm s}= 3\pm0.3\times 10^{-4}$ that agrees qualitatively well with the value $\varepsilon_{\rm s}=2\pm0.2\times 10^{-4}$ estimated from the gate-induced upshift of $\Omega_0$ (Fig.~ 1c and Supplementary Note 6). This agreement justifies our assumption that the Young's modulus of our drum is close to that of pristine graphene (see also Supplementary Note 5 for details on the drum effective mass).\\
Noteworthy, optical interference effects cause a large gate-dependent modulation of $I_{\rm G}$ and $I_{\rm{2D}}$ (Ref.~\onlinecite{Metten2014,Metten2016} and see normalisation factors in Fig.~\ref{Fig1}d).  Both strain-induced Raman shifts and Raman scattering intensity changes are exploited to consistently estimate that $\xi$ increases from about $30~\mathrm{nm}$ to $70~\mathrm{nm}$ when $\Vdc$ is varied from $-5~\mathrm V$ to $-9~\mathrm V$ (Supplementary Notes 2, 3 and 4).




~

\textbf{Non-linear mechanical response --}
We are now examining how the dynamically-induced strain  can be readout by means of Raman spectroscopy. First, to obtain a larger sensitivity towards static strain (Supplementary Note 3), we apply a sufficiently high $\Vdc$ to reach a sizeable $\xi$. $\Vac$ is then ramped up to yield large RMS amplitudes. After calibration of our setup (Supplementary Note 5), we estimate that resonant RMS amplitudes $z^0_{\rm{rms}}$ up to $\sim 10~\nm$ are attained in device 1 (Fig.~\ref{Fig2},\ref{Fig3}). In this regime, graphene is a strongly non-linear mechanical system that can be described to lowest order by a Duffing-like equation~\cite{NO_book,Davidovikj2017,Weber2014}:


\begin{figure*}[!htb]
\begin{center}
\includegraphics[width=14.3cm]{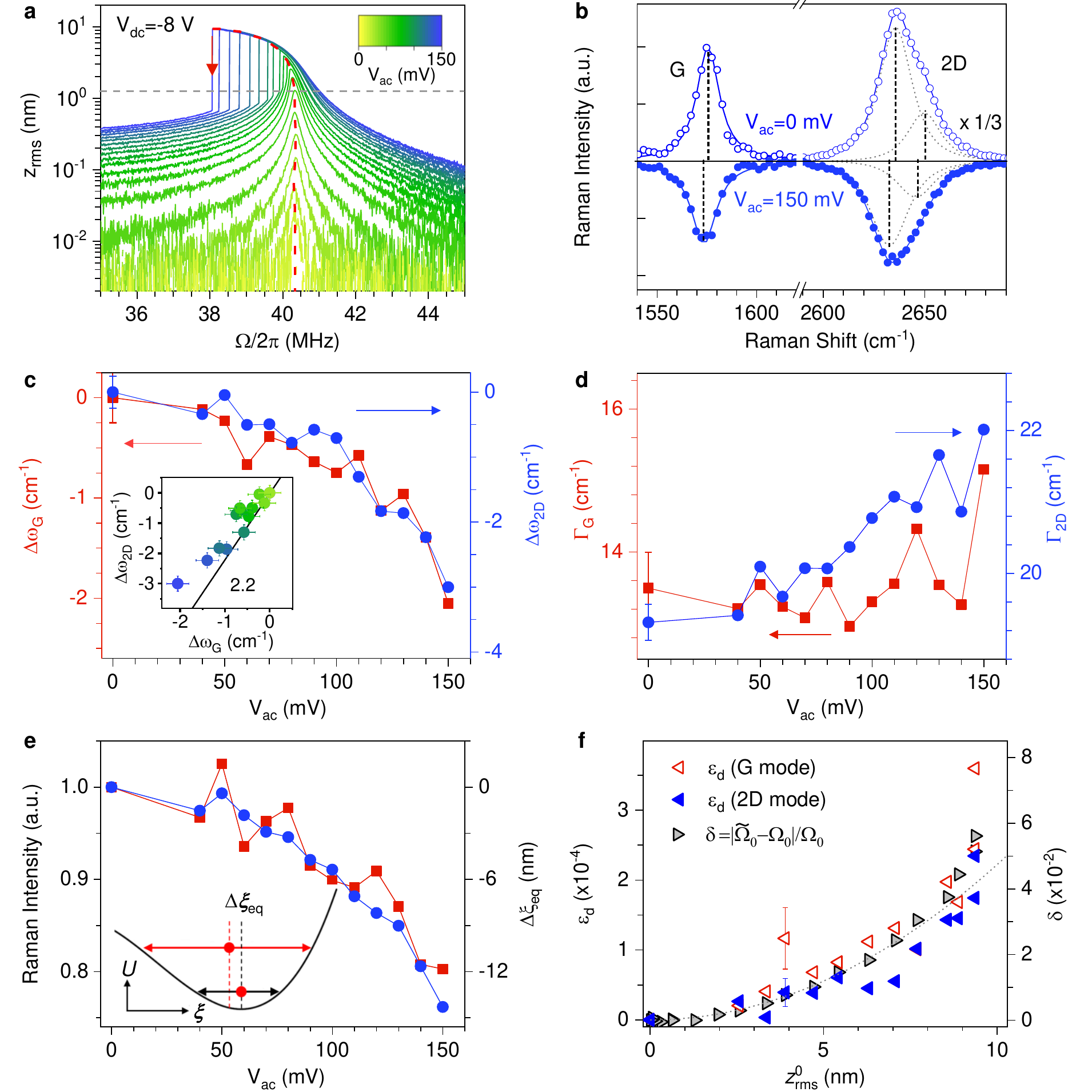}
\caption{{\bf Evidence for dynamically-enhanced strain in graphene.} All measurements are performed on device 1 at $\Vdc=-8~\V$. {\bf a}, Calibrated RMS mechanical amplitude at the drive frequency $\Omega/2\pi$ ($z_{\rm {rms}}$) recorded as the frequency is swept downwards, for $\Vac$ increasing from 0 to 150~mV. The red dashed line is the backbone curve evidencing non-linear resonance frequency softening. The red arrow indicates the  jump-down frequency at $\Vac=150~\mV$. The grey dashed line denotes the onset of non-linearity. {\bf b}, Raman spectra measured under $\Vac=0~\rm{mV}$ (open symbols and fit) and 150~mV (filled symbols and fit, vertically flipped for clarity). {\bf c,d}, G- and 2D-mode frequency shifts $\Delta\omega_{\mathrm{G,2D}}$ and full-width at half maximum variations ($\Delta\Gamma_{\mathrm{G,2D}}$), relative to the values at $\Vac=0~\mV$, as a function of $\Vac$. Inset in {\bf c}: correlation between $\Delta\omega_{\mathrm {2D}}$ and $\Delta\omega_{\mathrm{G}}$. The symbol color encodes the increase of $\Vac$ as in {\bf a}. The straight black line with a slope of 2.2 is a guide to the eye for strain-induced phonon softening. {\bf e}, Normalized integrated intensity of G- and 2D-mode features as a function of $\Vac$. The inset illustrates the equilibrium position shift ($\Delta \xi_{\rm{eq}}$ between the two red circles) in the non-linear regime, with $U(\xi)$ the potential energy. {\bf f}, Time-averaged dynamical strain $\ed$ extracted from the softening of G- and 2D-mode features (open red and filled blue triangles, respectively) as a function of the corresponding $z^0_{\rm {rms}}$. The right axis (grey triangles) shows the relative non-linear mechanical resonance frequency shift $\delta=\abs{\widetilde{\Omega}_0-\Omega_0}/\Omega_0$, where $\widetilde{\Omega}_0$ is the jump-down frequency in \textbf{a}. The grey dashed line is a parabolic fit (Supplementary Note 6). Error bars in {\bf {c,d,f}} are extracted from the fits of Raman spectra. Only one error bar is included in each plot for visibility.}
\label{Fig2}
\end{center}
\end{figure*}


\begin{figure*}[!ht]
\begin{center}
\includegraphics[width=14.3cm]{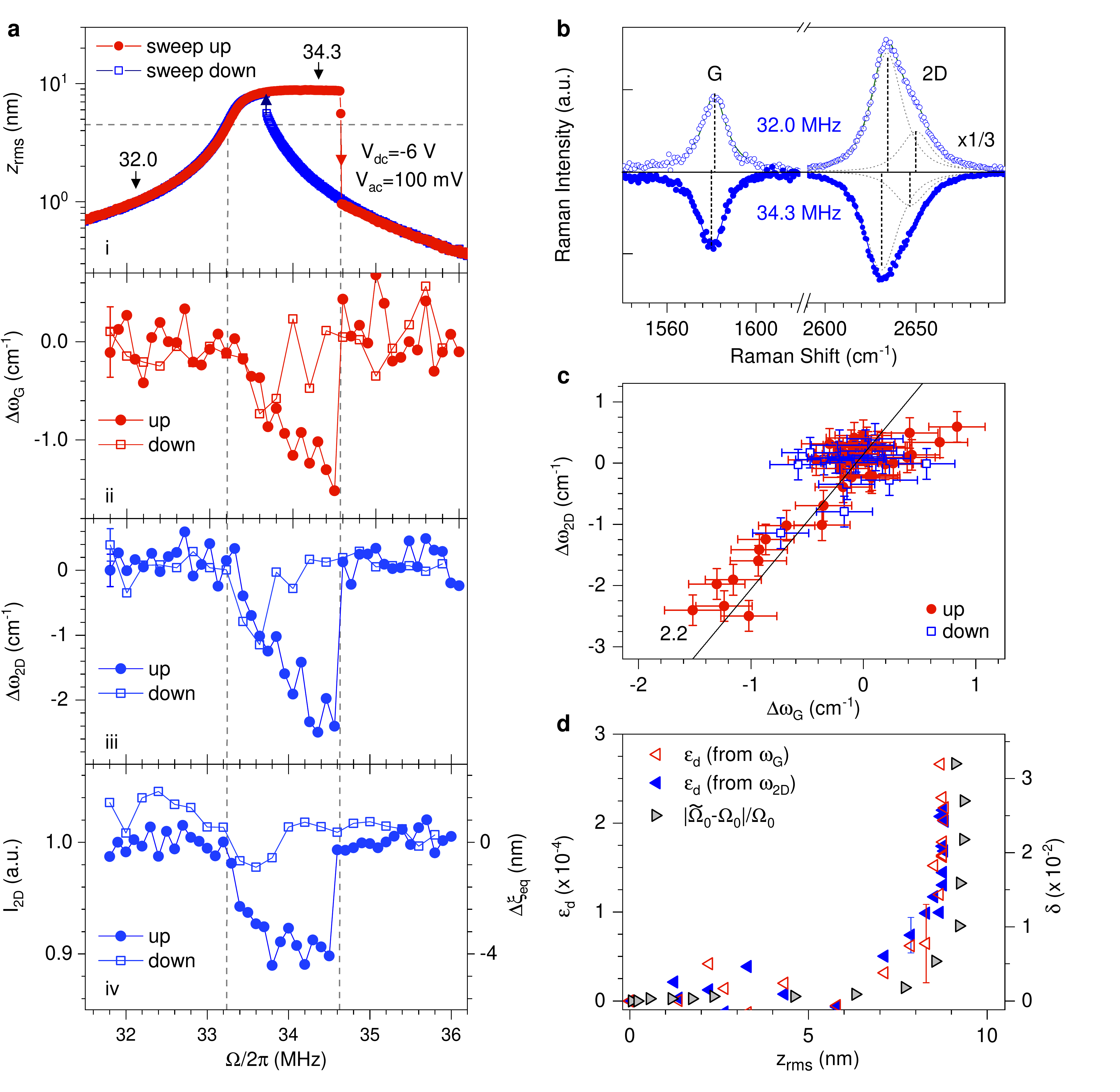}
\caption{{\bf Frequency-dependent dynamically-induced strain.} All measurements are performed on device 1 under $\Vdc=-6~\rm V$ {\bf a}, Panel i: RMS mechanical amplitude $z_{\rm {rms}}$ as a function of the drive frequency $\Omega/2\pi$ under $\Vac=100~\mathrm{mV}$. Red filled circles and blue open squares indicate upward and downward frequency sweeps. The blue and red arrows denote the jump-down and jump-up frequencies in the non-linear hardening region, respectively. The dashed lines are guides to the eye. Panels ii and iii: Raman frequency shifts $\Delta\omega_{\mathrm{G}}$ and $\Delta\omega_{\mathrm{2D}}$ as a function of $\Omega/2\pi$. Panel iv: integrated intensity of the 2D-mode feature as a function of $\Omega/2\pi$. Filled and open symbols in panels 2-4 correspond to upward and downward frequency sweeps, respectively. {\bf b}, Raman spectra recorded at $\Omega/2\pi=32~\mathrm{MHz}$ (open symbols and fit) and $34.3~\mathrm{MHz}$ (filled symbols and fit, vertically flipped for clarity), see arrows in {\bf a}. {\bf c} Correlation between the frequency shifts $\Delta\omega_{\mathrm G}$ and $\Delta\omega_{\mathrm{2D}}$, relative to the values recorded away from the mechanical resonance. The straight black line with a slope of 2.2 is a guide to the eye for the strain-induced phonon softening. {\bf d}, Time-averaged dynamical strain $\ed$ extracted from the softening of G- and 2D-mode features in {\bf a}-ii and {\bf a}-iii (open red and filled blue triangles, respectively) as a function of the RMS amplitude $z_{\rm{rms}}$. The right axis (grey triangles) shows the relative non-linear mechanical resonance frequency shift $\delta=\abs{\widetilde{\Omega}_0-\Omega_0}/\Omega_0$, where $\widetilde{\Omega}_0$ is the jump-down frequency (see \textbf{a} and Supplementary Note 6). Error bars in {\bf {a,c,d}} are extracted from the fits of Raman spectra. Only one error bar is included in {\bf a} and {\bf d} for visibility.}
\label{Fig3}
\end{center}
\end{figure*}



\begin{equation}
\ddot{z}+\frac{\Omega_0}{Q}\dot{z}+\Omega_0^2z+\widetilde{\alpha}_3z^3=\frac{ \widetilde{F}_{el}}{\widetilde{m}}\cos(\Omega t),
\end{equation}
where $z$ is the mechanical displacement at the membrane center relative to the equilibrium position $\xi$, $\Omega_0/2\pi$ is the resonance frequency in the linear regime, $Q$ is the quality factor and $\Omega_0/Q$ is the linear damping rate. The effective mass $\widetilde{m}$ and effective applied electrostatic force $\widetilde{F}_{\rm el}$ account for the mode profile of the fundamental resonance in a rigidly clamped circular drum~\cite{Hauer2013,Davidovikj2016} (see Methods and Supplementary Note 6). The linear spring constant is $\widetilde{m}\,\Omega_0^2$. Mechanical non-linearities are considered using an effective third-order term $\widetilde{\alpha}_3$ that changes sign at large enough $\xi$, leading to a transition from non-linear hardening to non-linear softening~\cite{Weber2014}. Such a behaviour is indeed revealed in our experiments, as shown in Fig.~\ref{Fig2}a  and Fig.\ref{Fig3}a, where non-linear softening and non-linear hardening are observed at $\Vdc=-8~\mathrm V$  and $\Vdc=-6~\mathrm V$, respectively. At $\Vdc=-7~\mathrm V$, we observe a $\Vac$-dependent softening-to-hardening transition (Supplementary Notes 6 and 7). 

~

\textbf{Dynamical optical phonon softening --} Figure~\ref{Fig2}c-e shows the frequencies,  linewidths  and integrated intensities of the Raman features  measured at $\Vdc=-8~\mathrm V$ (where $\xi\approx 60~\mathrm{nm}$), with $\Vac$ increasing from 0 to $150~\mathrm{mV}$ and applied at a drive frequency that tracks the $\Vac$-dependent non-linear softening of the mechanical resonance frequency $\widetilde{\Omega}_0/2\pi$, that is the so-called backbone curve in Fig.~\ref{Fig2}a,f (Supplementary Note 6). Both G- and 2D-mode features downshift as the drum is non-linearly driven. This phonon softening is accompanied by spectral broadening  by up to $\sim 10-15\;\%$ (Fig.~\ref{Fig2}d)~that increases with $z_{\rm{rms}}$. The correlation plot between $\omega_{\mathrm{2D}}$ and $\omega_{\mathrm{G}}$ reveals a linear slope near 2 (see also Supplementary Note 1), which is a characteristic signature of tensile strain~\cite{Metten2014,Lee2012a} that gets as high as $\approx 2.5\times10^{-4}$ for $z_{\rm{rms}}\approx 9~\rm{nm}$.


In Fig.~\ref{Fig3}a, we compare, on device 1, the frequency-dependence of $z_{\rm{rms}}$ to that of $\omega_{\mathrm {G,2D}}$ and $I_{2 \mathrm D}$, for upward and downward  sweeps under $\Vdc=-6~\rm V$ and $\Vac=100~\mV$. As in Fig.~\ref{Fig2}c,  sizeable G-mode and 2D-mode softenings are observed near the mechanical resonance (Fig.~\ref{Fig3}a-c) and assigned to tensile strain (see correlation plot in Fig.~\ref{Fig3}c).  Remarkably, the hysteretic behavior of the mechanical susceptibility, associated with non-linear hardening at $\Vdc=-6~\rm V$, is well-imprinted onto the frequency-dependence of $\omega_{\mathrm {G,2D}}$ and $I_{\mathrm {2D}}$. Looking further at Fig.~\ref{Fig3}a, we notice that while $z_{\rm{rms}}$ fully saturates at drive frequencies above 33.5~MHz and ultimately starts to decrease near the jump-down frequency, the tensile strain keeps increasing linearly up to $\approx 2.5\times 10^{-4}$ as $\Omega/2\pi$ is raised from $33.2~\mathrm{MHz}$ up to $34.5~\mathrm{MHz}$. 

~

\textbf{Equilibrium position shift --}
As our graphene drums are non-linearly driven, including beyond the Duffing regime (Fig.~\ref{Fig3}a and Supplementary Notes 6 and 7), the large strains revealed in Fig.~\ref{Fig2}~and~\ref{Fig3} could in part arise from an equilibrium position shift $\Delta\xi_{\rm{eq}}$ due to symmetry breaking non-linearities~\cite{NO_book,Eichler2013} (inset in Fig.~\ref{Fig2}e). This effect can be quantitatively assessed through analysis of $I_{\mathrm{G,2D}}$. As shown in Fig.~\ref{Fig2}e both $I_{\mathrm{2D}}$ and $I_{\mathrm{G}}$ decrease by about $\sim 20\%$ as $\Vac$ increases up to $150~\mV$.  These variations are assigned to optical interference effects (Ref.~\onlinecite{Metten2014,Metten2016}); in our experimental geometry they indicate an equilibrium position upshift $\Delta\xi_{\rm{eq}}$ by up to $\approx 12~\nm$ (Fig.~\ref{Fig2}e and Supplementary Note 4), that leads to a  reduction of the static tensile strain $\Delta\varepsilon_{\rm s}\approx 1\times 10^{-4}$, in stark contrast with the enhanced tensile strain unambiguously revealed in Fig.~\ref{Fig2}c. Similarly, the $\approx 10~\%$ drop in $I_{\mathrm {2D}}$ near the jump-down frequency at $34.5~\mathrm{MHz}$ indicates an equilibrium position upshift $\Delta\xi_{\rm{eq}}\approx 4~\rm{nm}$ that is qualitatively similar to the results in Fig.~\ref{Fig2}e. The  larger $\Delta\xi_{\rm{eq}}$ measured at $\Vdc=-8\rm V$ is consistent with our observation of non-linear mechanical resonance softening (Fig.~\ref{Fig2}a) due to an increased contribution from symmetry breaking non-linearities at large $\xi$ (Ref.~\onlinecite{NO_book,Eichler2013,Weber2014} and Supplementary Note 6). From these measurements, we conclude that the dynamical softening of $\omega_{\rm G}$ and $\omega_{\rm{2D}}$ is not due to an equilibrium position shift.

~


\begin{figure}[!h]
\begin{center}
\includegraphics[scale=0.7]{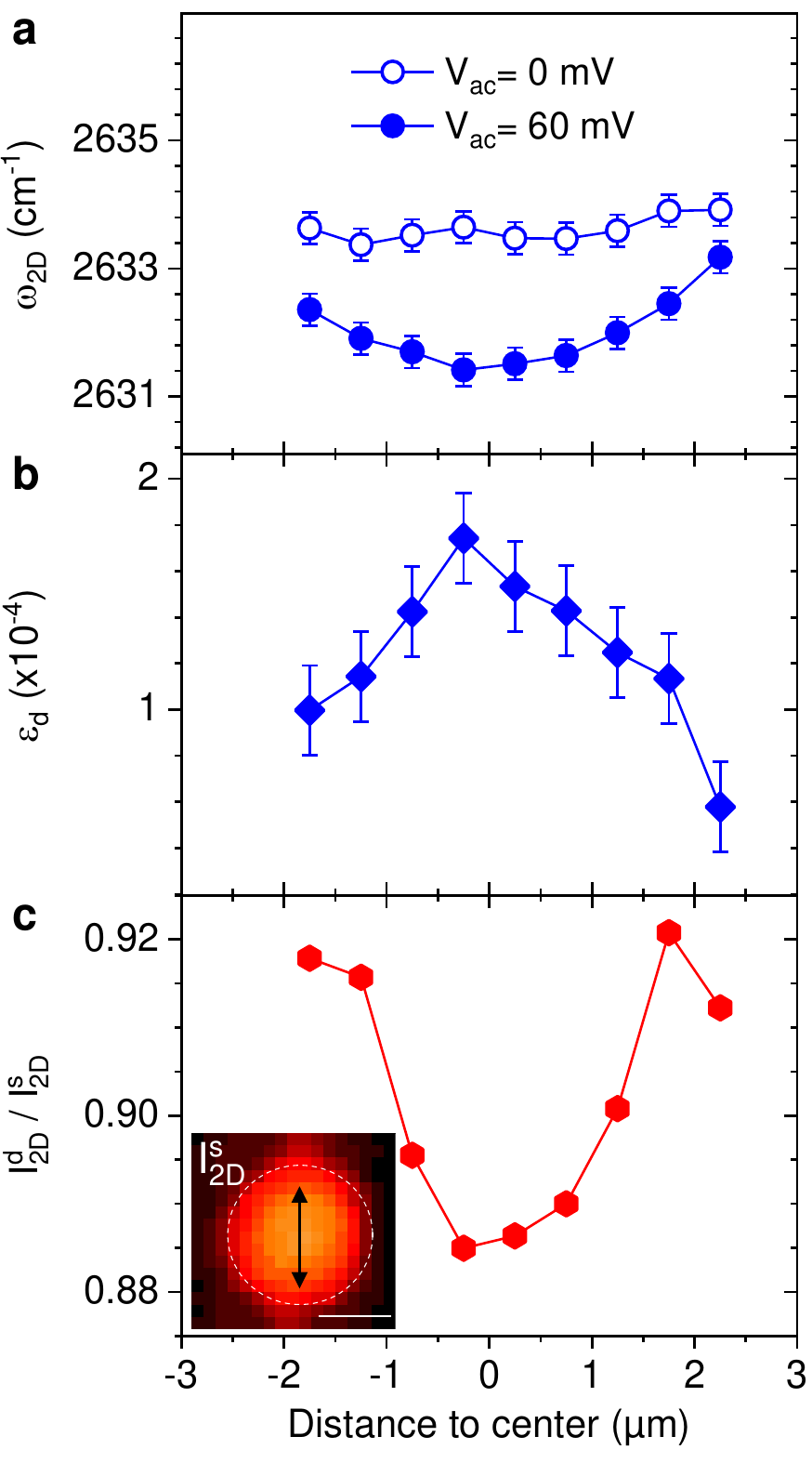}
\caption{\textbf{Mapping dynamically-induced strain.} \textbf{a}, Frequency of the Raman 2D mode along the cross-sections highlighted in \textbf{c} in a graphene drum (device 2, 3~$\mu \rm m$ radius) at $\Vdc=-6~\rm V$ and $\Vac=0~\rm {mV}$ (open symbols) and $\Vac=60~\rm {mV}$ (full symbols). \textbf{b}, Dynamical strain $\ed$ obtained from the difference of the data in \textbf{a}. \textbf{c}, Ratio of the  Raman 2D-mode intensity in the driven ($I_{2\rm D}^{\rm d}$)  and static ($I_{2\rm D}^{\rm s}$) cases. Inset: Map of the Raman 2D-mode intensity $I_{2\rm D}^{\rm s}$ recorded on the graphene drum (see white dashed contour), at $\Vdc=-6~\rm V$ and $\Vac=0~\rm V$. The double arrow indicates the location of the line scan. The scale bar is $3~\mu \rm m$. }
\label{Fig4}
\end{center}
\end{figure}


\textbf{Evidence for dynamically-induced strain --}
We therefore conclude that the tensile strain measured in device 1 is dynamically-induced (hereafter denoted $\ed$) and arises from the  time-averaged resonant vibrations of the graphene drum.  Starting from a reference recorded at $\Vdc=-8~\V$ and $\Vac=0~\mV$, $\ed$ recorded under resonant driving at $\Vac= 150 \rm {mV}$ (where $z_{\rm{rms}}\approx 9~\rm{nm}$) is as high as the static strain $\es$ induced  when ramping $\Vdc$ from $0~\V$ to $- 8~\V$ (where $\xi\approx 60~\rm {nm}$). Along these lines, the small yet observable broadenings $\Delta\Gamma_{\rm {G,2D}}$ of the Raman features (Fig.~\ref{Fig2}d) can be assigned to time-averaged Raman frequency shifts due to dynamical strain~\cite{Fandan2020}.  We have consistently observed dynamically-enhanced strain in three graphene drums with similar designs, denoted device 1,2,3. Complementary results are reported in Supplementary Note 9 for device 1 and in Supplementary Notes 10 and 11 devices 2 and 3, respectively.  In device 3, we have measured $\ed\approx 4\times 10^{-4}$ for $z_{\rm{rms}}\approx 14~\rm{nm}$.

~

\textbf{Spatially-resolved dynamically-induced strain --}
Interestingly, our diffraction-limited Raman readout enables local mapping of $\ed$. Fig.~\ref{Fig4} compares $\omega_{\rm {2D}}$ and $I_{\rm{2D}}$ recorded across the diameter of a graphene drum (device 2, similar to device 1) under $\Vdc=-6\rm V$ with and without resonant driving. Very similar results are observed when performing a line-scan along the perpendicular direction (Supplementary Note 9). In the undriven case, we find a nearly flat $\omega_{2\rm D}$ profile, which is consistent with the difficulty in resolving low-levels of static strain below $1\times10^{-4}$. In contrast, finite $\ed$ (Fig.~\ref{Fig4}b) and equilibrium position upshift (Fig.~\ref{Fig4}c) are observed at the centre of the drum, as in Fig.~\ref{Fig2} and  Fig.~\ref{Fig3}. We find that $\ed$ and the equilibrium position upshift decrease as they are probed away from the centre of the drum and the spatial profile of $\ed$ resembles the static tensile strain profile measured on bulged graphene blisters, where strain is biaxial at the centre of the drum and radial at the edges~\cite{Lee2012NL}.

~

\textbf{Dynamically-enhanced strain --}
It is instructive to compare the measured $\ed$ to $\edh=2/3 \left(z_{\rm{rms}}/a\right)^2$, with $a$ the drum radius, the time-averaged dynamically-induced strain estimated for an harmonic oscillation with RMS amplitude $z_{\rm{rms}}$ (Supplementary Note~7). For the largest $z_{\rm{rms}}\approx 9~\rm{nm}$ attained in device 1, $\edh\approx 6\times 10^{-6}$, i.e., about 40 times smaller than the measured $\ed$ (Fig.~\ref{Fig2}f and Fig.~\ref{Fig3}d). 
Under strong non-linear driving, we expect sizeable Fourier components of the mechanical amplitude at harmonics of the drive frequency, which could in part be responsible for the large discrepancy between $\ed$ and $\edh$. Harmonics are indeed observed experimentally in the displacement power spectrum of our drums (Supplementary Note 10, device 2) but display amplitudes significantly smaller than the linear component at the drive frequency. In addition, we do not observe any measurable fingerprint of internal resonances~\cite{DeAlba2016,Mathew2016,Guttinger2017} in the displacement power spectrum.

To get further insights into the unexpectedly large $\ed$ deduced from the G- and 2D-mode downshifts we plot $\ed$ as a function of the corresponding $z_{\rm{rms}}$ at the centre of the drum (Fig.~\ref{Fig2}f and Fig.~\ref{Fig3}d). This plot is directly compared to the backbone curves that connect the resonant $z_{\rm{rms}}$ to the non-linear relative resonance frequency shift $\delta=\abs{\widetilde{\Omega}_0-\Omega_0}/\Omega_0$, where $\widetilde{\Omega}_0$ is considered  equal to the measured jump-down frequency  (Fig.~\ref{Fig2}a,~\ref{Fig3}a and Supplementary Notes 6 and 7). Remarkably, $\ed$ grows proportionally to $\delta$, both in the case of non-linear softening and hardening, including when $z_{\rm{rms}}$ fully saturates (Fig.~\ref{Fig3}). This proportionality is expected from elasticity theory with a third order geometrical non-linearity~\cite{Schmid2016} and we experimentally show here that it still holds when symmetry breaking and higher-order non-linearities come into play (Supplementary Note 7).



\begin{center}
    \textbf{DISCUSSION}
\end{center}

The large values of $\ed\gg\edh$ reported in Fig.~\ref{Fig2}-\ref{Fig4} cannot be understood as a simple geometrical effect arising from the time-averaged harmonic oscillations of mode profile that remains smooth over the whole drum area.  Instead, the enhancement of $\ed$ could arise from so-called localisation of harmonics, a phenomenon recently observed in larger and thicker ($\sim500~ \mu\rm m$ wide, $\sim 500~\rm nm$ thick) SiN membranes~\cite{Yang2019} showing  RMS displacement saturation similar to Fig.~\ref{Fig3}a.  As the resonator enters the saturation regime, non-linearities (either intrinsic~\cite{Lee2008}, geometrical~\cite{Schmid2016,Cattiaux2019} or electrostatically-induced~\cite{Weber2014,Davidovikj2017,Banafsheh2017}) may lead to internal energy transfer towards harmonics of the driven mode (Supplementary Figure 17) and, crucially, to the emergence of ring-shaped patterns over length scales significantly smaller than the size of the membrane~\cite{Yang2019}. The large  displacement gradients associated with these profiles thus enhance $\ed$ (Supplementary Note 7). The mode profiles get increasingly complex as the driving force increases, explaining the rise of $\ed$ even when $z_{\rm{rms}}$ reaches a saturation plateau. Considering our study, with $\ed\sim 40 \,\edh$, we may roughly estimate that large mode profile gradients develop on a scale of $a/\sqrt{40}\approx 500~\rm{nm}$ that is smaller than our spatial resolution (see Methods). Finally, the fact that $\Delta\Gamma_{\rm {G,2D}}$ (Fig.~\ref{Fig2}d and Supplementary Figure 16) is smaller than the associated $\Delta\omega_{\mathrm{G,2D}}$ (Fig.~\ref{Fig2}c and ~\ref{Fig3}a)  suggests that the oscillations of $\ed(t)$  are rectified under strong non-linear driving, an effect that further increases the discrepancy between the time-averaged $\ed$ we measure and $\edh$. 

Combining multi-mode opto-mechanical tomography and hyperspectral Raman mapping on larger graphene drums (effectively leading to a higher spatial resolution) would allow us to test whether localisation of harmonics occurs in graphene and to possibly correlate this phenomenon to the dynamically-induced strain field. 
More generally, unravelling the origin of the anomalously large $\ed$ may require microscopic models that may go beyond elasticity theory~\cite{Atalaya2008} and  explicitly take into account the ultimate thinness and atomic structure of graphene~\cite{Ackerman2016,Kang2013}.

~


Concluding, we have unveiled efficient coupling between intrinsic microscopic degrees of freedom (here optical phonons) and macroscopic non-linear mechanical vibrations in monolayer graphene resonators. Room temperature resonant mechanical vibrations with $\approx 10~nm$ RMS amplitude induce unexpectedly large time-averaged tensile strains up to $\approx 4 \times 10^{-4}$. Realistic improvements of our setup, including phase-resolved Raman measurements~\cite{Xue2007,Pomeroy2008} could permit to probe dynamical strain in finer detail, including in the linear regime, where the effective coupling strength~\cite{Yeo2014} could be extracted. For this purpose, larger resonant displacements may be achieved at cryogenic temperatures. In addition, graphene drums, as a prototypical non-linear mechanical systems, can be engineered to favor mode coupling and frequency mixing, which in return can be readout through distinct modifications of their spatially-resolved Raman scattering response.

Our approach can be directly applied to a variety of 2D materials and related van der Waals heterostructures.  In few-layer systems, rigid layer shear and breathing Raman-active modes~\cite{Zhangx2015,Ferrari2013} could be used as invaluable probes of in-plane and out-of-plane dynamical strain, respectively. Strain-mediated coupling could also be employed to manipulate the rich excitonic manifolds in transition metal dichalcogenides~\cite{Wang2018} as well as the single photon emitters they can host~\cite{Palacios2017,Branny2017}. More broadly, light absorption and emission could be controlled electro-mechanically in nanoresonators made from custom-designed van der Waals heterostructures~\cite{Zhou2019}. Going one step further, with the emergence of 2D materials featuring robust magnetic order and topological phases~\cite{Gibertini2019}, that can be probed using optical spectroscopy, we foresee new possibilities to explore and harness phase transitions using nanomechanical resonators based on 2D materials~\cite{Siskins2020,Jiang2020}.


~

\begin{center}
    \textbf{METHODS}
\end{center}

{\bf Device fabrication --} Monolayer graphene flakes were deposited onto pre-patterned 285nm-SiO$_2$/Si substrates, using a thermally assisted mechanical exfoliation scheme as in Ref.~\onlinecite{Huang2018}.   The pattern is created by optical lithography followed by reactive ion etching and consists of hole arrays (5 and 6~$\mu\mathrm m$ in diameter and $250\pm5~\mathrm{nm}$ in depth) connected by $\sim 1\;\mu \mathrm m$-wide venting channels. Ti(3 nm)/Au(47 nm) contacts are evaporated using a transmission electron microscopy grid as a shadow mask~\cite{Metten2016} to avoid any contamination with resists and solvents. Our dry transfer method minimises rippling and crumpling effects~\cite{Nicholl2017}, resulting in graphene drums with intrinsic mechanical properties (see Ref.~\onlinecite{Metten2014} and Supplementary Note 5 for details). We could routinely obtain pristine monolayer graphene resonators with quality factors in excess of 1,500 at room temperature in high vacuum.

~

{\bf Optomechanical measurements --} Electrically connected graphene drums are mounted into a vacuum chamber (5$\times$10$^{-5}$ mbar). The drums are capacitively driven using the Si wafer as a backgate and a time-dependent gate bias $V_{\mathrm g}(t)=\Vdc+\Vac\cos{\Omega t}$ is applied as indicated in the main text. The applied force is given by $\epsilon_0 \pi a^2 \frac{V_{\mathrm g}^2\left(t\right)}{2d^2\left(\xi\right)}$, where $a$ is the drum radius, $\epsilon_0$ the vacuum dielectric constant, $d\left(\xi\right)=(d_{\mathrm{vac}}-\xi)+d_{\mathrm{SiO}_2}/\epsilon_{\mathrm{SiO}_2}$ the effective distance between graphene and the Si substrate, with $\xi$ the static displacement, $d_{\rm{vac}}$ the graphene-SiO$_2$ distance in the absence of any gate bias, $d_{\mathrm{SiO}_2}$ the thickness of the residual SiO$_2$ layer. This force contains a static component proportional to $\Vdc^2$, which sets the value of $\xi$ and a harmonic driving force proportional to $\Vdc\Vac \cos\left(\Omega t\right)$.  Note that since $\Vac \ll \left| \Vdc \right|$, we can safely neglect the force $\propto \Vac^2 \left(1+\cos\left(2\,\Omega t\right)\right)$ throughout our analysis.

A 632.8~nm HeNe continuous wave laser with a power of $\sim0.5~\rm {mW}$ is focused onto a $\sim~1.2~\mu \mathrm m$-diameter spot and is used both for optomechanical and Raman measurements.  Unless otherwise stated, (e.g., insets in Fig.\ref{Fig1}b and Fig.~\ref{Fig4}), measurements are performed at the centre of the drum.
The beam reflected from the Si/SiO$_2$/vacuum/graphene layered system is detected using an avalanche photodiode. In the driven regime, the mechanical amplitude at $\Omega/2\pi$ is readout using a lock-in amplifier. Mechanical mode mapping is implemented using a piezo scanner and a phase-locked loop. For amplitude calibration, the thermal noise spectrum is derived from the noise power spectral density of the laser beam reflected by the sample, recorded using a spectrum analyzer. Importantly, displacement calibration is performed assuming that the effective mass of our circular drums is $\widetilde{m}=0.27\,m_0$ (Ref.~\onlinecite{Hauer2013}), with $m_0$ the pristine mass of the graphene drum. As discussed in details in Supplementary Note 5, this assumption is validated by two other displacement calibration methods performed on a same drum. These calibrations are completely independent of $\widetilde{m}$. We therefore conclude that to experimental accuracy, our graphene drums are pristine and do not show measurable fingerprints of contamination by molecular adsorbates~\cite{Berciaud2009}, as expected for a resist-free fabrication process.
 
 ~

{\bf Micro-Raman spectroscopy --}  The Raman scattered light is filtered using a combination of a dichroic mirror and a notch filter. Raman spectra are recorded using a 500~mm monochromator equipped with 300 and 900 grooves/mm gratings, coupled to a cooled CCD array. In addition to electrostatically-induced strain, electrostatically-induced doping might in principle alter the Raman features of suspended graphene~\cite{Metten2016}. Pristine suspended graphene, as used here, is well-known to have minimal unintentional doping ($\lesssim 10^{11}~\mathrm{cm^{-2}}$) and charge inhomogeneity~\cite{Berciaud2009,Berciaud2013}. Considering our experimental geometry, we estimate a gate-induced induced doping level near $3 \times 10^{11}~\rm{cm^{-2}}$ at the largest $\abs{\Vdc}=10~\rm V$ applied here. Such doping levels are too small to induce any sizeable shift of the G- and 2D-mode features ~\cite{Pisana2007,Berciaud2009,Metten2016}. In the dynamical regime, the RMS modulation of the doping level induced by the application of $\Vac$ is typically two orders of magnitude smaller than the static doping level and can safely be neglected. Similarly, the reduction of the gate capacitance induced by equilibrium position upshifts discussed in Fig.~\ref{Fig2}e and Fig.~\ref{Fig3}a-iv do not induce measurable fingerprints of reduced doping on graphene.

Let us note that since the lifetime of optical phonons in graphene ($\sim 1\ \mathrm{ps}$)~\cite{Bonini2007} is more than three orders of magnitude shorter than the mechanical oscillation period, Raman scattering processes provide an instantaneous measurement of $\ed$. However, since our Raman measurements are performed under continuous wave laser illumination, we are dealing with time-averaged dynamical shifts and broadenings of the G-mode and 2D-mode features.  Raman G- and 2D-mode spectra are fit using one Lorentzian and two modified Lorentzian functions, as in Ref.~\onlinecite{Berciaud2013,Metten2014}, respectively (Supplementary Note 1). As indicated in the main text, Gr\"uneisen parameters of $\gamma_{\rm G}=1.8$ and $\gamma_{2\rm D}=2.4$ are used to estimate $\es$ and $\ed$. These values have been measured  in circular suspended graphene blisters under biaxial strain~\cite{Metten2014}. Considering a number of similar studies~\cite{Zabel2012a,Lee2012NL,Metten2014,Metten2016,Androulidakis2015}, we conservatively estimate that the values of $\es$ and $\ed$ are determined with a systematic error lower than $20~\%$. Such systematic errors have no impact whatsoever on our demonstration of dynamically-enhanced strain. Finally, the Raman frequencies and the associated $\es$ and $\ed$ are determined with fitting uncertainties represented by the errorbars in the figures.


~



~


%




~

{\bf Acknowledgements}

We thank T. Chen, A. Gloppe and G. Weick for fruitful discussions. We thank the StNano clean room staff (R. Bernard and S. Siegwald), M. Romeo, F. Chevrier, A. Boulard and the IPCMS workshop for technical support. This work has benefitted from support provided by the University of Strasbourg Institute for Advanced Study (USIAS) for a Fellowship, within the French national programme “Investment for the future” (IdEx-Unistra). We acknowledge financial support from the Agence Nationale de Recherche (ANR) under grants H2DH ANR-15-CE24-0016, 2D-POEM ANR-18-ERC1-0009, as well as the Labex NIE project ANR-11-LABX-0058-NIE.

~

\textbf{Author contributions}

 The project was originally proposed by S.B and P.V (GOLEM project, supported by USIAS). K.M. and D.M. built the experimental setup, with help from X.Z. X.Z. fabricated the samples, with help from H.M., D.M. and K.M. X.Z. carried out measurements, with help from K.M. and L.C. X.Z. and S.B. analysed the data with input from K.M., L.C., and P.V. X.Z. and S.B. wrote the manuscript with input from L.C. and P.V. S.B. supervised  the project.

~



~

{\bf Competing interests}

The authors declare no competing interests.

%


\onecolumngrid
\newpage
\begin{center}
{\Large\textbf{Supplementary Information for: \\ Dynamically-enhanced strain in atomically-thin resonators}}
\end{center}

\setcounter{equation}{0}
\setcounter{figure}{0}
\setcounter{section}{0}
\renewcommand{\thetable}{S\arabic{table}}
\renewcommand{\theequation}{S\arabic{equation}}
\renewcommand{\thefigure}{S\arabic{figure}}
\renewcommand{\thesection}{S\arabic{section}}
\renewcommand{\thesubsection}{S\arabic{section}\alph{subsection}}
\renewcommand{\thesubsubsection}{S\arabic{section}\alph{subsection}\arabic{subsubsection}}

\linespread{1.4}\selectfont

\bigskip

\tableofcontents

\clearpage

\clearpage

This Supplementary Information file is organised as follows. In \ref{Raman}, we provide details on the Raman scattering response of graphene and on our fitting procedure. In \ref{S1} and  \ref{S2}, we outline the sample design and discuss an elementary mechanical model, respectively, before discussing, in  \ref{SecInter}, how optical interference effects allow estimating the static displacement $\xi$ of a graphene drum and the static strain $\es$ it undergoes.
In \ref{Sec_cal}, we present a comprehensive displacement calibration scheme using three different methods that yield a consistent and accurate determination of the root mean square (RMS) displacement $z_{\rm{rms}}$ in the driven regime. These results also allow us to conclude that, within experimental accuracy, the effective mass of our drum is that of a pristine graphene monolayer. In \ref{SMech}, we present a basic modelling of the mechanical response of graphene both in the linear and non-linear regime, followed by a discussion on the links between dynamical strain and non-linearities in \ref{StrainNL}. Laser-induced heating effects are addressed in \ref{Sec_heat}. Finally, supplementary data on devices 1, 2 and 3 are presented in Supplementary Notes 9, 10 and 11, respectively. This material complements and/or bolster the data shown in the main text. Devices 1, 2 and 3 have similar designs.

\section{R\lowercase{aman scattering in graphene}}
\label{Raman}

\subsection*{The G mode and the 2D mode}

As introduced in the main text, our study focuses on the well-documented G mode and 2D modes in graphene~\cite{Ferrari2013}. Simplified sketches of the G- and 2D-mode processes are shown in Supplementary Fig.~\ref{FigS2}.  The G mode is a one phonon non-resonant process originating from in-plane (LO and TO) zero momentum optical phonons, that is at the centre ($\Gamma$ point) of the Brillouin zone. The G-mode feature is commonly described as a single, quasi-Lorentzian feature~\cite{Froehlicher2015}. The 2D-mode is a resonant, symmetry allowed two-phonon process involving a pair of near-zone edge TO phonons near the edges of the Brillouin zone (K and K$^\prime$ points)\cite{Maultzsch2004,Basko2008,Venezuela2011}. This 2D-mode frequency depends both on the electronic and phononic dispersion and hence on the incoming laser photon energy. The 2D mode-lineshape is \textit{a priori} very complex~\cite{Basko2008}. In the case of suspended graphene, this lineshape is phenomenologically fit to the sum of two modified Lorentzian profiles, as in Ref.~\onlinecite{Berciaud2013}.

\begin{figure*}[!htb]
\begin{center}
\includegraphics[width=12cm]{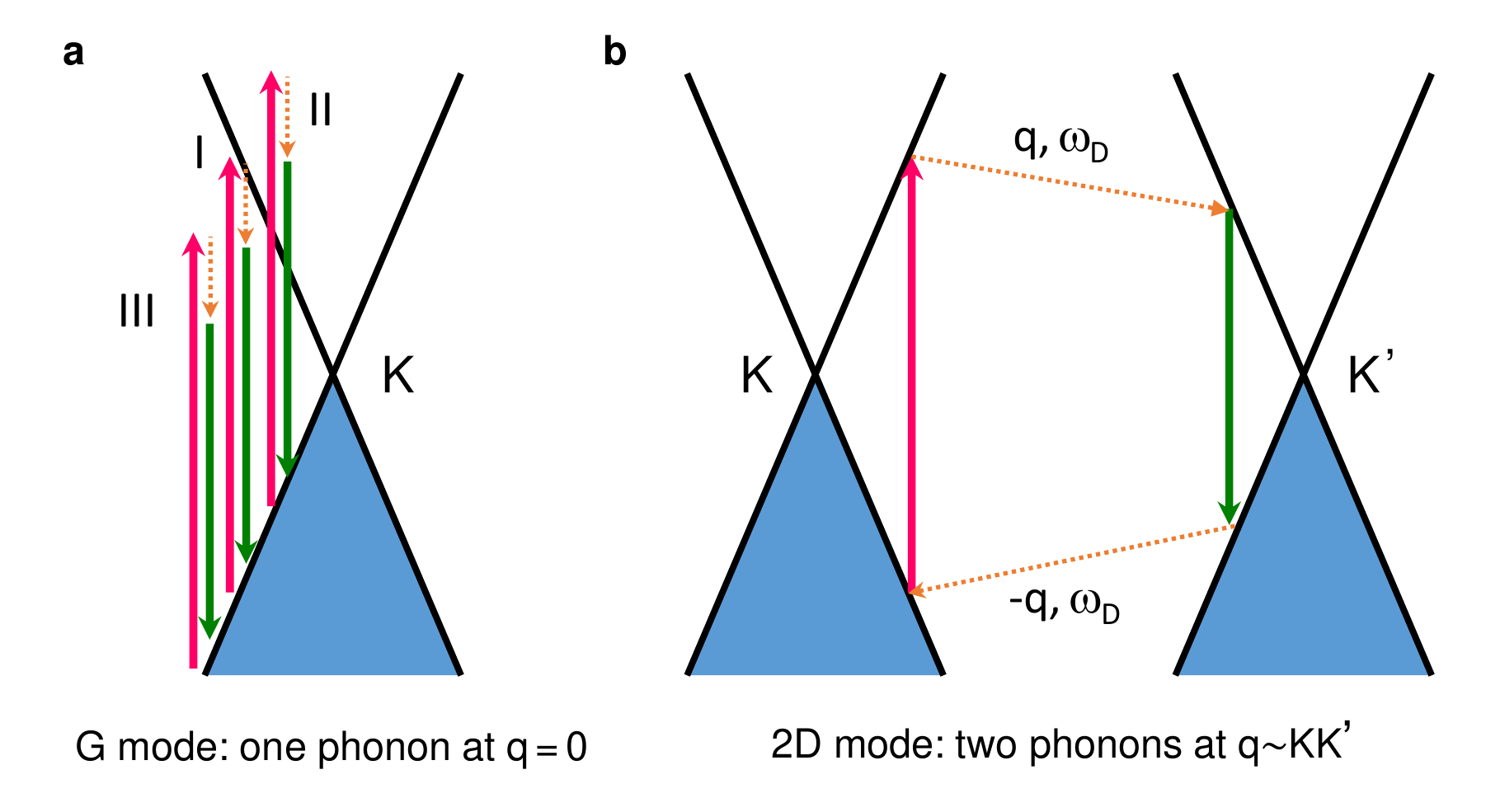}
\caption{ \textbf{Raman scattering processes in graphene.} The pink, green and dashed orange arrows in {\bf a} and {\bf b} indicate incoming photons, scattered photons and scattered phonons, respectively. The G mode ({\bf a}) is a one-phonon process involving zone-center optical phonons (LO and TO)~\cite{Ferrari2013}. Although resonant processes (\upperRomannumeral{1}) may contribute to the G-mode intensity, the G-mode feature arises for the most part from the quantum interference between non-resonant processes (\upperRomannumeral{2}, \upperRomannumeral{3}) across the whole Brillouin zone~\cite{Chen2011}. The 2D mode ({\bf b}) is a resonant inter-valley process involving a pair of near zone-edge TO phonons with opposite momenta $\pm q$. Here, for clarity, we only represent the so-called inner process involving phonons with momenta smaller than $\rm{KK^{\prime}}$ (Ref.~\onlinecite{Venezuela2011,Berciaud2013}).}
\label{FigS2}
\end{center}
\end{figure*}

\subsection*{Fitting the Raman 2D-mode spectra}
\label{Sec2D}
As discussed above and in Ref.~\onlinecite{Berciaud2013}, the 2D-mode lineshape in suspended graphene is asymmetric and best fit with the sum of two modified Lorentzian profiles, as exemplified in Supplementary Fig.~\ref{FigS3} and in Fig.~2b and~3b.  The  2D-mode frequency $\omega_{\rm{2D}}$ discussed in the main manuscript refers to the more intense 2D$^{-}$ sub-feature unless otherwise specified (Supplementary Fig.~\ref{FigS16}), while the 2D-mode intensity $I_{\rm{2D}}$ refers to the \textit{total} integrated intensity of both 2D$^{-}$ and 2D$^{+}$ sub-features. As we show in Supplementary Fig.~\ref{FigS16}, both low- and high-frequency 2D-mode sub-features are similarly affected in the driven regime.

\begin{figure*}[!htb]
\begin{center}
\includegraphics[width=12cm]{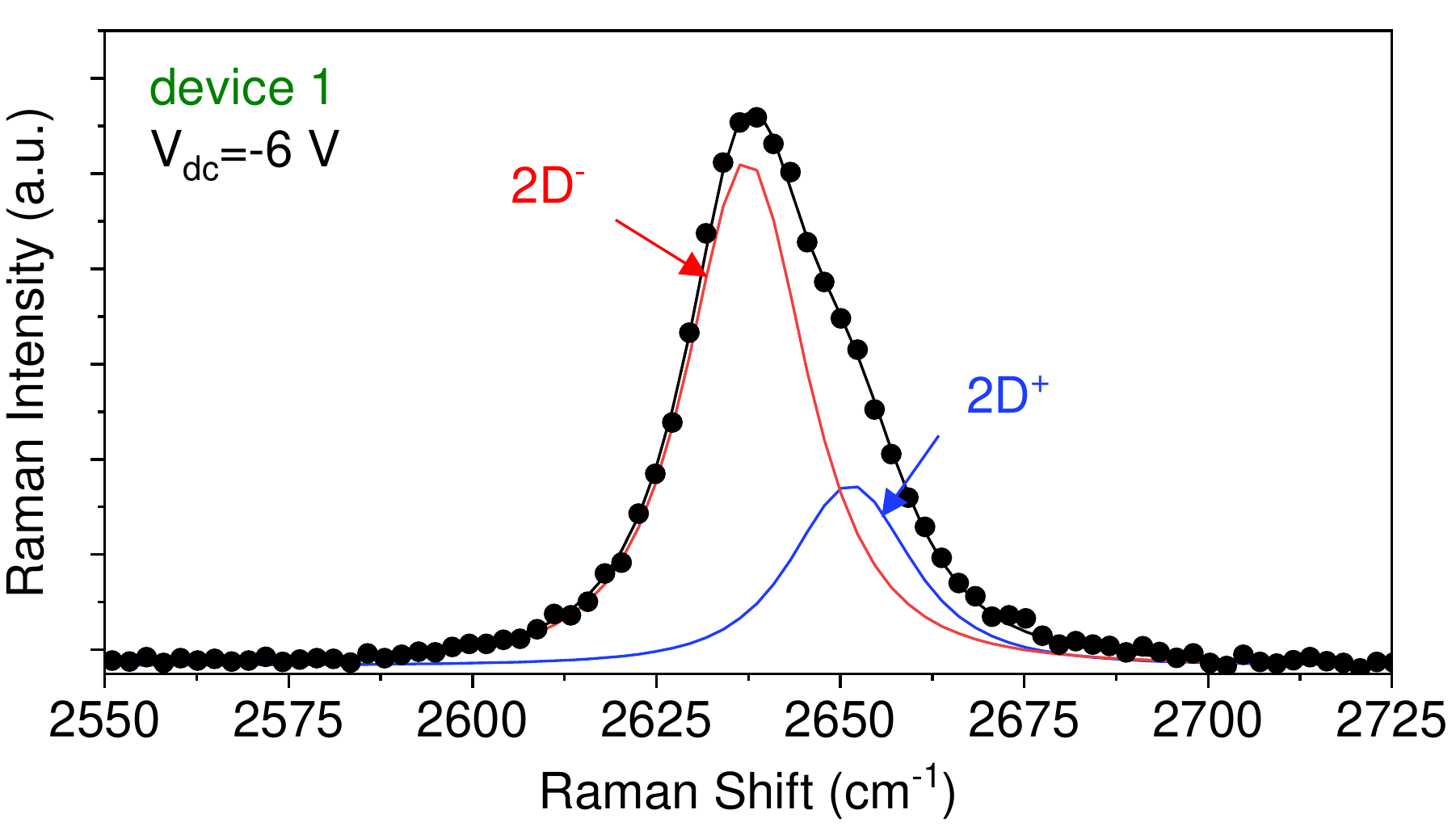}
\caption{\textbf{Fitting the Raman 2D-mode spectra in a suspended graphene drum}. Modified Lorentzian fit of the 2D-mode feature in suspended graphene using two sub-features, denoted 2D$^{-}$ and 2D$^{+}$, as in Ref.~\onlinecite{Berciaud2013}.}
\label{FigS3}
\end{center}
\end{figure*}

\begin{figure*}[!htb]
\begin{center}
\includegraphics[width=14cm]{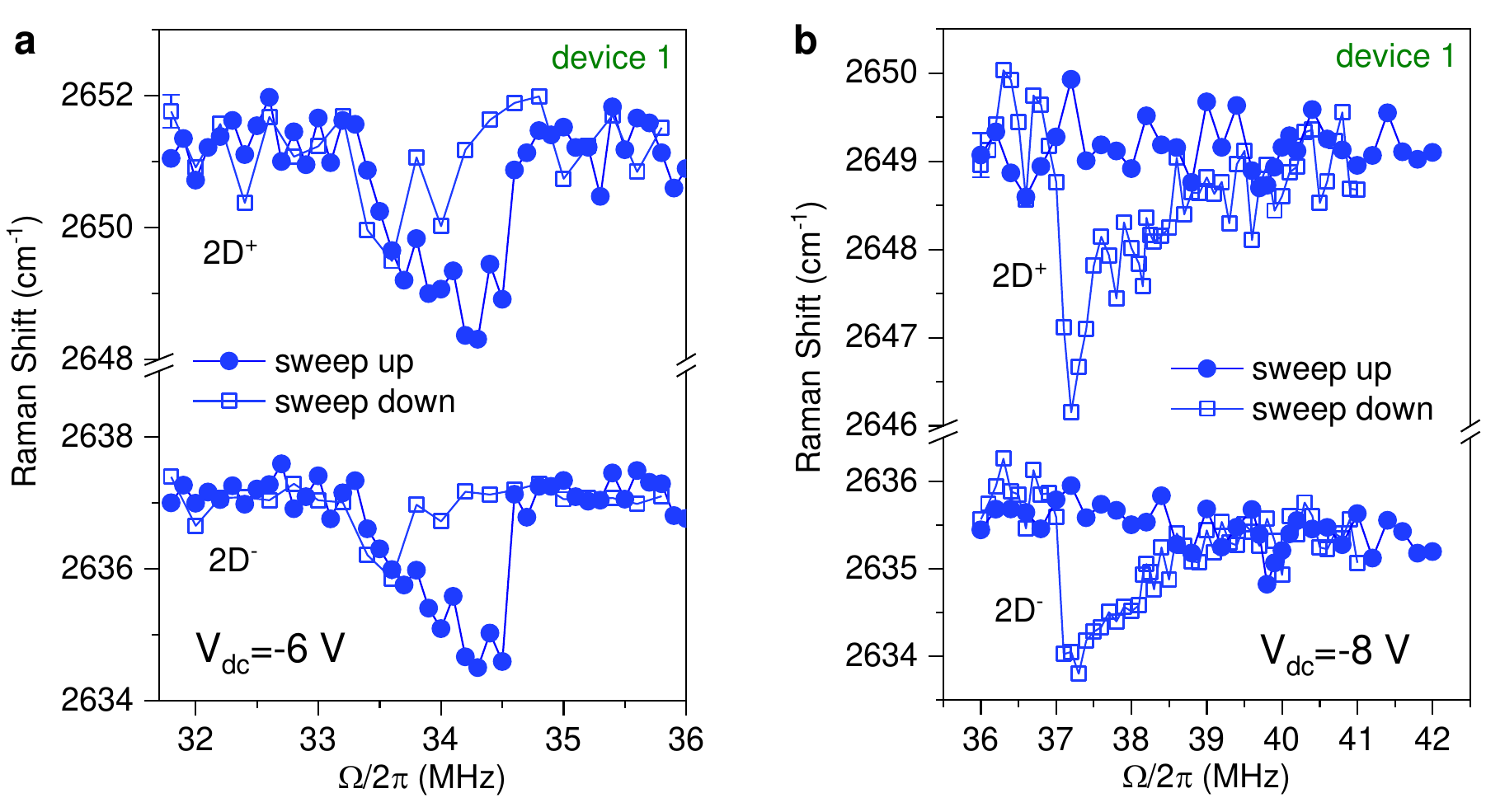}
\caption{\textbf{Fitting the Raman 2D-mode spectra in a resonantly-driven graphene drum}. Frequency of the two 2D mode subfeatures ($\omega_{2\rm D^{+}}$ and $\omega_{2\rm D^{-}}$) as a function of the drive frequency $\Omega/2\pi$ for both upward (circles) and downward (squares) drive frequency sweeps at $\Vdc=-8~\rm V$ ({\bf a}) and $\Vdc=-6~\rm V$ ({\bf b}) in device 1 (see also  Supplementary Fig.~\ref{FigS10} and Fig.~3, respectively). Only one error bar is included in each plot for clarity.}
\label{FigS16}
\end{center}
\end{figure*}

\begin{figure*}[!htb]
\begin{center}
\includegraphics[width=9cm]{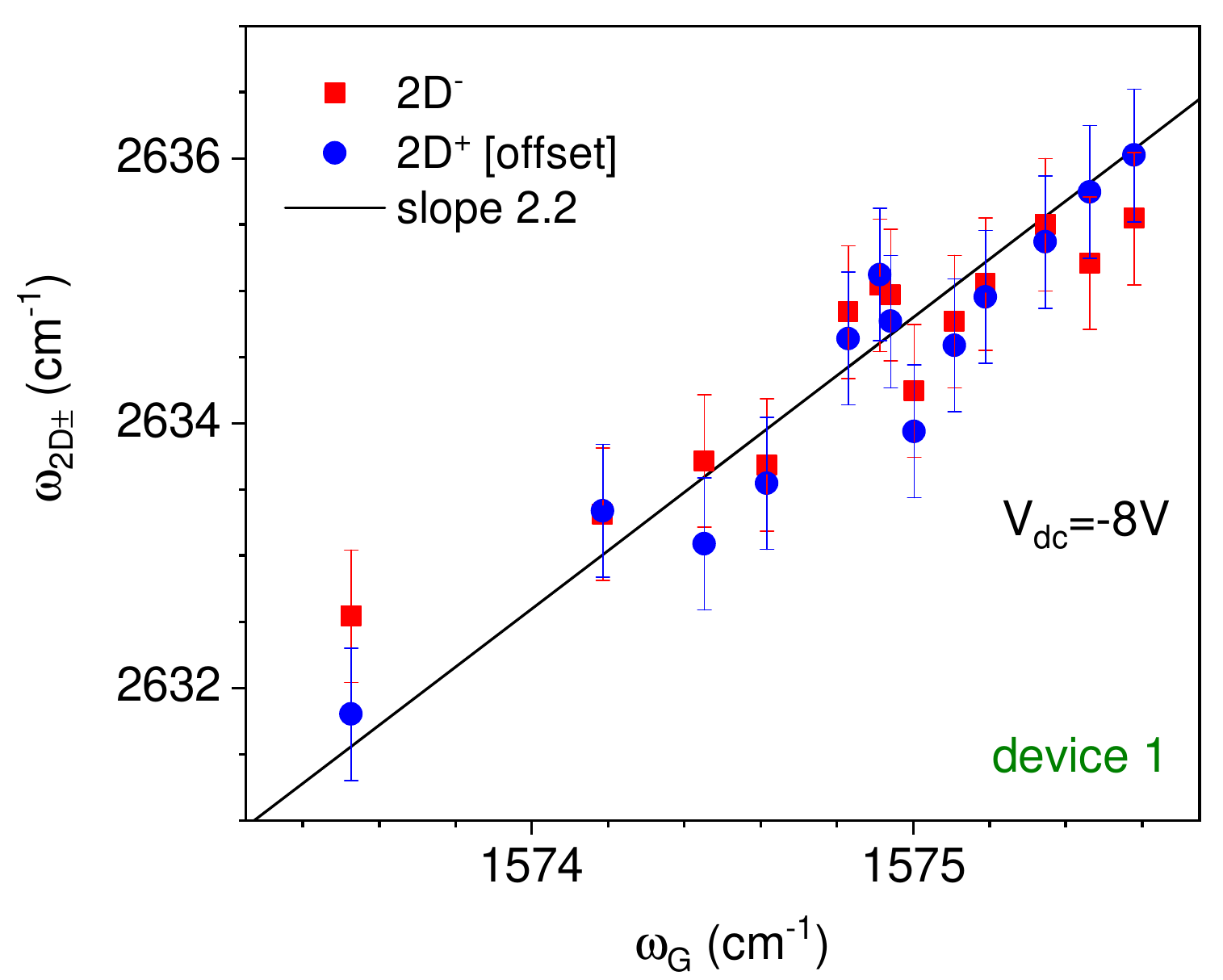}
\caption{\textbf{Correlation between the G- and 2D-mode frequencies}. This correlation plot is similar to Fig. 2c, except  that the two components ($2\rm D^{\pm}$) are shown. The $2\rm D^+$ component is  offset by $-13.7~\rm{cm}^{-1}$ for a clearer comparison. The straight black line with a slope of 2.2 is a guide to the eye showing the expected correlation for strain-induced phonon softening. The 2D$^-$ feature deviates slightly from this slope at large drive while the 2D$^+$ feature remains closer to guide to the eye. In spite of these slight deviations, the slopes $\frac{\partial\omega_{\rm 2D^{\pm}}}{\partial\omega_{\rm G}}$ remain close to the value expected under biaxial strain.}
\label{FigScorrG2Dpm}
\end{center}
\end{figure*}

\clearpage

\section{S\lowercase{ample design and interference effects}}
\label{S1}

\begin{figure*}[!htb]
\begin{center}
\includegraphics[width=14cm]{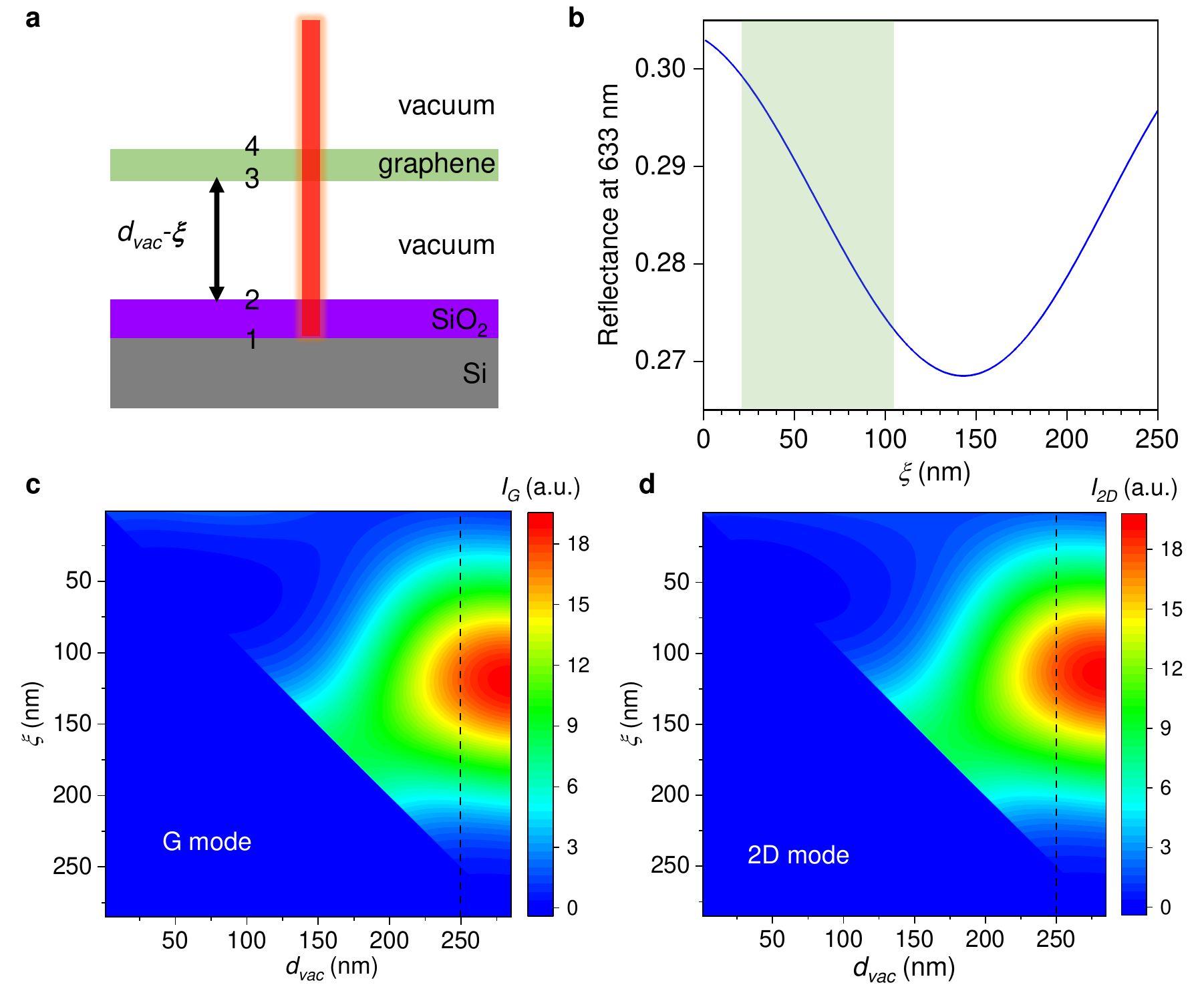}
\caption{\textbf{Sample geometry and interference effects.} {\bf a,} Multilayer model for our devices, where the labels $1-4$ represent the interfaces in the system. $d_{\rm{vac}}$-$\xi$ is the gap between suspended graphene, displaced by $\xi$ using a DC gate bias, and the SiO$_2$ surface. {\bf b,} Calculated reflectance as a function of $\xi$ in the case of a monolayer graphene for $d_{\rm{vac}}=250~\rm{nm}$ and a laser wavelength of 632.8~nm. The light-green area denotes the linear region, where a large and constant transduction coefficient allows interferometric readout of the mechanical vibrations. At a small $\xi$, the reflectance is  close to a maximum, resulting in a sharp decrease of the transduction coefficient. {\bf c,d,} Contour plots of the G- and 2D-mode intensity enhancement factors (\ref{SecInter}) as a function of $d_{\rm{vac}}$ and $\xi$  under optical excitation at 632.8~nm. The SiO$_2$ thickness is 285~nm. The black dashed lines highlight the results at $d_{\rm{vac}}=250~\rm{nm}$.}
\label{FigS1}
\end{center}
\end{figure*}

Figure \ref{FigS1}a shows the vacuum/graphene/vacuum/SiO$_2$/Si multilayered system discussed in the main text. Due to optical interference effects, the reflectance and Raman scattering intensity depend on the laser wavelength, hole depth ($d_{\rm {vac}}$) and residual SiO$_2$ thickness ($d_{\rm{SiO}_2}$). Starting from a given sample geometry, we have used well-established models to compute the interference enhancement factors allowing to quantitatively predict the dependence of the sample reflectance~\cite{Blake2007,Davidovikj2017} and Raman scattered intensity~\cite{Yoon2009,Metten2014,Metten2016}  as a function of the deflection of the graphene layer, denoted $\xi$. As we shall see in \ref{SecInter} and \ref{Sec_cal}, this modelling will allow us to accurately determine $\xi$ in graphene drums and to calibrate displacements in the driven regime.

We have optimized the sample geometry to provide both large transduction coefficient for displacement readout (see Methods) and sufficiently intense Raman scattering signal. First, 285nm-SiO$_2$/Si ($p$-doped) substrates are chosen to easily locate monolayer graphene flakes by optical microscopy~\cite{Blake2007}. With $d_{\rm{vac}}=250\pm5~\rm{nm}$ (correspondingly, $d_{\rm{SiO}_2}$=35 nm), the optical reflectance varies quasi-linearly with the static deflection of the membrane $\xi$ over the range $\xi$=30-100~nm, ensuring a constant transduction coefficient for optical readout of the root mean square (RMS) mechanical displacement around an equilibrium position (Supplementary Fig.~\ref{FigS1}b). At the same time, optical interferences lead to large enough Raman intensities, as shown in the calculated Raman enhancement factors~\cite{Yoon2009,Metten2014,Metten2016} in Supplementary Fig.~\ref{FigS1}c,d. Third, the hole diameters $2a=5~\mu\rm m$ and $6~\mu \rm m$, are chosen such that the resonance frequency of the fundamental flexural mode (\ref{SMech}) lies within the $50~\rm{MHz}$ bandwidth of our detection setup.

\clearpage

 \section{E\lowercase{lementary modelling of static strain}}
\label{S2}

Given the radial symmetry of our system, we will consider, for the sake of simplicity, a one-dimensional model system, of a doubly clamped beam (in the membrane limit)  with cross-sectional area $A$ and length $L=2a$. We denote $x$ the longitudinal coordinate, with $x=0$ corresponding to the middle of the beam. This model can be generalized to the case of a circular membrane of radius $a$ as in Ref.~\onlinecite{Cattiaux2019}. We assume that under an electrostatic pressure (here, a finite gate bias $\Vdc$), the membrane adopts a parabolic profile~\cite{Koenig2011,Metten2016}. The downward deflection $\xi(x)$ thus writes:

\begin{equation}
\xi(x)=\xi(0)\left(1-\frac{x^2}{a^2}\right),
\end{equation} 
where $\xi(0)$ is the static deflection at the membrane's center ($x=0$).

The  elongation $\Delta L$ is:
\begin{equation}
\Delta L=\int_{-a}^{a}\sqrt{1+[\xi^{\prime}(x)]^2}dx -2a
\label{eq2}
\end{equation} 
For small deflections, $i.e.$, $\xi(x) \ll a$, the static strain $\varepsilon_{\rm s}$ writes:

\begin{equation}
\varepsilon_{\rm s}=\frac{\Delta L}{2a}=\frac{2}{3}\left(\frac{\xi}{a}\right)^2.
\label{eq3}
\end{equation} 
In Eq.~\eqref{eq3} and in the following, $\xi(0)$ will be denoted $\xi$ for simplicity.

Besides, under biaxial strain, the Raman frequency shift ($\Delta \omega_{i}$, with $i = \rm G,~\rm {2D}$) relative to the unperturbed values $\omega_{i,0}$ are linked to $\varepsilon_{\rm s}$  by~\cite{Metten2014,Androulidakis2015}:
\begin{equation}
\Delta \omega_{i}=2\gamma_i \  \varepsilon_{\rm s}\  \omega_{i,0} 
\label{eq4}
\end{equation} 
with the Gr\"uneisen parameters $\gamma_{\rm{G}}=1.8$ and $\gamma_{\rm{2D}}=2.4$, as measured in similar circular graphene drums~\cite{Metten2014,Androulidakis2015}.  Eq.~\eqref{eq3} and \eqref{eq4} are combined to estimate $\xi$. Supplementary Fig.~\ref{FigS5} shows $\varepsilon_{\rm s}$ and $\Delta \omega_{\rm{2D}}$ as a function of $\xi$ for $a=3 ~\mu \rm m$. By comparing to the experimental data recorded on device 1 (Supplementary Fig.~\ref{FigS4} and Fig.~1d in the main text), we estimate $\xi\approx~42~\rm{nm}$ and $\xi\approx~63~\rm{nm}$ for $\Vdc= -6~\rm V$ and $\Vdc= -8~\rm V$, respectively. 

Starting from the estimated $\xi$ based on the $\Vdc$-dependent Raman mode frequencies, we can further cross-check our calibration by another method based on the dependence of Raman intensities ($I_{\rm G}$, $I_{\rm{2D}}$) on $\xi$ (\ref{SecInter} and Supplementary Fig.~\ref{FigS1}c,d  and \ref{FigS6}). The very good match between experimentally measured $I_{\rm G}$, $I_{\rm{2D}}$, their ratio ($I_{\rm{2D}}/I_{\rm G}$) and calculations based on an optical interference model~\cite{Yoon2009,Metten2014,Metten2016} allows us to further validate our calibration of $\xi$.




From Eq.~\eqref{eq3}, the strain sensitivity can be obtained:
\begin{equation}
\frac{\partial\varepsilon_{\rm s}}{\partial \xi}=\frac{4\xi}{3a^2}.
\end{equation} To obtain a larger sensitivity towards strain, dynamical Raman measurements were performed at sufficiently large $\Vdc$ to yield sizeable $\xi$, while at the same maintainting the graphene drum at reasonable distance ($\gtrsim 200~\rm{nm}$) from the Si/SiO$_2$ substrate and avoiding sample collapse and limiting electrostatic non-linearities~\cite{Davidovikj2017}.

\clearpage


\section{S\lowercase{tatic displacement and equilibrium position shift}}
\label{SecInter}

\subsection*{Determination of the static displacement}
\label{Secxi}
\begin{figure*}[!htb]
\begin{center}
\includegraphics[width=15.5cm]{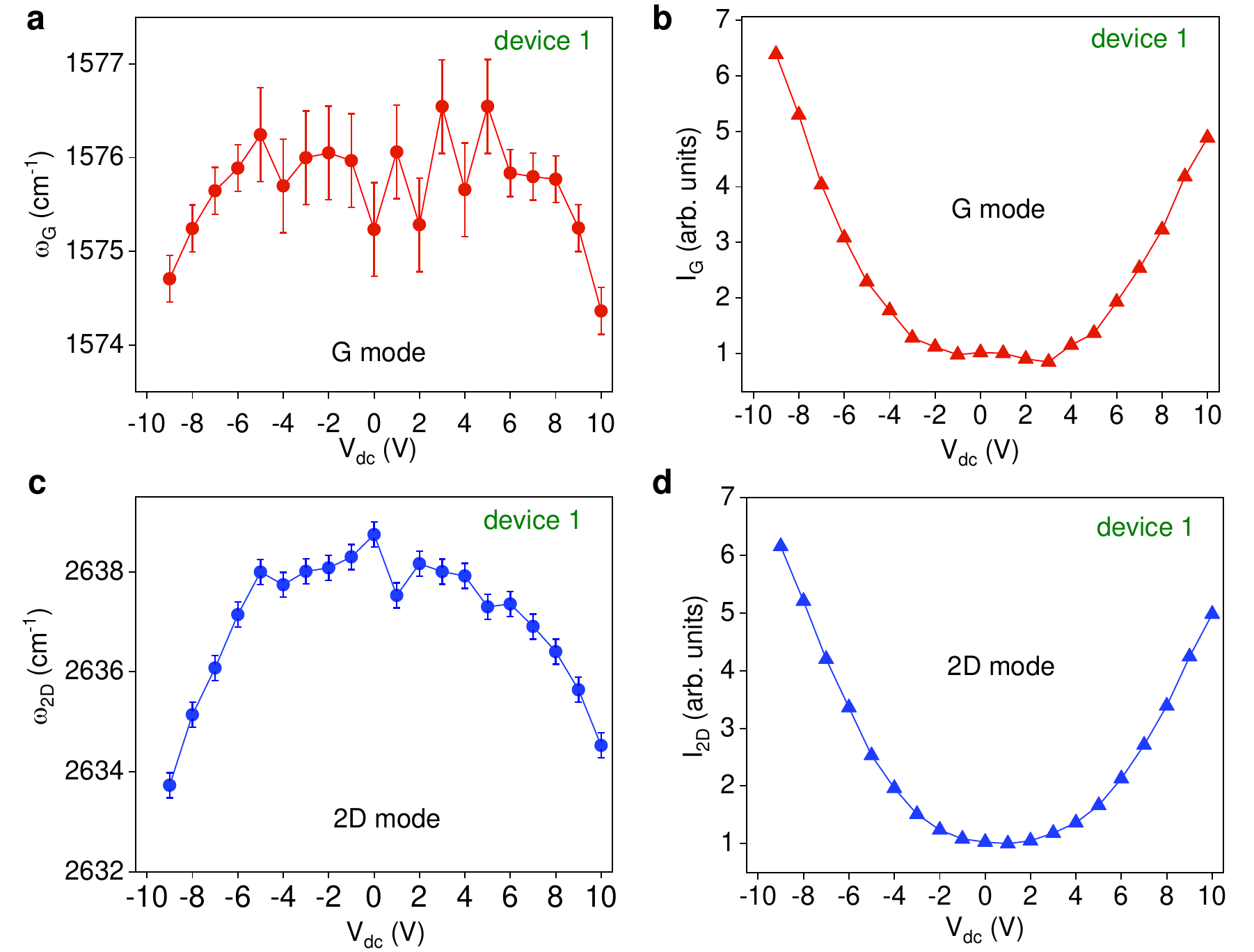}
\caption{\textbf{Probing static strain in an electrostatically gated graphene drum.} Frequency and integrated intensity of the G- ({\bf a} and {\bf b}, respectively) and 2D-mode ({\bf c} and {\bf d}, respectively) features as a function of $\Vdc$ in device 1 (see also Fig.~1d in the main manuscript for selected raw spectra). The integrated intensities are normalized with respect to the values measured at $\Vdc=1~\rm V$.}
\label{FigS4}
\end{center}
\end{figure*}

\begin{figure*}[!htb]
\begin{center}
\includegraphics[width=10cm]{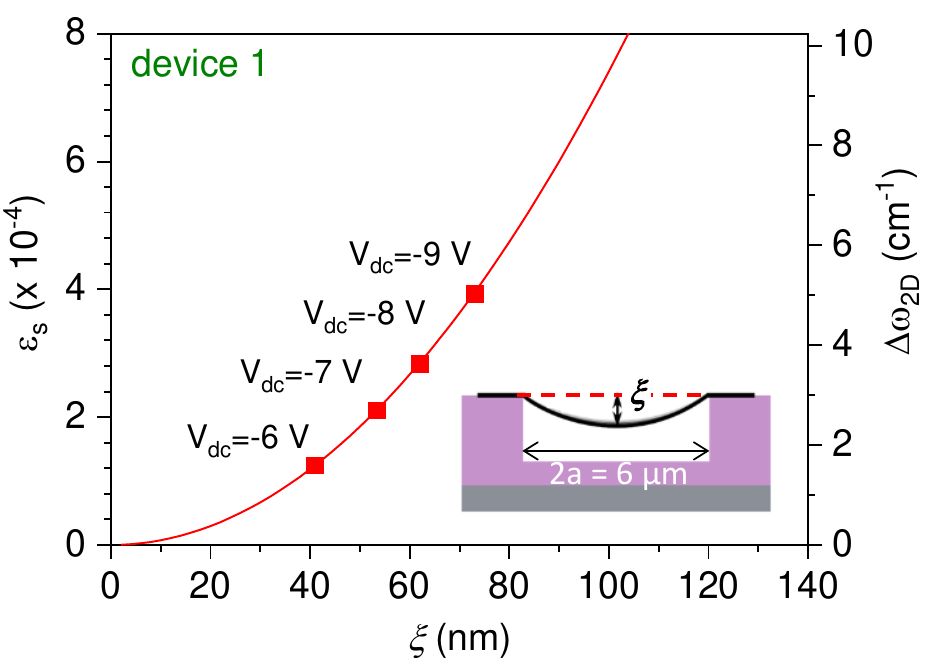}
\caption{\textbf{Determination of the static tensile strain and static deflection.} Calculated static strain $\varepsilon_{\rm s}$ using Eq.~\eqref{eq3} and corresponding frequency softening (\ref{S2}) of the 2D mode $\Delta\omega_{2\rm D}$ as a function of $\xi$ for a hole diameter $2a= 6~\mu \rm m$. The red symbols show $\Delta\omega_{2\rm D}$ (from Supplementary Fig.~\ref{FigS4}c) and the estimated $\xi$ for $\Vdc$ ranging from -5~V to -8~V. The device geometry is recalled as an inset.}
\label{FigS5}
\end{center}
\end{figure*}

 Using an multiple reflection model as in Ref.~\onlinecite{Yoon2009,Metten2014,Metten2016}, the intensity enhancement factors of the  G- and 2D-mode  features can be calculated as a function of $\xi$ for $d_{\rm{Si0}_2}=35~\rm{nm}$, $d_{\rm{vac}}=250~\rm{nm}$,  and a laser wavelength of 632.8~nm (Supplementary Fig.~\ref{FigS1}c,d). Both $I_{\rm G}$ and $I_{\rm{2D}}$ monotonically increase with $\xi$ for $\xi \lesssim 125~\rm{nm}$, above which they monotonically decrease after reaching an intensity maximum. In particular, $I_{\rm G} (\xi)$ and $I_{\rm {2D}} (\xi)$ can be approximated as linear in the range $\xi=20-70~\rm {nm}$ (Supplementary Fig.~\ref{FigS6}), which corresponds to the static displacements explored in our study. The data points in Supplementary Fig.~\ref{FigS6}a represent the equilibrium deflection ($\xi$) obtained at various $\Vdc$ and extracted from the measured Raman G- and 2D-mode frequencies (\ref{S2} and Supplementary Fig.~\ref{FigS4}-\ref{FigS5}). We can see that using an appropriate scaling factor that essentially accounts for the Raman susceptibilities of the G- and 2D-modes in the ``interference-free'' case~\cite{Metten2015,Metten2016}, the intensity ratio $I_{2\rm{D}}/I_{\rm{G}}$ matches very well with theoretical predictions (Supplementary Fig.~\ref{FigS6}b). This agreement validates our strain-based estimation of $\xi$ and provides a solid ground to calibrate the RMS displacements (\ref{Sec_cal}).

%

\begin{figure*}[!htb]
\begin{center}
\includegraphics[width=12cm]{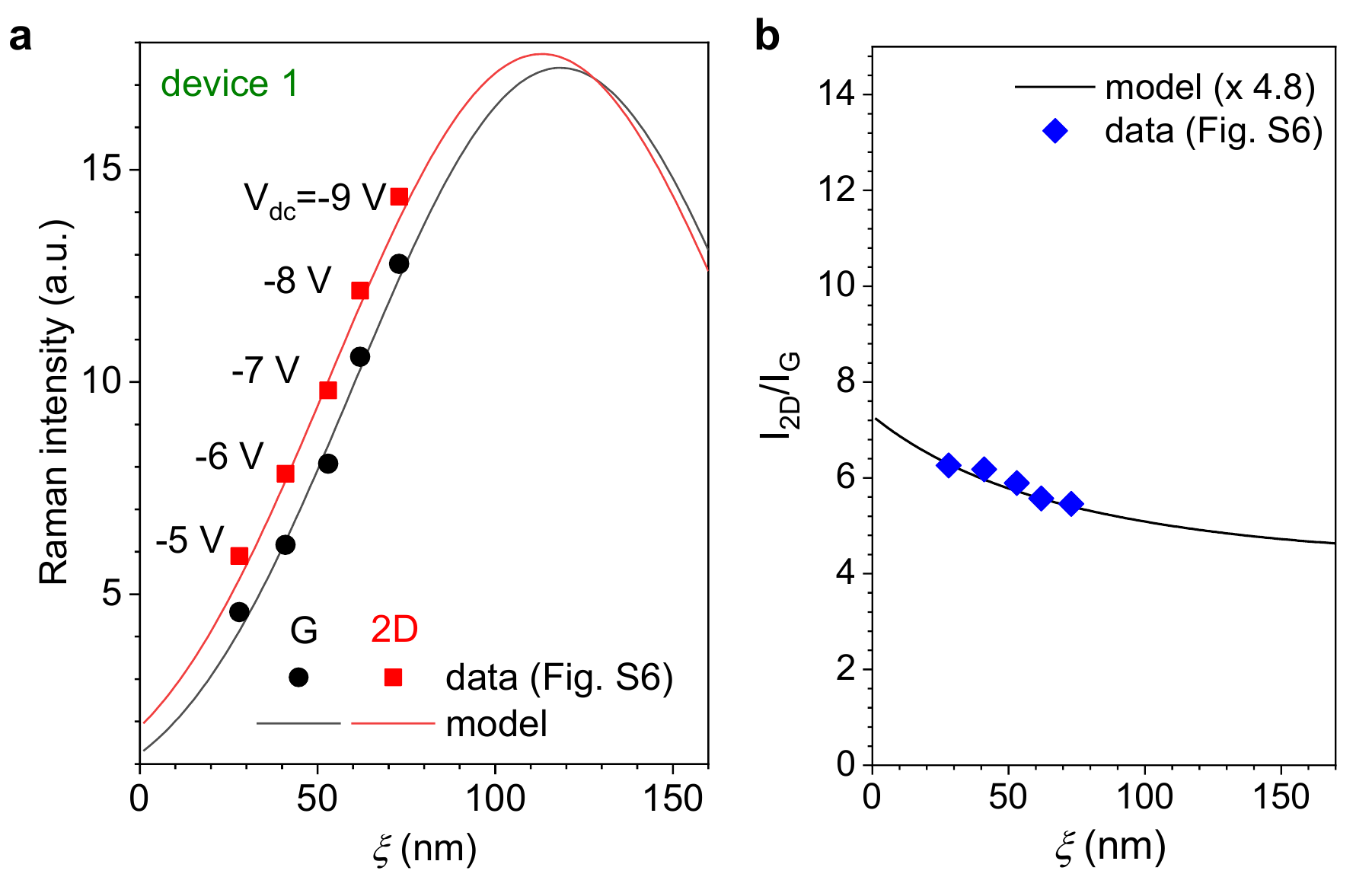}
\caption{\textbf{Raman scattering intensity as a function of the static deflection.} {\bf a,} Calculated  Raman intensity enhancement factors (solid lines) and measured  Raman intensities (symbols)  for the G (black) and 2D (red) modes as a function of $\xi$. The G- and 2D-mode intensities are extracted from Supplementary Fig.~\ref{FigS4}b and d, respectively and scaled by a constant factor to allow comparison with the Raman intensity enhancement factors (Supplementary Fig.~\ref{FigS1}c,d). The values of $\xi$ associated with the experimental data are deduced from the strain-induced 2D-mode softening (\ref{SecInter} and Supplementary Fig.~\ref{FigS5}). {\bf b,} Measured Raman intensity ratio ($I_{\rm{2D}}/I_{\rm {G}}$) as a function of $\xi$ (blue symbols). The solid line is the ratio of the Raman intensity enhancement factors multiplied by a scaling factor of 4.8 that corresponds to the ``interference-free case''~\cite{Metten2015}.}
\label{FigS6}
\end{center}
\end{figure*}

\subsection*{Equilibrium position shift in the driven regime}

In the linear regime,  a graphene drum vibrates harmonically in a symmetric potential $U\left(\xi\right)$ (inset in Fig.~2e in the main text) with respect to the static equilibrium displacement $\xi_{\rm {eq}}$ (Fig.~2e).  Under non-linear driving, the displacements are large enough such that the drum explores an asymmetric potential~\cite{NO_book,Eichler2013}. The drum now vibrates symmetrically with respect to an equilibrium position shifted by $\Delta \xi_{\rm{eq}}$, for which the Raman intensity enhancement factor (Supplementary Fig.~\ref{FigS1},~\ref{FigS4},~\ref{FigS6}) is different. As a result, the measured Raman intensities become dependent on the driving force as the graphene drum is driven non-linearly, as evidenced in Fig.~2e and 3a, where Raman intensity drops by $\sim\, 20\%$ (at $\Vdc=-8 ~\rm V$) and $\sim\,10\%$ (at $\Vdc=-6 ~\rm V$)  are consistently observed. As discussed in the main text, these intensity drops correspond to an upshift of the equilibrium position (Supplementary Fig.~\ref{FigS6}). Similar equilibrium position upshifts are discussed in device 3 (Fig.~4).



\clearpage

\section{D\lowercase{isplacement calibration}}
\label{Sec_cal}

\subsection*{Calibration Methods}
A careful displacement calibration is essential to make sure that our assumption of a constant optomechanical transduction coefficient remains valid at the largest displacements attained in the non-linear regime. In addition, displacement calibration permits an estimation of the effective mass (see below) and allow demonstrating the pristine character of our samples and the generality of our findings.


The RMS displacements $z_{\rm{rms}}$ of our monolayer graphene drums are calibrated using three distinct methods described in the following subsections. The transduction coefficients $\beta_i \ \rm (nm/mV)$ (with $i=1,2,3$) that connect the RMS voltage measured with our lock-in amplifier to $z_{\rm{rms}}$ are found to be very similar for the 3 methods and are summarized  in Table~\ref{TableCalib} for device 2 at $\rm Vdc = -8~V$. 

\begin{table}[h!]
\begin{tabular}{|l|c|r}
\hline 
Calibration method & $\beta_i \ \rm (nm/mV)$, $i=1,2,3$ \\
\hline
$C_1:$ Thermal noise & $1.1 \pm 0.15$ \\
$C_2:$ DC reflectance and Raman spectroscopy & $1.0 \pm 0.10$ \\
$C_3:$ DC reflectance and interference model & $1.2 \pm 0.20$  \\
\hline 
\end{tabular}
\caption{\textbf{Diplacement calibration methods.} Transduction coefficient $\beta_i \ \rm (nm/mV)$, $i=1,2,3$ connecting the  measured RMS voltage on our lock-in amplifier to the measured RMS displacement $z_{\rm{rms}}$ of a driven graphene drum for three calibration methods ($C_i$, $i=1,2,3$). Measurements were performed on device 2 at $\rm Vdc = -8~V$.}
\label{TableCalib}
\end{table}
 
\subsubsection*{$C_1$: Thermal noise}
\label{SecTN}

The mechanical oscillations of the graphene drum its thermal noise power spectral density (PSD) are related via~\cite{Hauer2013}:
\begin{equation}
\langle z^2_n(t)\rangle=\int_{0}^{\infty}dfS_{zz}(f)
\end{equation}
where  $f=\Omega/2\pi$ is the mechanical frequency, $\langle z^2_n(t)\rangle$ is the mean-square amplitude of vibration of the $n$-th mode, which one-sided displacement spectral density $S_{zz}(f)$ writes:
\begin{equation}
S_{zz}(f)=\frac{k_{\rm B}Tf_n}{2\pi^3\widetilde{m}_nQ_n\left[(f^2-f^2_n)^2+(ff_n/Q_n)^2\right]}
\label{eqSzz}
\end{equation}
where $k_{\rm B}$, $T$, $f_n$, $Q_n$ and  $\widetilde{m}_n$  are the Boltzmann constant, the temperature (here taken equal to the ambient temperature), the resonance frequency, the quality factor and the effective mass of the $n$-th mode, respectively. Importantly, the surface mass density of our drum  is assumed to be equal to that of pristine monolayer graphene (see below for a discussion on the relevance of this assumption). In the following, we will focus on the fundamental mechanical mode discussed in the main text.

\begin{figure*}[!htb]
\begin{center}
\includegraphics[width=16cm]{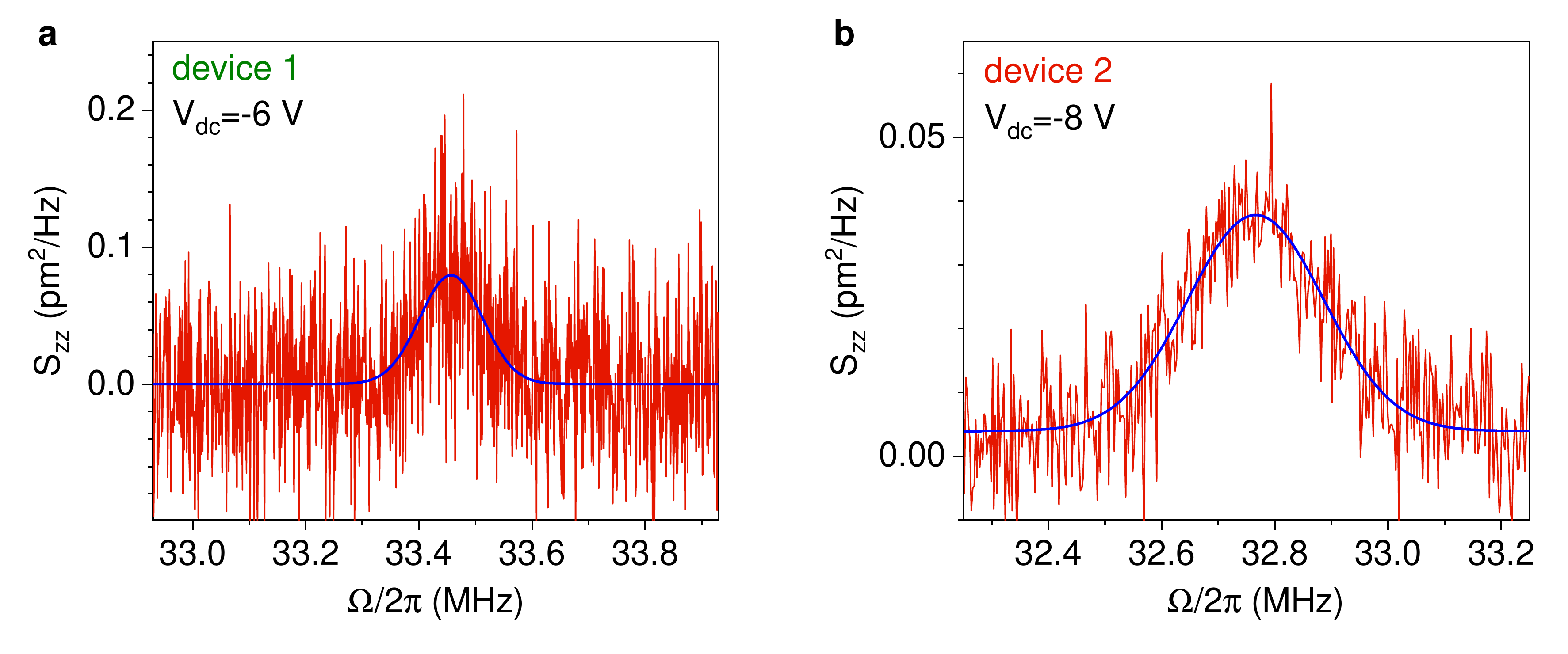}
\caption{\textbf{Displacement calibration method $C_1$.} Thermal noise power spectral density (PSD) of the fundamental mode of device 1 at $\Vdc=-6\ \rm V$ ({\bf a}) and device 2 at $\Vdc=-8\ \rm V$  ({\bf b}). The blue curves are the fit of the PSD using Eq.~\eqref{eqSzz}.}
\label{FigS7_VLeo}
\end{center}
\end{figure*}

The thermal noise PSD $S_{zz}(f)$ of the graphene drum is determined from the spectrum $V(f)$ of the output voltage of our avalanche photodiode, measured using a spectrum analyser. The resulting PSD is $S_{VV}(f)=V(f)^2/\Delta f$, where $\Delta f$ is the resolution bandwidth (typically in the $10^2-10^3~\rm{Hz}$ range). The measured signal includes a flat noise floor ($S^w_{VV}$) due to the dark current noise of the photodiode and other sources of white noise and is connected to $S_{zz}(f)$ through:
\begin{equation}
S_{VV}(f)=S^w_{VV}+\eta S_{zz}(f),
\label{eqSVV}
\end{equation}
where $\eta$ is another transduction coefficient expressed in $\rm V^2/\rm m^2$. $\eta$ is obtained by fitting the measured $S_{VV}(f)$ by Eq.~\eqref{eqSVV}, as in Supplementary Supplementary Fig.~\ref{FigS7_VLeo}. Finally, to calibrate the mechanical amplitude of the driven graphene drum measured using our lock-in amplifier, we simultaneously record the mechanical amplitude in the linear regime (typically with $\Vac= 1 ~\rm{mV}$) using the spectrum analyser and our lock-in amplifier and deduce $\beta_1$ (Table~\ref{TableCalib}). This calibration method was applied to all the devices studied in this work at various $\Vdc$. 


\subsubsection*{DC reflectance-based methods}
\label{calDC}

The following two methods rely on a measurement of the DC  reflectance of the sample (proportional to the intensity of the 632.8~nm laser beam reflected by the sample, see Supplementary Fig.~\ref{FigS1}) as a function of $\Vdc$,  combined with a calibration of the gate-dependent static deflection $\xi$ (Supplementary Fig.~\ref{FigS1} and Supplementary Fig.~\ref{FigS5}). Both methods connect the DC reflectance to $\xi$ and  yield the transduction coefficients $\beta_2$ and $\beta_3$.

\paragraph*{$C_2$: DC reflectance and Raman spectroscopy.}

With calibration $C_2$, $\xi$ is estimated through the gate-dependent spectral shifts of the Raman G and 2D modes as discussed in \ref{S2}, \ref{SecInter} and Supplementary Fig.~\ref{FigS4}-\ref{FigS5}). Coincidentally, the gate-induced changes of the DC reflectance are monitored with our lock-in amplifier. 

\begin{figure*}[!htb]
\begin{center}
\includegraphics[width=9.5cm]{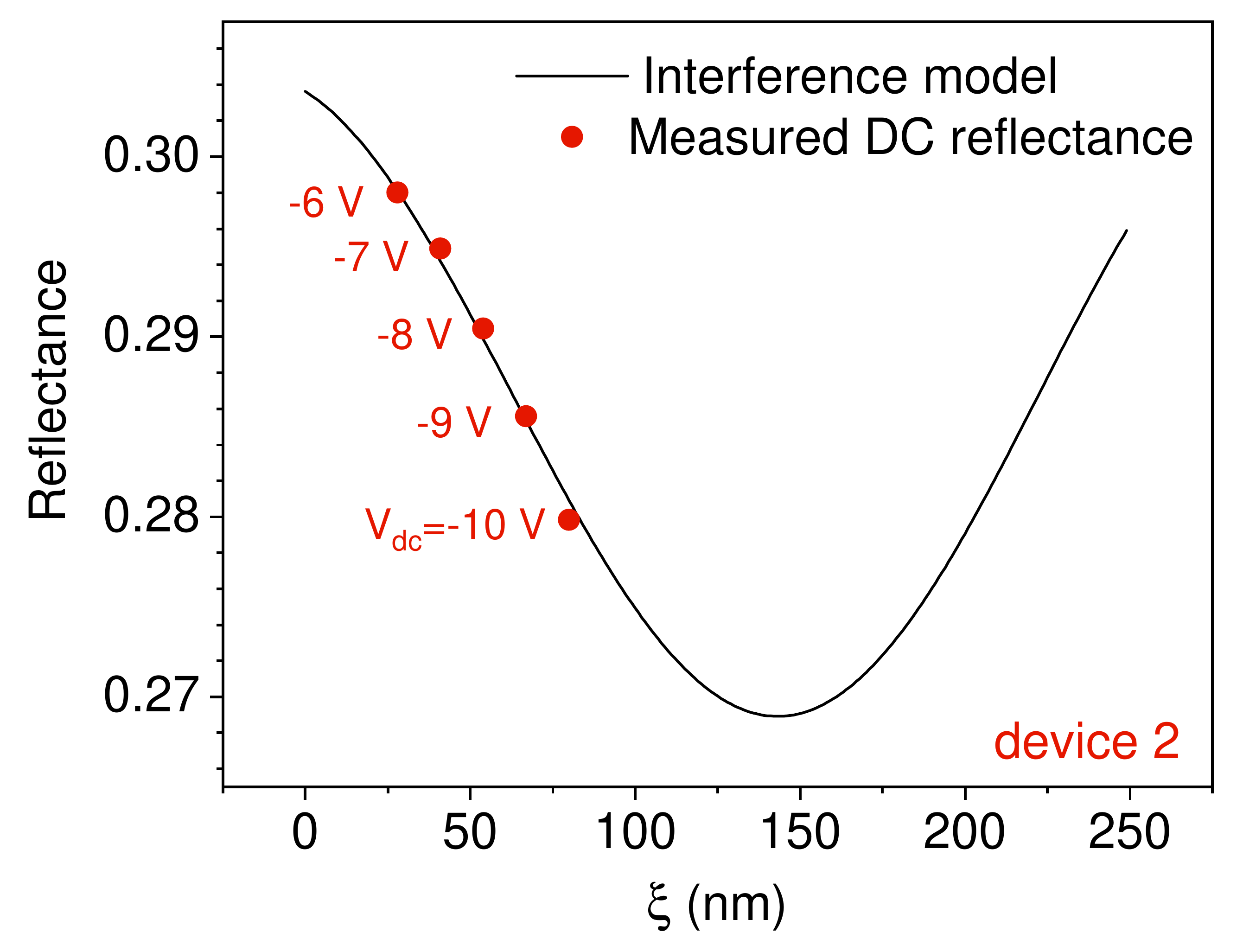}
\caption{\textbf{Displacement calibration method $C_3$.} Calculated (solid line) and measured (symbols) sample reflectance at 632.8~nm as a function of the static displacement $\xi$.}
\label{Figure_reflectanceModel}
\end{center}
\end{figure*}

\paragraph*{$C_3$: DC reflectance and interference model.}

As discussed in \ref{S1} and Supplementary Fig.~\ref{FigS1}, an interference calculation~\cite{Yoon2009,Blake2007} can be applied to obtain the reflectance of our samples as a function of $\xi$. Supplementary Fig.~\ref{Figure_reflectanceModel} shows the calculated reflectance together with our measurements of the reflected laser intensity vs $\Vdc$, scaled to match the simulated values. 



\subsection*{Discussion on the effective mass of graphene drums}
\label{Disc_cal}

The calibration of the displacement of a nanomechanical system with thermal noise measurements ($C_1$) requires accurate knowledge of its effective mass. Here, we have considered the surface mass density of pristine monolayer graphene ($\approx 7.5\times 10^{-7}~\rm{kg/m^2}$). For the fundamental mechanical mode of circular drum, the rest mass of graphene has to be scaled by a factor $\approx 0.27$ (Ref.~\onlinecite{Hauer2013}), such that the \textit{effective} mass of our $6~\mu \rm m-$diameter drum is $\widetilde{m}_0\approx 5.7\times 10^{-18}\;\rm{kg}$.
Calibration $C_2$ and $C_3$ are totally independent of $\widetilde{m}$ and yield transduction coefficients $\beta_{2,3}$ that are, within experimental accuracy, equal the coefficient $\beta_1$ obtained using thermal noise measurements considering $\widetilde{m}_0$ (see values and associated errorbars in Table~\ref{TableCalib}).  This key result justifies our assumption that $\widetilde{m}=\widetilde{m}_0$.



Following previous reports, we could have expected that $\widetilde{m}$ would \textit{a priori} exceed $\widetilde{m}_0$  due to the presence of molecular adsorbates and other sources of contamination~\cite{Weber2014}. In addition, graphene drums and blisters, in particular when made from wet-transfer of graphene layers grown by chemical vapor deposition (CVD), are known to exhibit rippling and crumpling~\cite{Nicholl2015}. The resulting hidden area effects lead to discrepancies between the levels of stain determined through Raman and interferometric measurements~\cite{Nicholl2017} and thus affect our displacement and strain calibration. Here, the excellent agreement between calibration methods $C_2$ and $C_3$ demonstrates that our graphene drums are immune from hidden area effects, as previously observed in our blister test on pristine suspended graphene, where a Young's modulus matching that of bulk graphite was found~\cite{Metten2014}.

Our devices are made from freshly exfoliated natural graphite flakes using a dry, resist-free transfer method and then held in high vacuum. Such freely suspended graphene membranes have consistently shown intrinsic electronic~\cite{Bolotin2008} and optical~\cite{Berciaud2009,Berciaud2013,Berciaud2014} properties. Our study also demonstrates that the same holds for their mechanical figures of merit. 

Let us note in closing that assuming $\widetilde{m}>\widetilde{m}_0$ when using method $C_1$  would lead to smaller calibrated displacements than those estimated assuming $\widetilde{m}_0$. Smaller displacements would lead to smaller values of $\edh$ calculated through Eq.~\eqref{EMM} and to a larger discrepancy between $\edh$ and the enhanced $\ed$ determined from our Raman measurements in resonantly driven graphene drums.



\clearpage

\section{M\lowercase{echanical response of driven graphene drums}}
\label{SMech}

 The displacement of our graphene drums can be modeled as that of a driven  non-linear oscillator by~\cite{NO_book}:
\begin{equation}
\ddot{z}+\frac{\Omega_0}{Q}\dot{z}+\Omega_0^2z+\alpha_2z^2+\alpha_3z^3=\frac{ \widetilde{F}_{el}}{\widetilde{m}}\cos(\Omega t)
\label{eqmech1}
\end{equation} 
where $z$ is the mechanical displacement at the membrane center, $\Omega_0/2\pi$ is the resonance frequency in the linear regime, $Q$ is the quality factor and $\Omega_0/Q$ is the linear damping rate, $\alpha_2$, $\alpha_3$ are the quadratic and the cubic spring constant, respectively. Finally, $\widetilde{m}=0.27\,m_0$ (with $m_0$ the rest mass of the graphene drum) is  the effective mass with a correction factor that accounts for the mode shape of the fundamental resonance of a clamped circular membrane~\cite{Hauer2013,Davidovikj2016,Davidovikj2017,Weber2014} and $\widetilde{F}_{el}$ is the effective applied electrostatic force. \footnote{In principle, Eq.~(S9) could include other non-linear contributions, and in particular a non-linear damping term $(\propto \dot{z}z^2)$~\cite{Eichler2011a,Imboden2013}. Non-linear damping may broaden the frequency-dependent mechanical susceptibility of our drums, reduce its resonant amplitude and may thus act against the enhancement of $\ed$. As a result, more pronounced dynamically-induced strain enhancement could be achieved provided non-linear damping in minimized.}

\subsection*{Linear response}

In the linear response regime, $\alpha_{2,3}=0$, Eq.~\eqref{eqmech1} is the well-known differential equation of a driven harmonic oscillator. Assuming a harmonic solution $z(t)=z_0 e^{i\Omega t}$, one gets:

\begin{equation}
z_0=\frac{\widetilde{F}_{\rm{el}}/\widetilde{m}}{\Omega^2_0-\Omega^2+\rm i\,\Omega_0\Omega/Q}.
\label{eqmech2}
\end{equation}

For the fundamental mechanical mode of a thin circular membrane resonator under a sufficiently high built-in tension $T_0$ (as is the case for our graphene drums) $\Omega_0$ writes~\cite{Schwarz-thesis}
\begin{equation}
\Omega_0=2\pi f_0=u_{01}\sqrt{\frac{T_0}{\rho_{1\rm{LG}}\,a^2}},
\label{Omega0}
\end{equation} 
where $\rho_{1\rm{LG}}\approx 7.5\times10^{-7}~\rm{kg/m^2}$ is the surface mass density of pristine graphene and $u_{01}\approx2.405$ is the first zero of the zero-order Bessel function. Therefore, $T_0$ writes
\begin{equation}
T_0=0.69\,\pi^2f_0^2\,\rho_{1\rm{LG}}\,a^2,
\label{eqT0}
\end{equation} 
where $T_0=E_{\rm{1LG}}\;\varepsilon_{\rm s}/\left(1-\nu\right)$ (Ref.~\onlinecite{Schwarz-thesis}), with $E_{\rm{1LG}}=340~\rm{Nm^{-1}}$ and $\nu=0.16$ the Young's modulus and Poisson ratio of pristine monolayer graphene, respectively~\cite{Lee2008}. Eq.~\eqref{eqT0} is then used to compute the built-in and the gate-induced static strain discussed in the text. These strain values can be compared with estimates from the G- and 2D-mode softenings.

From Eq.~\eqref{eqmech2}, we get

\begin{equation}
\lvert z_0\rvert^2=\frac{\left(\widetilde{F}_{\rm{el}}/\widetilde{m}\right)^2}{(\Omega^2_0-\Omega^2)^2+(\Omega\Omega_0/Q)^2},
\label{eqmech3}
\end{equation} 
Eq.~\eqref{eqmech3} can be used to fit the frequency-response curve in the linear response region and extract $Q$, as in Fig.~1. Furthermore, near resonance ($\left|\Omega-\Omega_0\right|\ll\Omega_0$) and for $Q\gg1$,  Eq.~\eqref{eqmech3} simplifies as
\begin{equation}
\lvert z_0\rvert^2=\frac{ \widetilde{F}_{el}^2}{4\widetilde{m}^2\Omega^2_0}\frac{1}{(\Omega-\Omega_0)^2+\Omega_0^2/4Q^2},
\label{eqmech4}
\end{equation} which is a Lorentzian lineshape with full width at half maximum (FWHM) $\Omega_0/Q$. 

\subsection*{Non-linear response}
\label{SecNLrep}

Eq.~\eqref{eqmech1} can be rewritten by introducing an effective cubic spring constant\cite{NO_book,Weber2014} given by

\begin{equation}
\widetilde{\alpha}_3=\alpha_3-\frac{10\alpha_2^2}{9\Omega^2_0},
\label{eqalpha3eff}
\end{equation} 

such that a Duffing-like equation can still be written as

\begin{equation}
\ddot{z}+\frac{\Omega_0}{Q}\dot{z}+\Omega_0^2z+\widetilde{\alpha}_3z^3=\frac{ \widetilde{F}_{el}}{\widetilde{m}}\cos(\Omega t).
\label{eqmech5}
\end{equation}

\begin{figure*}[!th]
\begin{center}
\includegraphics[width=16cm]{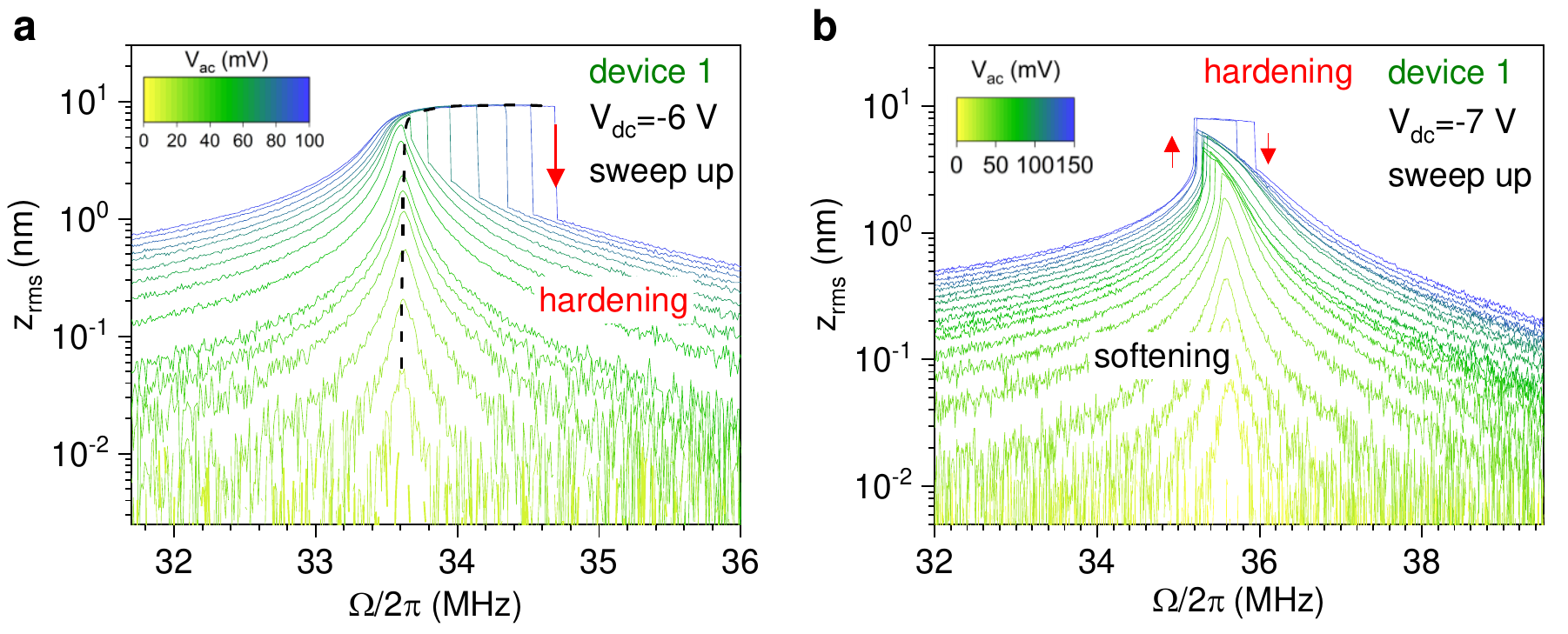}
\caption{\textbf{Mechanical non-linearities in graphene drums (1).} Frequency-response curves obtained by sweeping the drive frequency upward at $V_{dc}=-6~\rm V$ ({\bf a}) and $V_{dc}=-7~\rm V$ ({\bf b}) on device 1. At $V_{dc}=-7~\rm V$, a nonlinear softening to hardening transition is revealed above $\Vac=130~\rm{mV}$.}
\label{FigS8}
\end{center}
\end{figure*}

To obtain $\widetilde{\alpha}_3$, we can approximate the solution of Eq.~\ref{eqmech5} by a truncated Fourier series, restricted here to first order. This  approach allows establishing the analytical expression of the so-called \textit{backbone curve} that connects the maximum amplitude $z_0$ to the drive frequency $\widetilde{\Omega}_0/2\pi$ at which it is obtained. Following Refs.~\onlinecite{NO_book,Davidovikj2017}, we get
\begin{equation}
\widetilde{\Omega}_0=\Omega_0+\frac{3}{8}\frac{\widetilde{\alpha}_3}{\Omega_0} z_0^2.
\label{eqbackbone}
\end{equation} 

\begin{figure*}[!th]
\begin{center}
\includegraphics[width=14cm]{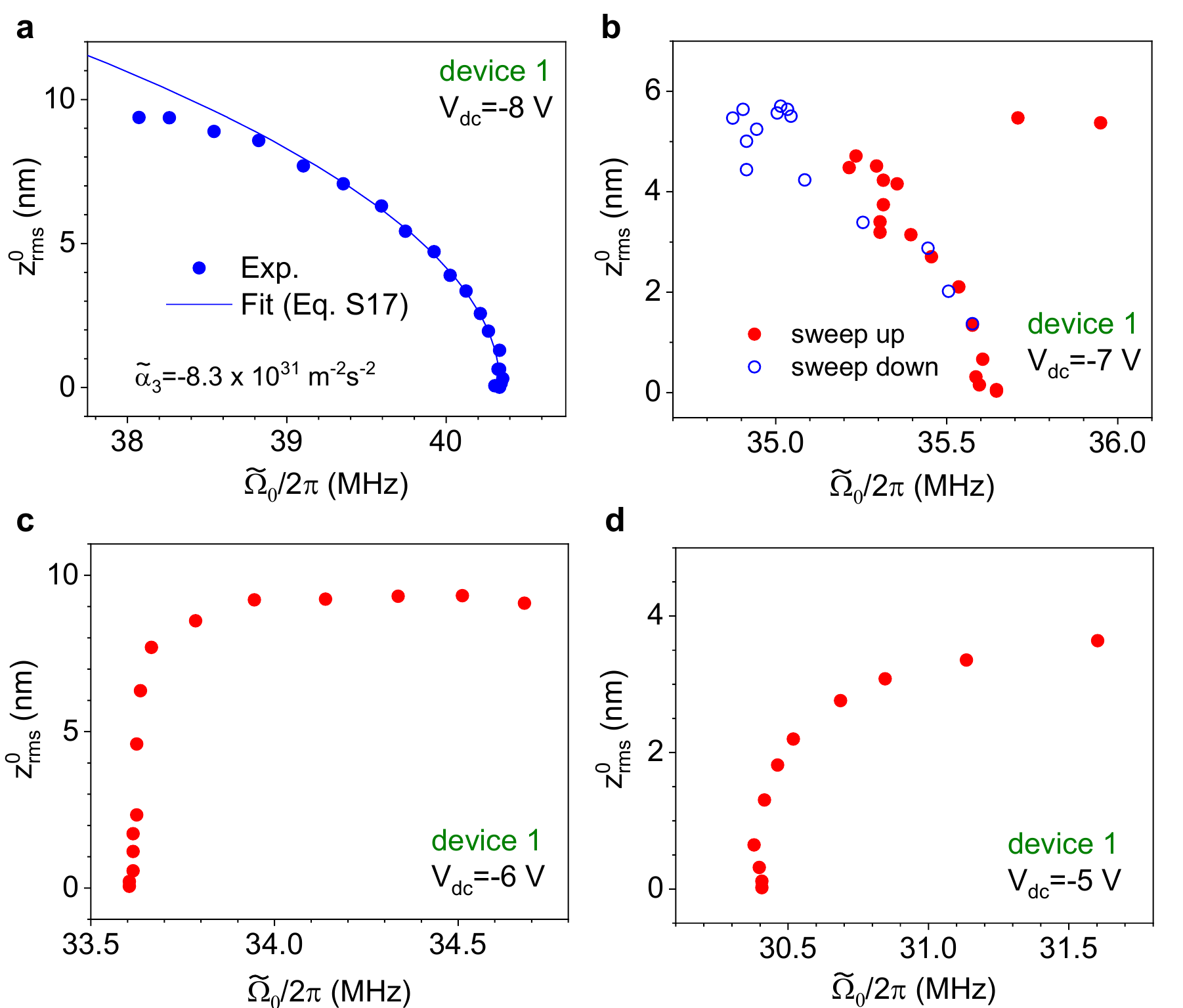}
\caption{\textbf{Mechanical non-linearities in graphene drums (2).} Backbone curves (see \ref{SMech}) recorded  on device 1 at four distinct gate biases ($\Vdc=-8,\,-7,\,-6,\,-5~\rm{V}$ in \textbf{a}, \textbf{b}, \textbf{c}, \textbf{d}, respectively) . At $\Vdc=-8~\rm V$, a fit using Eq.~\eqref{eqbackbone} allows to extract the non-linear coefficient $\widetilde{\alpha}_3$ in Eq.~\eqref{eqmech5}. The data at $\Vdc=-8~\rm V$ and $\Vdc=-6~\rm V$ are plotted in Fig.~2f and 3d, respectively.}
\label{FigS9}
\end{center}
\end{figure*}

As the driving force is increased, the onset of third-order non-linearities leads to resonance frequency hardening for $\widetilde{\alpha}_3>0$ (see data at $\Vdc=-6 ~ \rm V$ in Fig.~3 and Supplementary Fig.~\ref{FigS8} and at $\Vdc=-5 ~ \rm V$ in Supplementary Fig.~\ref{FigS9}), and to resonance frequency softening for $\widetilde{\alpha}_3<0$ (see Fig.~2 in the main text, for $\Vdc=-8 ~\rm V$), respectively. As expected from Eq.~\eqref{eqbackbone}, a parabolic backbone curve is observed at $\Vdc=-8 ~\rm V$ and to a lesser extent at  $\Vdc=-5 ~\rm V$ (Supplementary Fig.~\ref{FigS9}). However, the backbone curve fully saturates at $\Vdc=-6~\rm V$ for $\Vac>40~ \rm{mV}$ (Fig.~3a, Fig.~3d and Supplementary Fig.~\ref{FigS9}). In this strongly non-linear regime, sizeable Fourier components are expected at harmonics of the drive frequency, as experimentally verified on device 2 in Supplementary Fig.~\ref{FigS14}, and the first order expansion is insufficient.
Non-linearities can be either be i) intrinsic to graphene, e.g. due to its cubic spring constant~\cite{Lee2008} but also ii) electrostatically-induced by the dependence of the gate capacitance on the distance between the vibrating graphene drum and the Si backgate~\cite{Davidovikj2017} or iii) geometrically induced by the displacement-dependent tension induced by the vibrations of the drum~\cite{Schmid2016}. For instance, using Eq. (12) and (26) in the supplementary information of Ref.~\onlinecite{Davidovikj2017}, we can estimate that the ratio between the third order intrinsic stiffness of graphene and the gate-induced third order softening term is close to 3 at $\Vdc=-6~\rm V$ and near unity at $\Vdc=-8~\rm V$. At the same time, we estimate that the gate-induced second order spring constant ($\alpha_2$) is large enough such that Eq.~\eqref{eqalpha3eff} yields $\widetilde{\alpha}_3\approx-\frac{10\alpha_2^2}{9\Omega^2_0}\approx -1\times 10^{32}~\rm{m^2s^{-2}}$ at $\Vdc=-8~\rm V$. This value is in good agreement with the experimental value extracted from a fit of the backbone curve in Supplementary Fig.~\ref{FigS9}a. At this point, geometrical non-linearities have not been considered and are discussed below.

\clearpage

\section{D\lowercase{ynamical strain and non-linearities}}
\label{StrainNL}
In this Supplementary Note, we provide insights into the origin of the enhanced dynamical strain observed in our experiments.

\subsection*{Dynamical strain induced by harmonic vibrations}

Let us return to the simple one-dimensional model introduced in \ref{S2}. We first consider a given RMS amplitude $z_{\rm{rms}}$ and compare the values of dynamically-induced strain $\ed$ measured under strong non-linear driving to the values expected with harmonic oscillations. For simplicity, we assume that under the application of a sinusoidal driving force at frequency $\Omega/2\pi$, the drum maintains a parabolic mode shape and that the time-dependent displacement at the membrane center writes ${\xi(t)=\xi+\sqrt{2}\,z_{\rm{rms}}\cos(\Omega t+\varphi)}$, with  $\varphi$ the phase difference between the drive and the mechanical response (\ref{SMech}). The time-averaged \textit{harmonic} dynamical strain $\edh$ can be estimated by inserting $\xi(t)$ into Eq.~\eqref{eq3} and averaging over one oscillation period. Since the crossed term $2\sqrt{2}\,\xi z_{\rm{rms}}\,\cos(\Omega t+\varphi)$ averages out to zero, we obtain
\begin{equation}
\edh=\frac{2}{3}\left(\frac{z_{\rm rms}}{a}\right)^2.
\label{EMM}
\end{equation}
Eq.~\eqref{EMM} is then used with the measured RMS displacements $z_{\rm{rms}}$ to compare $\edh$ with the measured $\ed$ in Fig.~2-4 in the main manuscript. With $z_{\rm{rms}}= 9~\rm{nm}$ and $a=3~\rm{\mu m}$, Eq.~\eqref{EMM} yields $\edh= 6\times 10^{-6}$, a value that is about 40 times smaller than the measured $\ed$ obtained when $z_{\rm{rms}}$ reaches 9~nm (Fig. 2f). This obvious discrepancy suggests that non-linearities result in anharmonic oscillations and complex mode profiles, leading to enhanced $\ed$, as further discussed below.

\subsection*{Geometrical non-linearities}

We now provide additional insights into the key observation in Fig.~2f and 3d that the non-linear frequency shift $\delta=\frac{\widetilde{\Omega}_0-\Omega_0}{\Omega_0}$ (Eq.~\eqref{eqbackbone}) is proportional to the dynamical strain $\ed$. 

 For the sake of simplicity, the static displacement profile introduced above will not be explicitly considered in the following discussion. For a given transverse vibrational mode (whose mode index $n$ will be omitted in the following), the time and space-dependent displacement of the resonator  writes $u(x,t)=z(t)\phi(x)$, with $\phi(x)$ the dimensionless mode profile (defined such that $\phi(0)\equiv 1$) and $z(t)$ the displacement introduced in Eq.~\eqref{eqmech1}.
With the reasonable assumption that $\abs{\phi^{\prime}(x)}\, a\ll 1$, the time-averaged longitudinal dynamical strain writes
\begin{equation}
    \ed=\frac{z_{\rm{rms}}^2}{4a}\int_{-a}^a \left[\phi^{\prime}(x)\right]^2 dx.
    \label{epsilond_int}
\end{equation}

We will restrict ourselves to the simple case of a third order geometrical non-linearity, and consider a Duffing-like equation (i.e., Eq.~\eqref{eqmech1} with $\alpha_2=0$ and $\alpha_3\neq0$). The effective mass, the linear and non-linear spring constants associated with the mechanical mode under study can be written, respectively as~\cite{Schmid2016}
\begin{subequations}
\begin{align}
\widetilde{m}=\frac{m_0}{2a}\int_{-a}^{a} \phi^2_n(x) dx\\
k_1=\widetilde{m}\Omega_0^2=\sigma A\int_{-a}^a \left[\phi^{\prime}(x)\right]^2 dx \label{GNL1}\\
k_3=\widetilde{m}\alpha_{3}=\frac{EA}{4a}\left(\int_{-a}^a \left[\phi^{\prime}(x)\right]^2 dx \right) ^2 \label{GNL2}
\end{align}
\end{subequations}

where $\sigma$ and $E$ are the initial stress and bulk Young's modulus.
Eq.~\eqref{eqbackbone} can be recast as
\begin{equation}
    \delta=\frac{3}{8}\frac{k_3}{k_1}z_0^2=\frac{3}{32}\frac{z_0^2}{a}\frac{E}{\sigma}\int_{-a}^a \left[\phi^{\prime}(x)\right]^2 dx.
        \label{eqdelta}
\end{equation}

Using Eq.~\eqref{eqdelta} and \eqref{epsilond_int}, and assuming that $z_{\rm{rms}}^2\approx z_0^2/2 $, we obtain
\begin{equation}
    \delta\approx\frac{3}{4}\frac{E}{\sigma}\ed.
    \label{epsilond2}
\end{equation}

Eq.~\eqref{epsilond2} thus establishes the proportionality between $\delta$ and $\ed$, in qualitative agreement with the results in Fig.~2f and Fig.~3d. As indicated in the main manuscript and in Supplementary Fig.~\ref{FigS5}, for the values of $\Vdc$ used in our study (see also Fig.~2 and 3), the gate-induced static strain $\es\approx \sigma/E \approx 2\times 10^{-4}$  is close to these values of $\ed$ attained as $z_{\rm{rms}}$ saturates (Fig.~2f and 3d). With these values, Eq.~\eqref{epsilond2} would yield $\delta\sim 1$, in obvious contradiction with Fig.~2f, 3d, and~\ref{FigS9} that show that $\abs{\delta}$ does hardly exceed 5\%. To explain this discrepancy, one should keep in mind that Eq.~\eqref{epsilond2} has been derived using solely third order geometrical non-linearities (Eq.~\eqref{GNL2}) to describe the Duffing coefficient and hence ignoring other intrinsic and electrostatically-induced non-linearities as discussed in \ref{SMech}. These various non-lineartites lead to amplitude saturation and may cause the emergence of non-trivial mode profiles, with large gradients ($\phi^{\prime}(x)$), as recently observed experimentally~\cite{Yang2019}. From Eq.~\eqref{epsilond_int}, it is clear that sharp changes in the mode profiles will enhance $\ed$. At the same time, non-linearities may lead to mechanical mode hardening (as in the case of a geometrical Duffing non-linearity described by Eq.~\eqref{GNL2}) or  softening, as exemplified in Fig.~2 and discussed above (Eq.~\eqref{eqalpha3eff}, see also Supplementary Fig.~\ref{FigS8} and Supplementary Fig.~\ref{FigS9}). All in all, the measured values of $\delta$ result from the interplay between several sources of non-linearity listed above~\cite{Schmid2016,Davidovikj2017,Banafsheh2017}. One may thus observe $\abs{\delta}$ of a few $\%$ together with non-linearly enhanced $\ed$ that gets as large as $\es$. We conclude that our results strongly suggest that $\phi^{\prime}(x)$ takes on large values on length scales that are significantly smaller than $a$ that cannot be resolved using our diffraction-limited setup (see main text for details). 





\clearpage

\section{E\lowercase{ffect of laser-induced heating}}
\label{Sec_heat}
In our measurements, the laser spot is typically around $1.2~\mu\rm{m}$ in diameter~\cite{Metten2016} and the laser power was set to $P_{\rm{laser}}\sim~500 ~\mu\rm W$ for the measurements in Fig.~1-3 and $P_{\rm{laser}}\sim~200 ~\mu\rm W$ for the measurements in Fig.~4. These values corresponds to a reasonable trade off to obtain a sufficiently large Raman signal without being perturbed by softening of the Raman modes due to laser-induced heating~\cite{Calizo2007}. However, the photon flux on the suspended drum is sufficient to induce photothermal effects on its mechanical susceptibility~\cite{Barton2012}. As shown in Supplementary Fig.~\ref{FigS15}, at $\Vdc=-6~\rm V$ the resonance frequency $\Omega_0/2\pi\approx 30.6~\rm{MHz}$ is nearly independent on the laser power below a threshold $P_{\rm{laser}}\approx200~\mu \rm W$, above which a linear increase in $\Omega_0$ is found, as in previous reports~\cite{Barton2012}. To estimate the temperature $(T)$ increase caused by laser heating, we extracted the thermally induced strain $\varepsilon_T$  from the experimental data in Supplementary Fig.~\ref{FigS15}a using Eq.~\eqref{Omega0}
\begin{equation}
\varepsilon_T=\frac{1-\nu}{E_{\rm{1LG}}}\times 0.69\,\pi^2f(T)^2\,\rho_{1\rm{LG}}\,a^2.
\end{equation}
As shown in Supplementary Fig.~\ref{FigS15}b, above $P_{\rm{laser}}\sim200~\mu \rm W$, the obtained values of $\varepsilon_T$ increase linearly with $P_{\rm{laser}}$. Using a thermal expansion coefficient $\kappa_T\approx -8\times 10^{-6} ~\rm K^{-1}$ (Ref.~\onlinecite{Yoon2011}), we estimate a temperature increase $\Delta T=-\varepsilon_T/\kappa_T\approx 2.5~\rm K$ at $P_{\rm{laser}}=500~\mu\rm W$, a value that is about two orders of magnitude too small to account for the dynamical Raman frequency softenings discussed in the main text.

\begin{figure*}[!htb]
\begin{center}
\includegraphics[width=15cm]{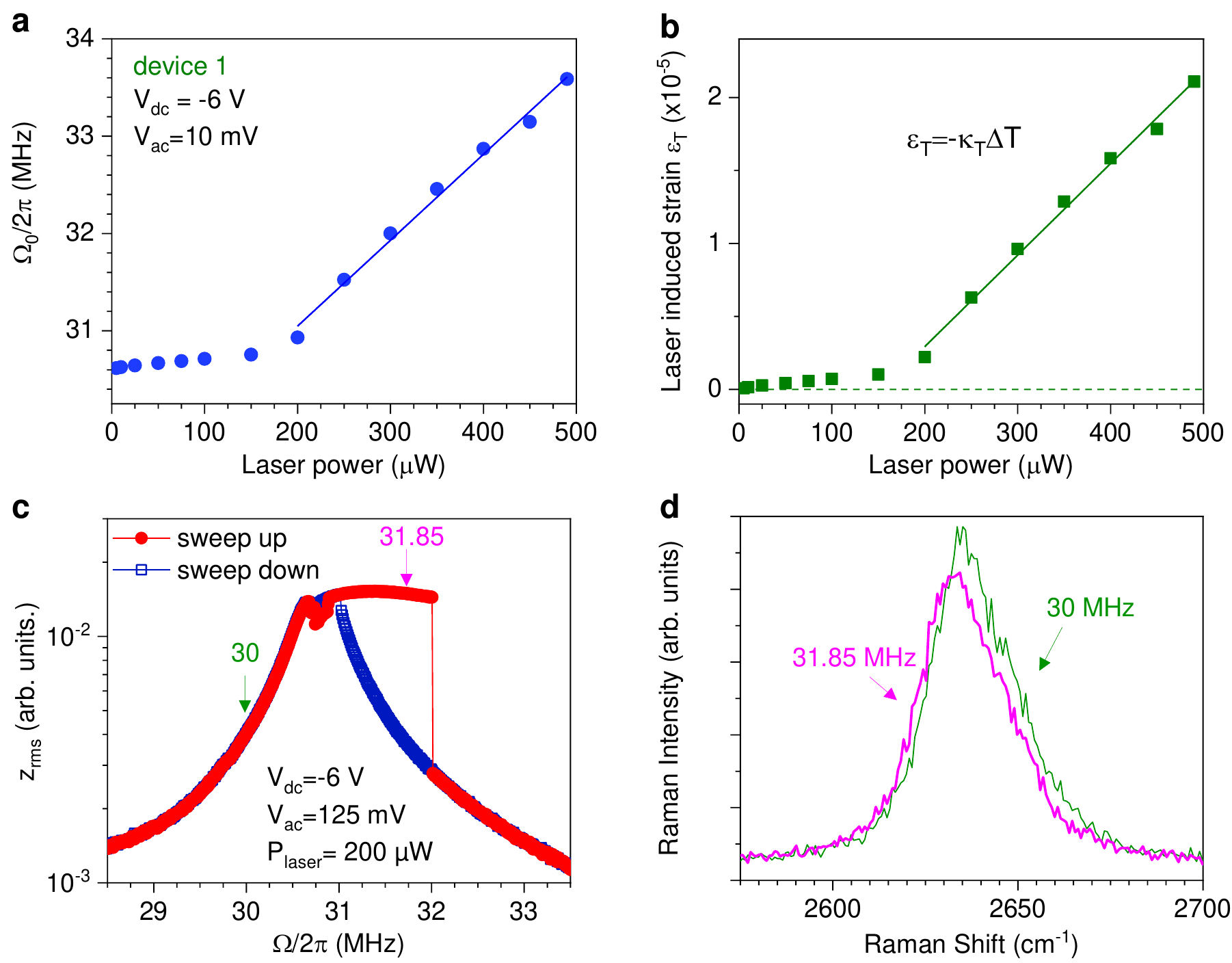}
\caption{\textbf{Effects of laser-induced heating  on the mechanical response of graphene drums.} {\bf a,} Resonance frequency $\Omega_0/2\pi$ measured as a function of the laser power in device 1. The blue line is a linear fit. {\bf b,} Extracted stain ($\varepsilon_T$) as a function of laser power. The solid line is a linear fit (\ref{Sec_heat}). {\bf c,} Frequency-dependent RMS displacement $z_{\rm{rms}}$ at $\Vdc=-6~\rm V$ and $V_{ac}=150~\rm {mV}$ using a laser power $200~\mu\rm W$. {\bf d,} Dynamical Raman spectra recorded in the aforementioned conditions under two distinct drive frequencies indicated by the green and pink arrows in ({\bf c}).}
\label{FigS15}
\end{center}
\end{figure*}

To further rule out laser-induced Raman frequency softening, we repeated the Raman measurements in driven graphene drums at $P_{\rm {laser}}= 200~\mu \rm W$, a value that is low enough to neglect photothermal effects on the mechanical resonance frequency (Supplementary Fig.~\ref{FigS15}a,c). Supplementary Fig.~\ref{FigS15}d shows the Raman 2D-mode spectra recorded under $\Vdc=-6 \rm V$ and $\Vac=125~\rm{mV}$ at near-resonant and off-resonant drive frequencies. Raman frequency softening under resonant driving akin to Fig.~3 of the main text is clearly observed.


\clearpage

\section{S\lowercase{upplementary data on device 1}}

\label{SecD1}

\begin{figure*}[!htb]
\begin{center}
\includegraphics[width=13 cm]{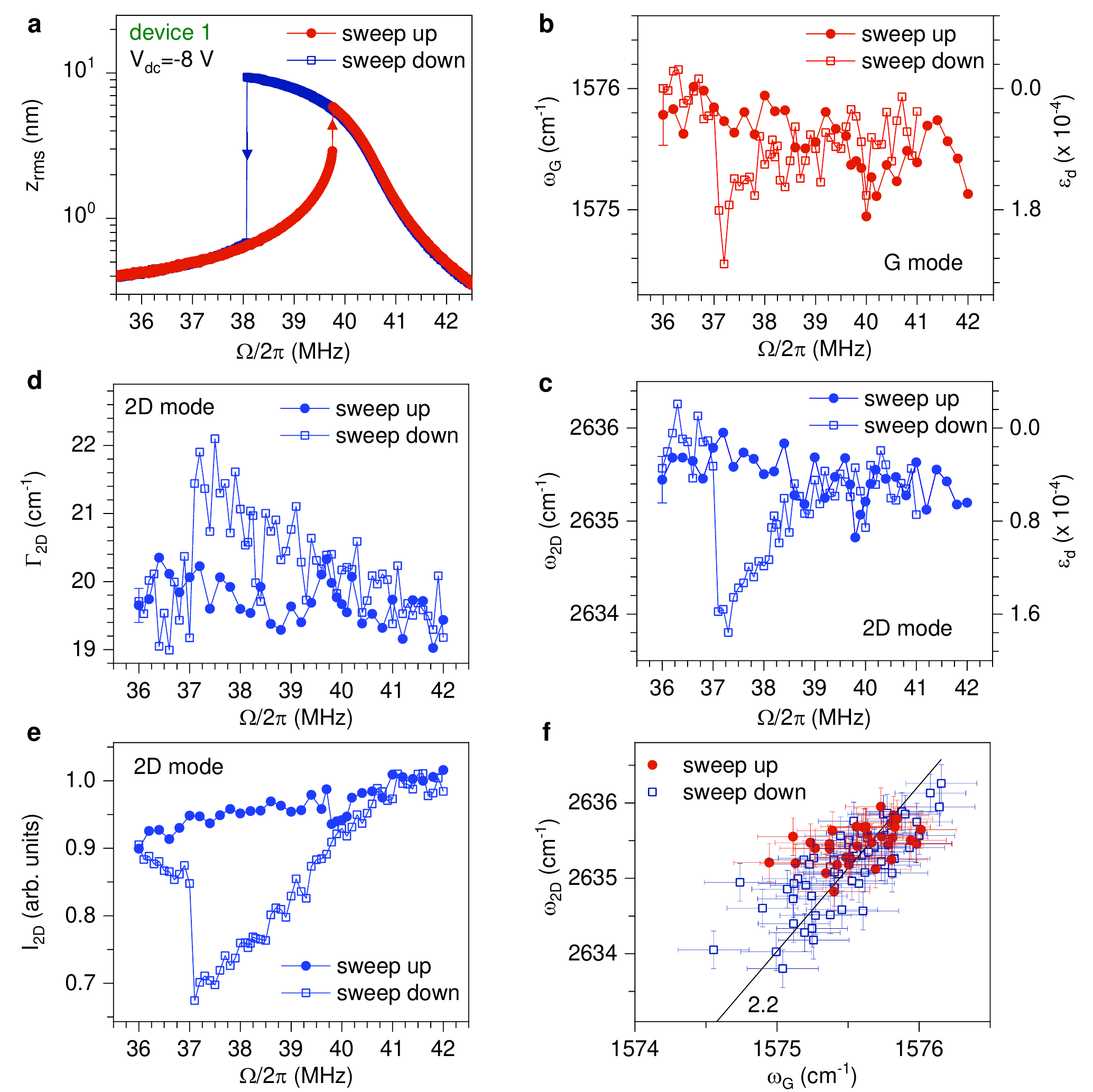}
\caption{\textbf{Frequency-dependent dynamically-induced strain at $\Vdc=-8~\rm V$ in device 1.} {\bf a,} Frequency-response curves on device 1 at $\Vdc$=-8 V and $\Vac=$150~mV. The arrows denote the jump-up and jump-down frequencies. Frequencies of the Raman G mode ({\bf b}) and 2D mode ({\bf c}) as a function of $\Omega/2\pi$. FWHM ({\bf d}) and integrated Raman intensity ({\bf e}) of the 2D-mode feature as a function of $\Omega/2\pi$. {\bf f,} Correlation between G- and 2D-mode frequencies. A straight black line with slope of 2.2 is a guide to the eye showing the expected correlation in the case of strain-induced phonon softening~\cite{Metten2014}. Only one error bar is included in ({\bf b,d,e}) for clarity. The jump frequencies appear at drive frequencies that are slightly redshifted by $\sim 1~\rm{MHz}$ relative to the frequency-response curves in ({\bf a}). This effect is attributed to photothermally induced mechanical frequency downshift (\ref{Sec_heat} and Supplementary Fig.~\ref{FigS15}).}
\label{FigS10}
\end{center}
\end{figure*}


\begin{figure*}[!htb]
\begin{center}
\includegraphics[width=13cm]{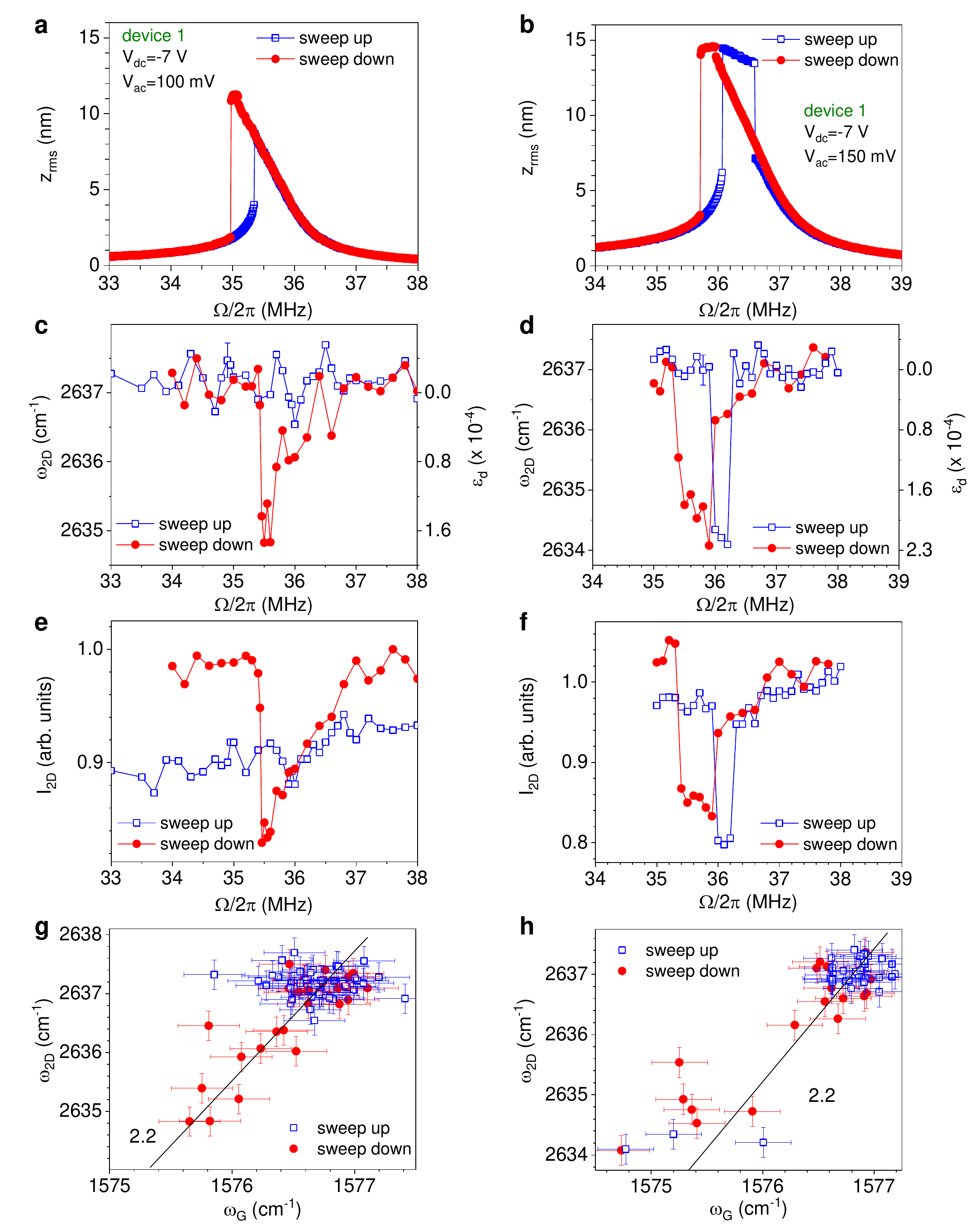}
\caption{\textbf{Frequency-dependent dynamically-induced strain at $\Vdc=-7~\rm V$ in device 1.} Frequency-response curves measured on device 1 at $\Vdc=-7 ~\rm V$ with $\Vac=100~\rm{mV}$ ({\bf a}) and $\Vac=150~\rm{mV}$ ({\bf b}). Raman 2D mode frequency $\omega_{2\rm D}$ as a function of $\Omega/2\pi$ under $\Vac=100~\rm{mV}$ ({\bf c}) and $\Vac=150~\rm{mV}$ ({\bf d}), respectively. Corresponding integrated intensity  $I_{2\rm D}$ ({\bf e}, {\bf f}) and correlations between the G- and 2D-mode frequencies {\bf g, h}. The straight black line with a slope of 2.2 is a guide to the eye  showing the expected correlation for strain-induced phonon softening. The jump frequencies appear at drive frequencies that are slightly shifted by $\lesssim1~\rm{MHz}$ relative to the frequency-response curves in ({\bf a},{\bf b}). This effect is attributed to photothermally induced mechanical frequency downshift (\ref{Sec_heat} and Supplementary Fig.~\ref{FigS15}).}
\label{FigS11}
\end{center}
\end{figure*}


\begin{figure*}[!htb]
\begin{center}
\includegraphics[width=8cm]{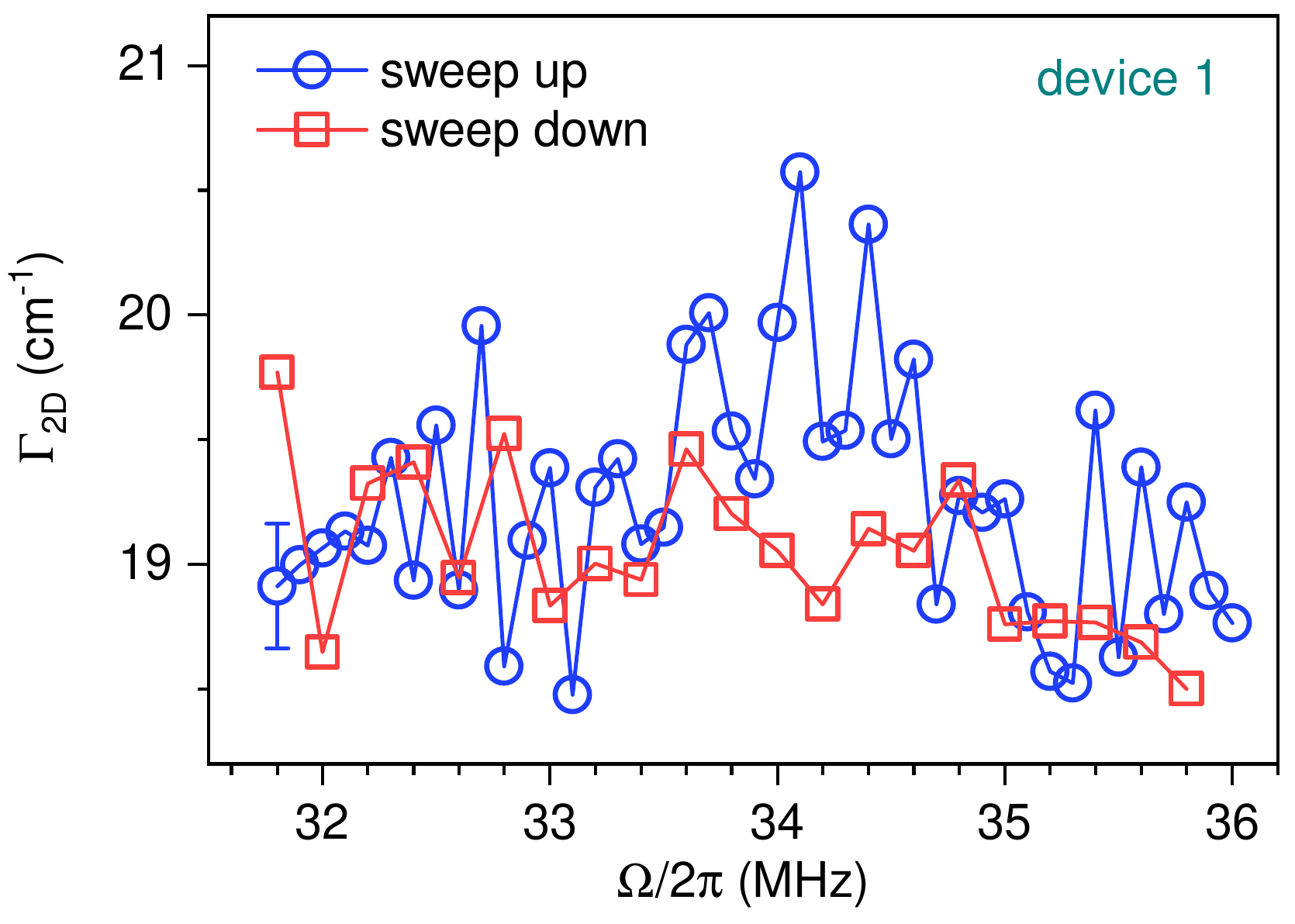}
\caption{\textbf{Dynamically-induced 2D-mode broadening at $\Vdc=-6~\rm V$ in device 1.} Full width at half maximum of the 2D mode feature $\gamma_{2\rm D}$ (we considered the 2D$^-$ component, see \ref{Raman}) as a function of the drive frequency. This data is extracted from the measurements show in Fig.~3 of the main manuscript. A slight broadening is observable as the RMS amplitude saturates near $\Omega/2\pi=34~\rm{MHz}$ (see Fig.~3a).}
\label{FigSGamma2D}
\end{center}
\end{figure*}

\clearpage
\section{S\lowercase{upplementary data on device 2}}
\label{SecD2}
\begin{figure*}[!htb]
\begin{center}
\includegraphics[width=16cm]{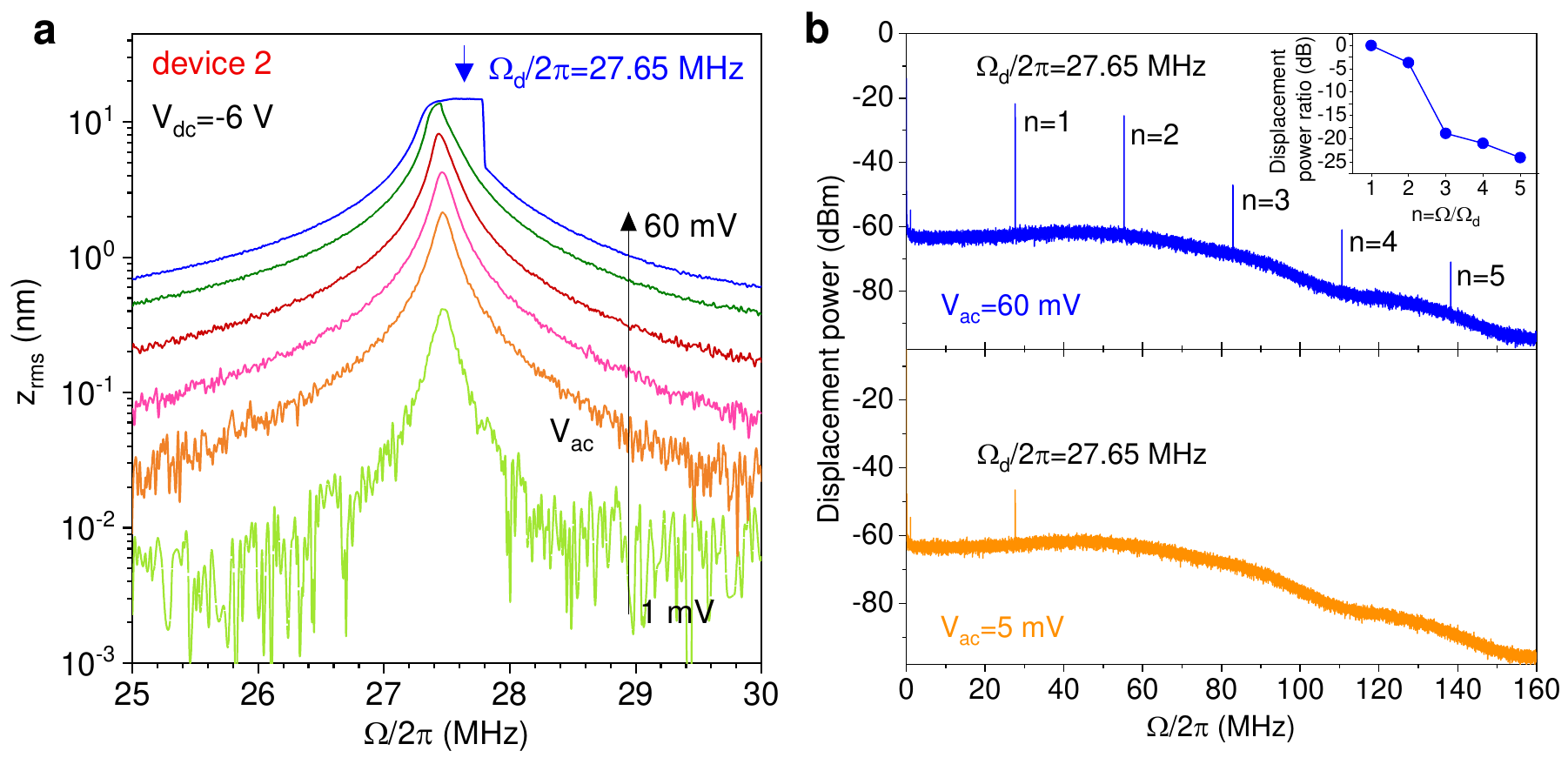}
\caption{ \textbf{Harmonic generation under non-linear mechanical driving.} {\bf a}, Frequency-response curve measured at $\Vdc=-6~\rm V$  with $\Vac$ ranging from $5~\rm{mV}$ up to  $60~\rm{mV}$ in device 2, a graphene drum similar to devices 1 and 2. The blue arrow denotes the drive frequency $\Omega_{\rm d}/2\pi$ used in ({\bf b}). {\bf b}, Broadband displacement power spectral density under $\Omega_{\rm d}/2\pi=27.65~\rm{MHz}$. Bottom panel, with $\Vac=5~\rm{mV}$; top panel, with $\Vac=60~\rm{mV}$. Sizeable high-order harmonic components (here up to $5\;\Omega_{\rm d}/2\pi$) are revealed when the drum is resonantly driven with large amplitude. The 50~MHz bandwidth of our avalanche photodiode is clearly visible. Inset: Displacement power of the harmonics relative to the displacement power at $\Omega_{\rm d}$.}
\label{FigS14}
\end{center}
\end{figure*}

\begin{figure*}[!htb]
\begin{center}
\includegraphics[width=16.5cm]{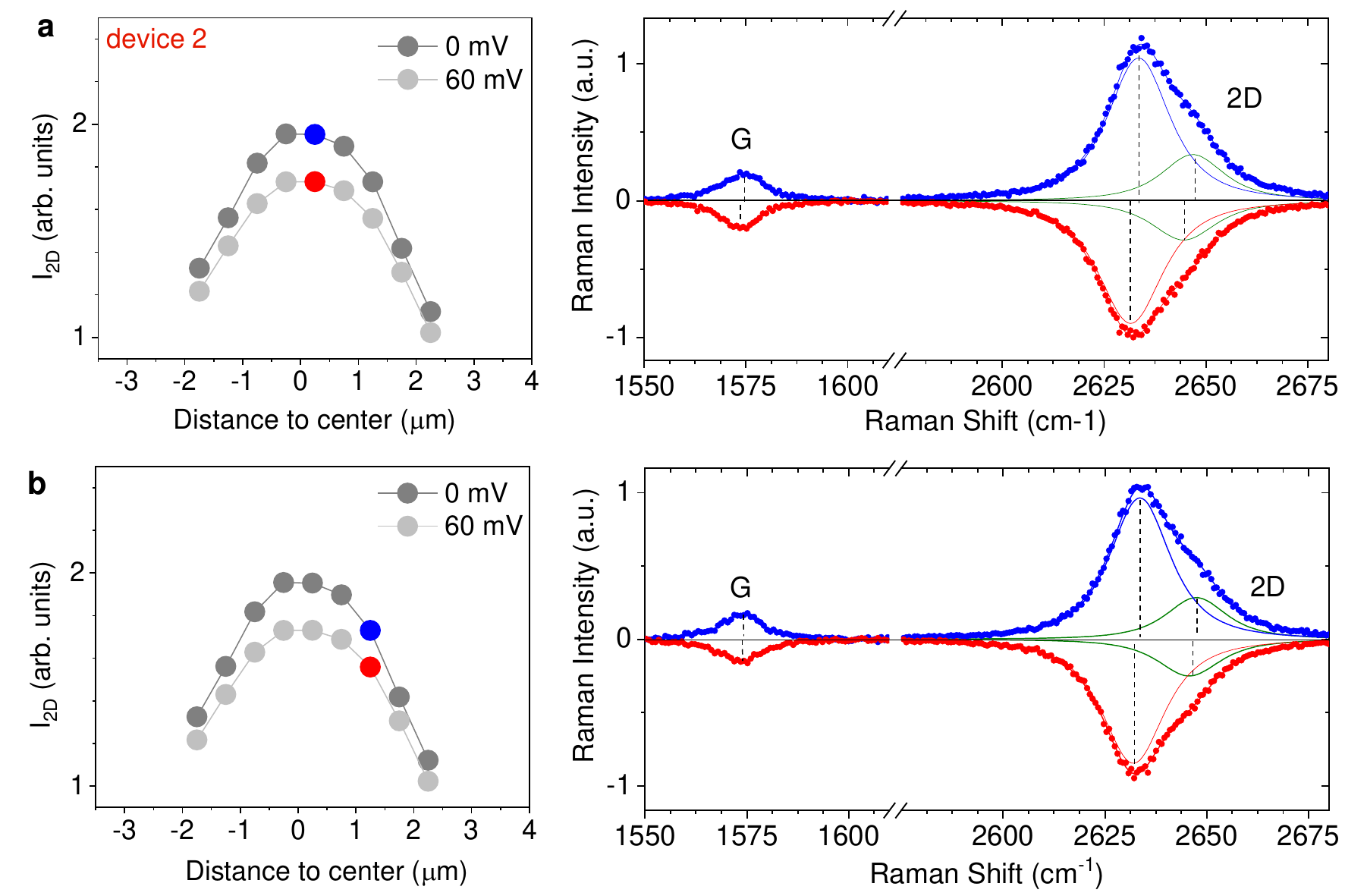}
\caption{\textbf{Spatially-revolved Raman spectroscopy in device 2.} Selected Raman spectra taken at the centre of device 2 ({\bf a}) and 2~$\mu\rm m$ away from the centre ({\bf b}) under $\Vdc=-6~\rm V$ and $\Vac=0$ (data in blue) and $\Vdc=-6~\rm V$ and $\Vac=60~\rm{mV}$ (data in red). See also Fig.~4 in the main text and related discussion.}
\label{FigS17}
\end{center}
\end{figure*}

\begin{figure*}[!htb]
\begin{center}
\includegraphics[width=7cm]{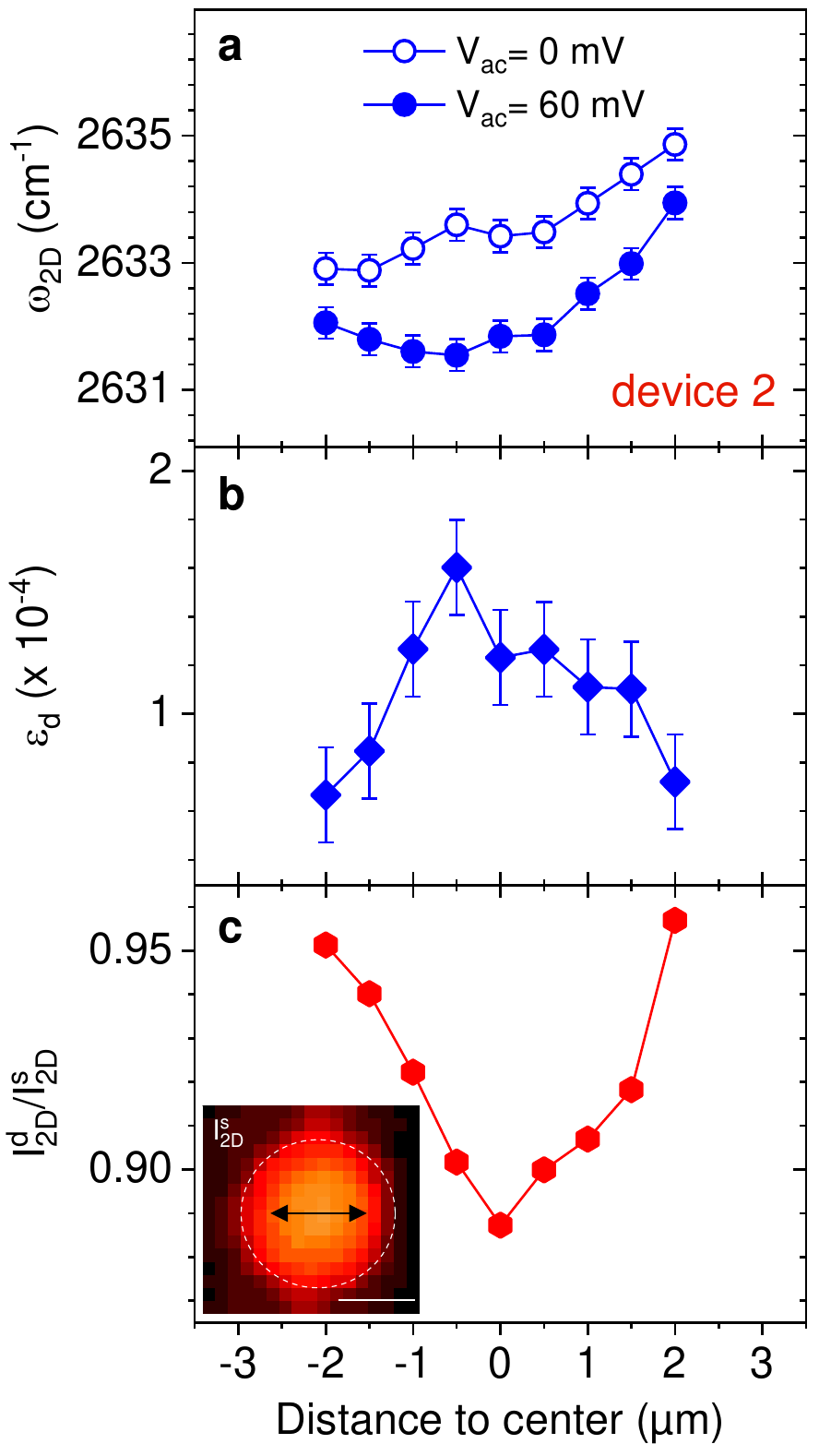}
\caption{\textbf{Spatially-revolved dynamically-induced strain in device 2.} \textbf{a}, Frequency of the Raman 2D mode along the cross-sections highlighted in \textbf{c} in a graphene drum (device 2, radius 3~$\mu \rm m$) at $\Vdc=-6~\rm V$ and $\Vac=0~\rm {mV}$ (open symbols) and $\Vac=60~\rm {mV}$ (full symbols). \textbf{b}, Dynamical strain $\ed$ obtained from the difference of the data in \textbf{a}. \textbf{c}, Ratio of the  Raman 2D-mode intensity in the driven ($I_{2\rm D}^{\rm d}$)  and static ($I_{2\rm D}^{\rm s}$) cases. Inset: Map of the Raman 2D-mode intensity $I_{2\rm D}^{\rm s}$ recorded on the graphene drum (see white dashed contour), at $\Vdc=-6~\rm V$ and $\Vac=0~\rm V$. The double arrow indicates the location of the line scan. The scale bar is $3~\mu \rm m$. See also Figure 4 in the main text and related discussion.}
\label{FigS19}
\end{center}
\end{figure*}

\begin{figure*}[!htb]
\begin{center}
\includegraphics[width=16.5cm]{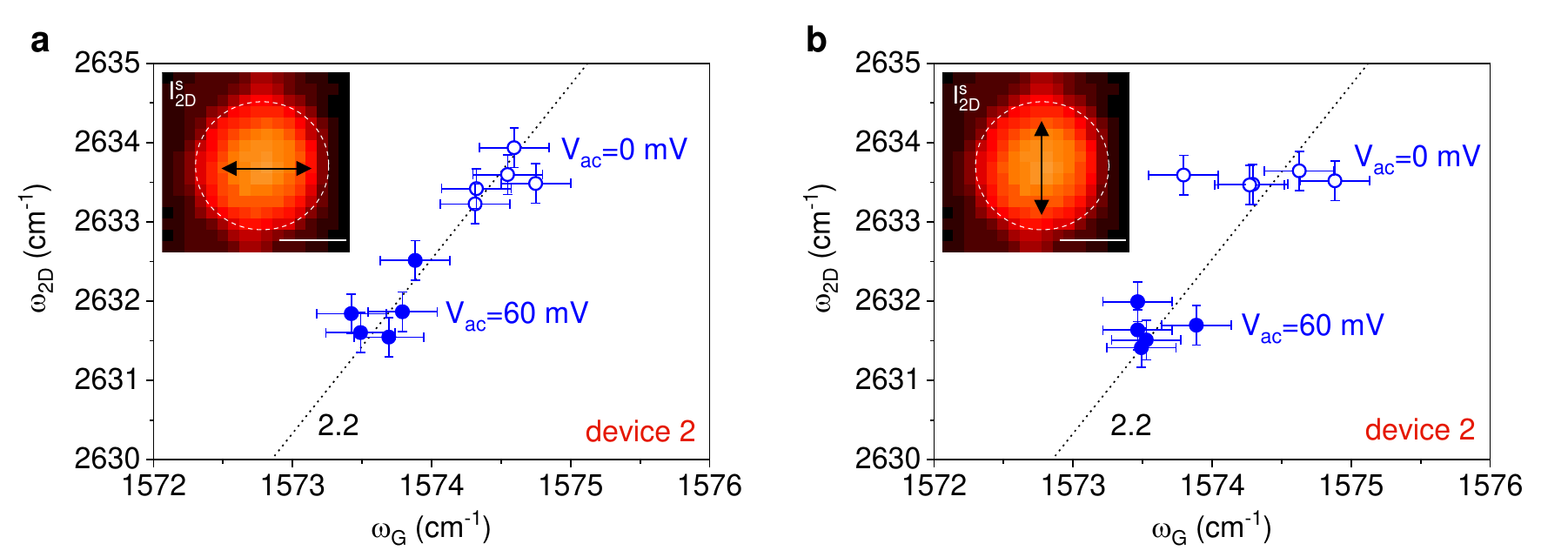}
\caption{\textbf{Correlation plot of the frequencies of the G-mode and 2D-mode frequencies in device 2.} The plots in \textbf{a} and \textbf{b}  are made from the data in Supplementary Fig.~\ref{FigS19} and in Fig.~4, respectively. The dashed lines with a slope of 2.2 are guides to the eye showing the expected correlation in the case of strain-induced phonon softening. See also Figure 4 in the main text and related discussion.}
\label{FigS20}
\end{center}
\end{figure*}

%
\clearpage 
\section{S\lowercase{upplementary data on device 3}}
\label{SecD3}

\begin{figure*}[!htb]
\begin{center}
\includegraphics[width=13.2cm]{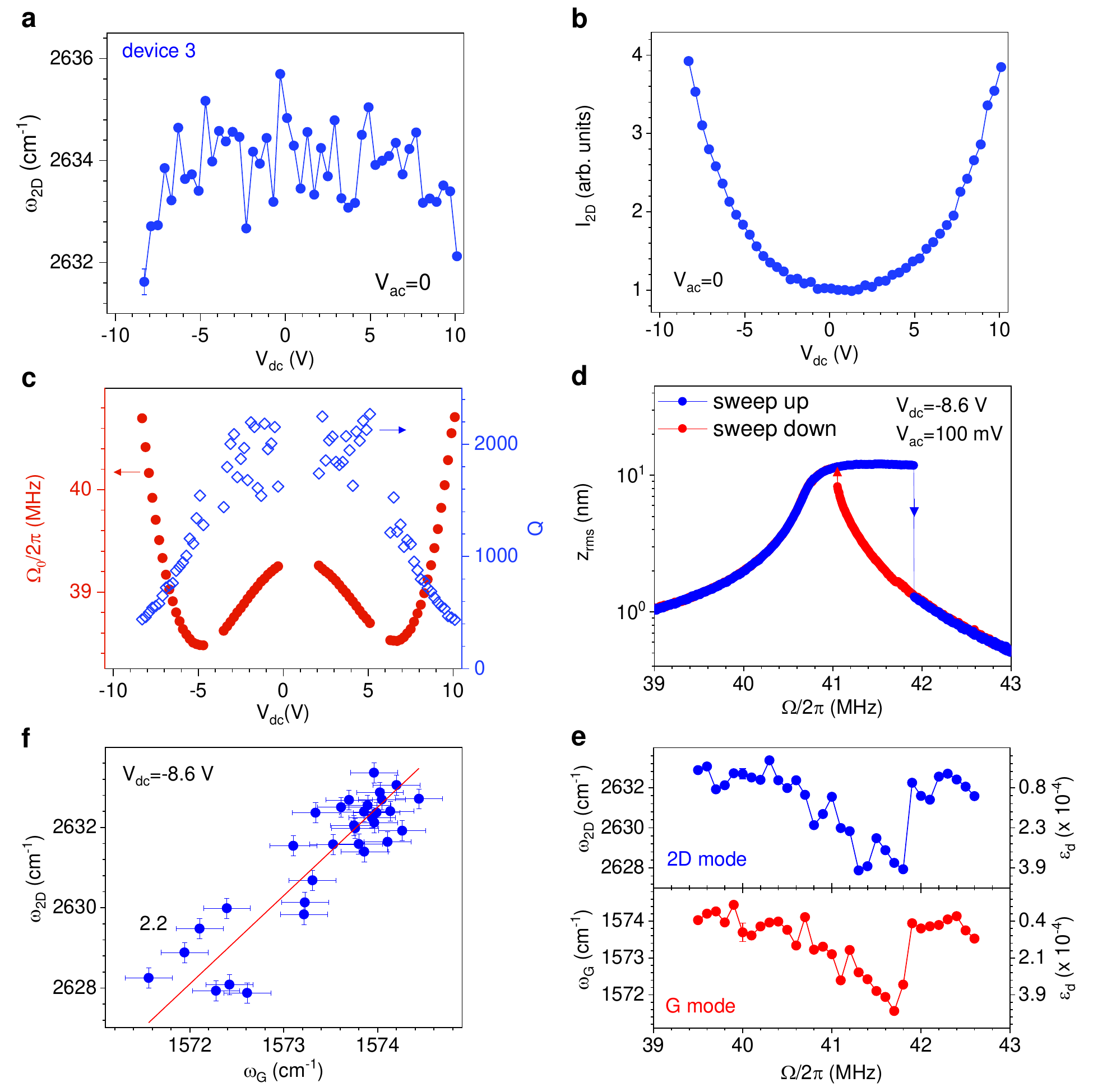}
\caption{\textbf{Dynamically-induced strain in device 3.} {\bf a,b}, Frequency and Raman intensity of the 2D mode as a function of $\Vdc$ with $\Vac=0$ for another graphene drum (device 2). This device exhibits larger built in-tension ($\varepsilon_0\approx 0.014\,\%$, estimated from Eq.~\eqref{eqT0}) and thus reduced gate-tunability as compared to device 1. The limited tunability ($\sim 5\,\%$ over the range of $\Vdc$ explored here) is due to a negative spring effect that competes with the gate-induced tension, leading to a ``W-shaped'' characteristics~\cite{Singh2010,Lee2018,Weber2014}. {\bf c}, Mechanical frequency and corresponding $Q$-factor as a function of $\Vdc$. {\bf d}, Frequency-response curves at $\Vdc=-8.6~\rm V$ with $\Vac=100~\rm{mV}$. {\bf e,} Corresponding dynamically-induced G- and 2D-mode downshifts and estimated dynamical strain $\ed$. {\bf f,} Correlation between the G- and 2D-mode frequencies. The straight black line with a slope of 2.2 is a guide to the eye showing the expected correlation for strain-induced phonon softening.}
\label{FigS12}
\end{center}
\end{figure*}

\begin{figure*}[!htb]
\begin{center}
\includegraphics[width=9cm]{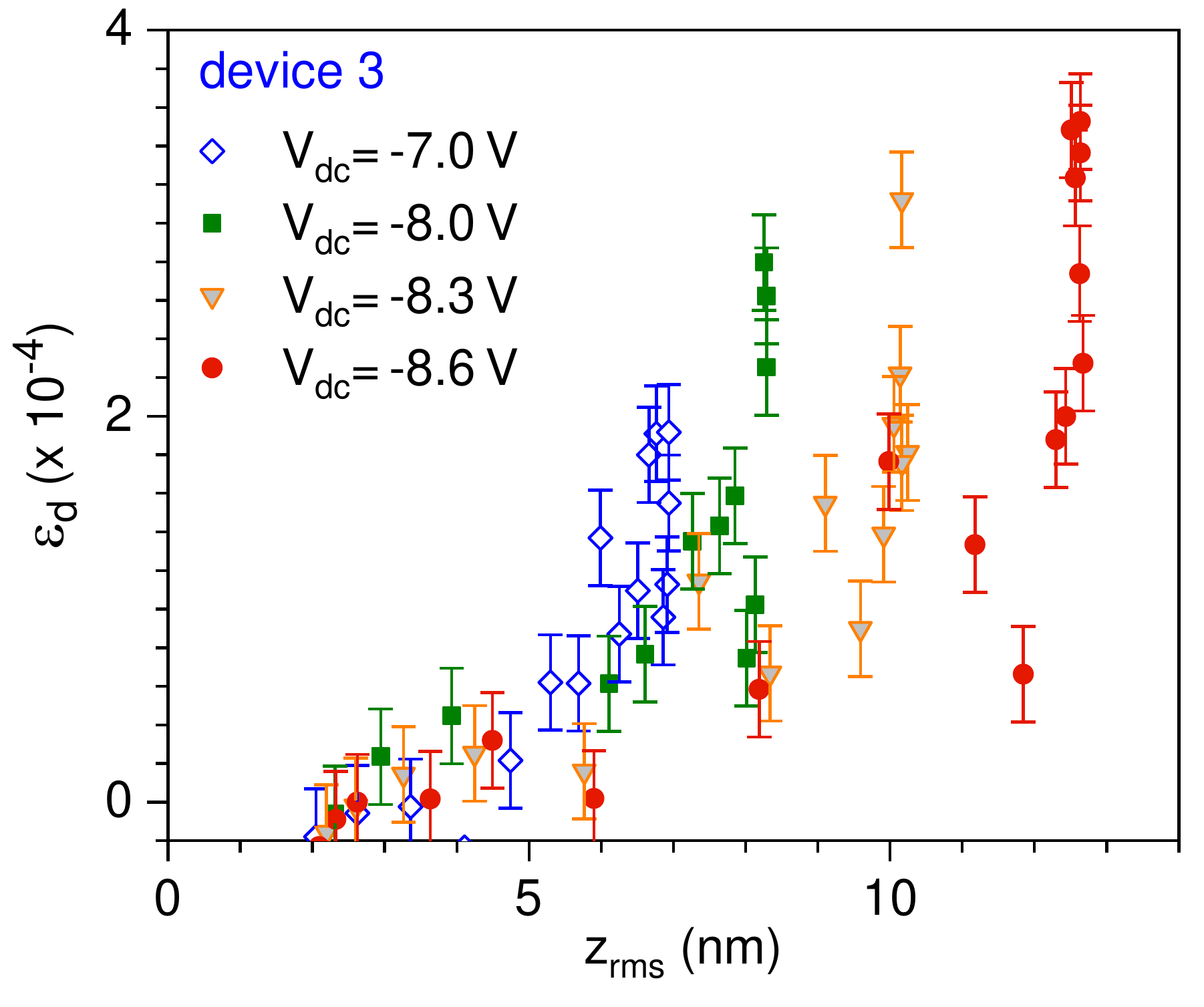}
\caption{\textbf{Gate-bias dependent dynamically-induced strain $\ed$ measured in device 3.} $\ed$ is obtained from the 2D-mode softening recorded during frequency sweeps (akin to Supplementary Fig.~\ref{FigS12} and Fig.~3) and plotted as a function of $z_{\rm{rms}}$, for $\Vdc=-7 ~\rm V$, $-8.0~\rm V$, -8.3~\rm V and $-8.6~\rm V$. The error bars come from the standard deviation of the fits of the Raman spectra.}
\label{FigS13}
\end{center}
\end{figure*}



\begin{thebibliography}{85}%
\makeatletter
\providecommand \@ifxundefined [1]{%
 \@ifx{#1\undefined}
}%
\providecommand \@ifnum [1]{%
 \ifnum #1\expandafter \@firstoftwo
 \else \expandafter \@secondoftwo
 \fi
}%
\providecommand \@ifx [1]{%
 \ifx #1\expandafter \@firstoftwo
 \else \expandafter \@secondoftwo
 \fi
}%
\providecommand \natexlab [1]{#1}%
\providecommand \enquote  [1]{``#1''}%
\providecommand \bibnamefont  [1]{#1}%
\providecommand \bibfnamefont [1]{#1}%
\providecommand \citenamefont [1]{#1}%
\providecommand \href@noop [0]{\@secondoftwo}%
\providecommand \href [0]{\begingroup \@sanitize@url \@href}%
\providecommand \@href[1]{\@@startlink{#1}\@@href}%
\providecommand \@@href[1]{\endgroup#1\@@endlink}%
\providecommand \@sanitize@url [0]{\catcode `\\12\catcode `\$12\catcode
  `\&12\catcode `\#12\catcode `\^12\catcode `\_12\catcode `\%12\relax}%
\providecommand \@@startlink[1]{}%
\providecommand \@@endlink[0]{}%
\providecommand \url  [0]{\begingroup\@sanitize@url \@url }%
\providecommand \@url [1]{\endgroup\@href {#1}{\urlprefix }}%
\providecommand \urlprefix  [0]{URL }%
\providecommand \Eprint [0]{\href }%
\providecommand \doibase [0]{http://dx.doi.org/}%
\providecommand \selectlanguage [0]{\@gobble}%
\providecommand \bibinfo  [0]{\@secondoftwo}%
\providecommand \bibfield  [0]{\@secondoftwo}%
\providecommand \translation [1]{[#1]}%
\providecommand \BibitemOpen [0]{}%
\providecommand \bibitemStop [0]{}%
\providecommand \bibitemNoStop [0]{.\EOS\space}%
\providecommand \EOS [0]{\spacefactor3000\relax}%
\providecommand \BibitemShut  [1]{\csname bibitem#1\endcsname}%
\let\auto@bib@innerbib\@empty
\bibitem [{\citenamefont {Bunch}\ \emph {et~al.}(2007)\citenamefont {Bunch},
  \citenamefont {van~der Zande}, \citenamefont {Verbridge}, \citenamefont
  {Frank}, \citenamefont {Tanenbaum}, \citenamefont {Parpia}, \citenamefont
  {Craighead},\ and\ \citenamefont {McEuen}}]{Bunch2007}%
  \BibitemOpen
  \bibfield  {author} {\bibinfo {author} {\bibfnamefont {J.~S.}\ \bibnamefont
  {Bunch}}, \bibinfo {author} {\bibfnamefont {A.~M.}\ \bibnamefont {van~der
  Zande}}, \bibinfo {author} {\bibfnamefont {S.~S.}\ \bibnamefont {Verbridge}},
  \bibinfo {author} {\bibfnamefont {I.~W.}\ \bibnamefont {Frank}}, \bibinfo
  {author} {\bibfnamefont {D.~M.}\ \bibnamefont {Tanenbaum}}, \bibinfo {author}
  {\bibfnamefont {J.~M.}\ \bibnamefont {Parpia}}, \bibinfo {author}
  {\bibfnamefont {H.~G.}\ \bibnamefont {Craighead}}, \ and\ \bibinfo {author}
  {\bibfnamefont {P.~L.}\ \bibnamefont {McEuen}},\ }\href {\doibase
  10.1126/science.1136836} {\bibfield  {journal} {\bibinfo  {journal}
  {Science}\ }\textbf {\bibinfo {volume} {315}},\ \bibinfo {pages} {490}
  (\bibinfo {year} {2007})}\BibitemShut {NoStop}%
\bibitem [{\citenamefont {Castellanos-Gomez}\ \emph {et~al.}(2015)\citenamefont
  {Castellanos-Gomez}, \citenamefont {Singh}, \citenamefont {van~der Zant},\
  and\ \citenamefont {Steele}}]{Castellanos-Gomez2015}%
  \BibitemOpen
  \bibfield  {author} {\bibinfo {author} {\bibfnamefont {A.}~\bibnamefont
  {Castellanos-Gomez}}, \bibinfo {author} {\bibfnamefont {V.}~\bibnamefont
  {Singh}}, \bibinfo {author} {\bibfnamefont {H.~S.~J.}\ \bibnamefont {van~der
  Zant}}, \ and\ \bibinfo {author} {\bibfnamefont {G.~A.}\ \bibnamefont
  {Steele}},\ }\href {\doibase 10.1002/andp.201400153} {\bibfield  {journal}
  {\bibinfo  {journal} {Annalen der Physik}\ }\textbf {\bibinfo {volume}
  {527}},\ \bibinfo {pages} {27} (\bibinfo {year} {2015})}\BibitemShut
  {NoStop}%
\bibitem [{\citenamefont {Geim}\ and\ \citenamefont
  {Grigorieva}(2013)}]{Geim2013}%
  \BibitemOpen
  \bibfield  {author} {\bibinfo {author} {\bibfnamefont {A.~K.}\ \bibnamefont
  {Geim}}\ and\ \bibinfo {author} {\bibfnamefont {I.~V.}\ \bibnamefont
  {Grigorieva}},\ }\href {\doibase 10.1038/nature12385} {\bibfield  {journal}
  {\bibinfo  {journal} {Nature}\ }\textbf {\bibinfo {volume} {499}},\ \bibinfo
  {pages} {419} (\bibinfo {year} {2013})}\BibitemShut {NoStop}%
\bibitem [{\citenamefont {Chen}\ \emph {et~al.}(2009)\citenamefont {Chen},
  \citenamefont {Rosenblatt}, \citenamefont {Bolotin}, \citenamefont {Kalb},
  \citenamefont {Kim}, \citenamefont {Kymissis}, \citenamefont {Stormer},
  \citenamefont {Heinz},\ and\ \citenamefont {Hone}}]{Chen2009}%
  \BibitemOpen
  \bibfield  {author} {\bibinfo {author} {\bibfnamefont {C.}~\bibnamefont
  {Chen}}, \bibinfo {author} {\bibfnamefont {S.}~\bibnamefont {Rosenblatt}},
  \bibinfo {author} {\bibfnamefont {K.~I.}\ \bibnamefont {Bolotin}}, \bibinfo
  {author} {\bibfnamefont {W.}~\bibnamefont {Kalb}}, \bibinfo {author}
  {\bibfnamefont {P.}~\bibnamefont {Kim}}, \bibinfo {author} {\bibfnamefont
  {I.}~\bibnamefont {Kymissis}}, \bibinfo {author} {\bibfnamefont {H.~L.}\
  \bibnamefont {Stormer}}, \bibinfo {author} {\bibfnamefont {T.~F.}\
  \bibnamefont {Heinz}}, \ and\ \bibinfo {author} {\bibfnamefont
  {J.}~\bibnamefont {Hone}},\ }\href {\doibase 10.1038/nnano.2009.267}
  {\bibfield  {journal} {\bibinfo  {journal} {Nat. Nanotechnol.}\ }\textbf
  {\bibinfo {volume} {4}},\ \bibinfo {pages} {861} (\bibinfo {year}
  {2009})}\BibitemShut {NoStop}%
\bibitem [{\citenamefont {Weber}\ \emph {et~al.}(2014)\citenamefont {Weber},
  \citenamefont {G{\"{u}}ttinger}, \citenamefont {Tsioutsios}, \citenamefont
  {Chang},\ and\ \citenamefont {Bachtold}}]{Weber2014}%
  \BibitemOpen
  \bibfield  {author} {\bibinfo {author} {\bibfnamefont {P.}~\bibnamefont
  {Weber}}, \bibinfo {author} {\bibfnamefont {J.}~\bibnamefont
  {G{\"{u}}ttinger}}, \bibinfo {author} {\bibfnamefont {I.}~\bibnamefont
  {Tsioutsios}}, \bibinfo {author} {\bibfnamefont {D.~E.}\ \bibnamefont
  {Chang}}, \ and\ \bibinfo {author} {\bibfnamefont {A.}~\bibnamefont
  {Bachtold}},\ }\href {\doibase 10.1021/nl500879k} {\bibfield  {journal}
  {\bibinfo  {journal} {Nano Lett.}\ }\textbf {\bibinfo {volume} {14}},\
  \bibinfo {pages} {2854} (\bibinfo {year} {2014})}\BibitemShut {NoStop}%
\bibitem [{\citenamefont {Davidovikj}\ \emph {et~al.}(2016)\citenamefont
  {Davidovikj}, \citenamefont {Slim}, \citenamefont {Cartamil-Bueno},
  \citenamefont {{Van Der Zant}}, \citenamefont {Steeneken},\ and\
  \citenamefont {Venstra}}]{Davidovikj2016}%
  \BibitemOpen
  \bibfield  {author} {\bibinfo {author} {\bibfnamefont {D.}~\bibnamefont
  {Davidovikj}}, \bibinfo {author} {\bibfnamefont {J.~J.}\ \bibnamefont
  {Slim}}, \bibinfo {author} {\bibfnamefont {S.~J.}\ \bibnamefont
  {Cartamil-Bueno}}, \bibinfo {author} {\bibfnamefont {H.~S.~J.}\ \bibnamefont
  {{Van Der Zant}}}, \bibinfo {author} {\bibfnamefont {P.~G.}\ \bibnamefont
  {Steeneken}}, \ and\ \bibinfo {author} {\bibfnamefont {W.~J.}\ \bibnamefont
  {Venstra}},\ }\href {\doibase 10.1021/acs.nanolett.6b00477} {\bibfield
  {journal} {\bibinfo  {journal} {Nano Lett.}\ }\textbf {\bibinfo {volume}
  {16}},\ \bibinfo {pages} {2768} (\bibinfo {year} {2016})}\BibitemShut
  {NoStop}%
\bibitem [{\citenamefont {Davidovikj}\ \emph {et~al.}(2017)\citenamefont
  {Davidovikj}, \citenamefont {Alijani}, \citenamefont {Cartamil-Bueno},
  \citenamefont {van~der Zant}, \citenamefont {Amabili},\ and\ \citenamefont
  {Steeneken}}]{Davidovikj2017}%
  \BibitemOpen
  \bibfield  {author} {\bibinfo {author} {\bibfnamefont {D.}~\bibnamefont
  {Davidovikj}}, \bibinfo {author} {\bibfnamefont {F.}~\bibnamefont {Alijani}},
  \bibinfo {author} {\bibfnamefont {S.~J.}\ \bibnamefont {Cartamil-Bueno}},
  \bibinfo {author} {\bibfnamefont {H.~S.~J.}\ \bibnamefont {van~der Zant}},
  \bibinfo {author} {\bibfnamefont {M.}~\bibnamefont {Amabili}}, \ and\
  \bibinfo {author} {\bibfnamefont {P.~G.}\ \bibnamefont {Steeneken}},\ }\href
  {\doibase 10.1038/s41467-017-01351-4} {\bibfield  {journal} {\bibinfo
  {journal} {Nat. Commun.}\ }\textbf {\bibinfo {volume} {8}},\ \bibinfo {pages}
  {1253} (\bibinfo {year} {2017})}\BibitemShut {NoStop}%
\bibitem [{\citenamefont {Lee}\ \emph {et~al.}(2018)\citenamefont {Lee},
  \citenamefont {Wang}, \citenamefont {He}, \citenamefont {Yang}, \citenamefont
  {Shan},\ and\ \citenamefont {Feng}}]{Lee2018}%
  \BibitemOpen
  \bibfield  {author} {\bibinfo {author} {\bibfnamefont {J.}~\bibnamefont
  {Lee}}, \bibinfo {author} {\bibfnamefont {Z.}~\bibnamefont {Wang}}, \bibinfo
  {author} {\bibfnamefont {K.}~\bibnamefont {He}}, \bibinfo {author}
  {\bibfnamefont {R.}~\bibnamefont {Yang}}, \bibinfo {author} {\bibfnamefont
  {J.}~\bibnamefont {Shan}}, \ and\ \bibinfo {author} {\bibfnamefont
  {P.~X.-L.}\ \bibnamefont {Feng}},\ }\href {\doibase 10.1126/sciadv.aao6653}
  {\bibfield  {journal} {\bibinfo  {journal} {Science Advances}\ }\textbf
  {\bibinfo {volume} {4}},\ \bibinfo {pages} {eaao6653} (\bibinfo {year}
  {2018})}\BibitemShut {NoStop}%
\bibitem [{\citenamefont {Weber}\ \emph {et~al.}(2016)\citenamefont {Weber},
  \citenamefont {G\"{u}ttinger}, \citenamefont {Noury}, \citenamefont
  {Vergara-Cruz},\ and\ \citenamefont {Bachtold}}]{Weber2016}%
  \BibitemOpen
  \bibfield  {author} {\bibinfo {author} {\bibfnamefont {P.}~\bibnamefont
  {Weber}}, \bibinfo {author} {\bibfnamefont {J.}~\bibnamefont
  {G\"{u}ttinger}}, \bibinfo {author} {\bibfnamefont {A.}~\bibnamefont
  {Noury}}, \bibinfo {author} {\bibfnamefont {J.}~\bibnamefont {Vergara-Cruz}},
  \ and\ \bibinfo {author} {\bibfnamefont {A.}~\bibnamefont {Bachtold}},\
  }\href {\doibase 10.1038/ncomms12496} {\bibfield  {journal} {\bibinfo
  {journal} {Nat. Commun.}\ }\textbf {\bibinfo {volume} {7}},\ \bibinfo {pages}
  {12496} (\bibinfo {year} {2016})}\BibitemShut {NoStop}%
\bibitem [{\citenamefont {Barton}\ \emph {et~al.}(2012)\citenamefont {Barton},
  \citenamefont {Storch}, \citenamefont {Adiga}, \citenamefont {Sakakibara},
  \citenamefont {Cipriany}, \citenamefont {Ilic}, \citenamefont {Wang},
  \citenamefont {Ong}, \citenamefont {McEuen}, \citenamefont {Parpia},\ and\
  \citenamefont {Craighead}}]{Barton2012}%
  \BibitemOpen
  \bibfield  {author} {\bibinfo {author} {\bibfnamefont {R.~A.}\ \bibnamefont
  {Barton}}, \bibinfo {author} {\bibfnamefont {I.~R.}\ \bibnamefont {Storch}},
  \bibinfo {author} {\bibfnamefont {V.~P.}\ \bibnamefont {Adiga}}, \bibinfo
  {author} {\bibfnamefont {R.}~\bibnamefont {Sakakibara}}, \bibinfo {author}
  {\bibfnamefont {B.~R.}\ \bibnamefont {Cipriany}}, \bibinfo {author}
  {\bibfnamefont {B.}~\bibnamefont {Ilic}}, \bibinfo {author} {\bibfnamefont
  {S.~P.}\ \bibnamefont {Wang}}, \bibinfo {author} {\bibfnamefont
  {P.}~\bibnamefont {Ong}}, \bibinfo {author} {\bibfnamefont {P.~L.}\
  \bibnamefont {McEuen}}, \bibinfo {author} {\bibfnamefont {J.~M.}\
  \bibnamefont {Parpia}}, \ and\ \bibinfo {author} {\bibfnamefont {H.~G.}\
  \bibnamefont {Craighead}},\ }\href {\doibase 10.1021/nl302036x} {\bibfield
  {journal} {\bibinfo  {journal} {Nano Lett.}\ }\textbf {\bibinfo {volume}
  {12}},\ \bibinfo {pages} {4681} (\bibinfo {year} {2012})}\BibitemShut
  {NoStop}%
\bibitem [{\citenamefont {Morell}\ \emph {et~al.}(2019)\citenamefont {Morell},
  \citenamefont {Tepsic}, \citenamefont {Reserbat-Plantey}, \citenamefont
  {Cepellotti}, \citenamefont {Manca}, \citenamefont {Epstein}, \citenamefont
  {Isacsson}, \citenamefont {Marie}, \citenamefont {Mauri},\ and\ \citenamefont
  {Bachtold}}]{Morell2019}%
  \BibitemOpen
  \bibfield  {author} {\bibinfo {author} {\bibfnamefont {N.}~\bibnamefont
  {Morell}}, \bibinfo {author} {\bibfnamefont {S.}~\bibnamefont {Tepsic}},
  \bibinfo {author} {\bibfnamefont {A.}~\bibnamefont {Reserbat-Plantey}},
  \bibinfo {author} {\bibfnamefont {A.}~\bibnamefont {Cepellotti}}, \bibinfo
  {author} {\bibfnamefont {M.}~\bibnamefont {Manca}}, \bibinfo {author}
  {\bibfnamefont {I.}~\bibnamefont {Epstein}}, \bibinfo {author} {\bibfnamefont
  {A.}~\bibnamefont {Isacsson}}, \bibinfo {author} {\bibfnamefont
  {X.}~\bibnamefont {Marie}}, \bibinfo {author} {\bibfnamefont
  {F.}~\bibnamefont {Mauri}}, \ and\ \bibinfo {author} {\bibfnamefont
  {A.}~\bibnamefont {Bachtold}},\ }\href {\doibase
  10.1021/acs.nanolett.9b00560} {\bibfield  {journal} {\bibinfo  {journal}
  {Nano Lett.}\ }\textbf {\bibinfo {volume} {19}},\ \bibinfo {pages} {3143}
  (\bibinfo {year} {2019})}\BibitemShut {NoStop}%
\bibitem [{\citenamefont {{De Alba}}\ \emph {et~al.}(2016)\citenamefont {{De
  Alba}}, \citenamefont {Massel}, \citenamefont {Storch}, \citenamefont
  {Abhilash}, \citenamefont {Hui}, \citenamefont {McEuen}, \citenamefont
  {Craighead},\ and\ \citenamefont {Parpia}}]{DeAlba2016}%
  \BibitemOpen
  \bibfield  {author} {\bibinfo {author} {\bibfnamefont {R.}~\bibnamefont {{De
  Alba}}}, \bibinfo {author} {\bibfnamefont {F.}~\bibnamefont {Massel}},
  \bibinfo {author} {\bibfnamefont {I.~R.}\ \bibnamefont {Storch}}, \bibinfo
  {author} {\bibfnamefont {T.~S.}\ \bibnamefont {Abhilash}}, \bibinfo {author}
  {\bibfnamefont {A.}~\bibnamefont {Hui}}, \bibinfo {author} {\bibfnamefont
  {P.~L.}\ \bibnamefont {McEuen}}, \bibinfo {author} {\bibfnamefont {H.~G.}\
  \bibnamefont {Craighead}}, \ and\ \bibinfo {author} {\bibfnamefont {J.~M.}\
  \bibnamefont {Parpia}},\ }\href {\doibase 10.1038/nnano.2016.86} {\bibfield
  {journal} {\bibinfo  {journal} {Nat. Nanotechnol.}\ }\textbf {\bibinfo
  {volume} {11}},\ \bibinfo {pages} {741} (\bibinfo {year} {2016})}\BibitemShut
  {NoStop}%
\bibitem [{\citenamefont {Mathew}\ \emph {et~al.}(2016)\citenamefont {Mathew},
  \citenamefont {Patel}, \citenamefont {Borah}, \citenamefont {Vijay},\ and\
  \citenamefont {Deshmukh}}]{Mathew2016}%
  \BibitemOpen
  \bibfield  {author} {\bibinfo {author} {\bibfnamefont {J.~P.}\ \bibnamefont
  {Mathew}}, \bibinfo {author} {\bibfnamefont {R.~N.}\ \bibnamefont {Patel}},
  \bibinfo {author} {\bibfnamefont {A.}~\bibnamefont {Borah}}, \bibinfo
  {author} {\bibfnamefont {R.}~\bibnamefont {Vijay}}, \ and\ \bibinfo {author}
  {\bibfnamefont {M.~M.}\ \bibnamefont {Deshmukh}},\ }\href {\doibase
  10.1038/nnano.2016.94} {\bibfield  {journal} {\bibinfo  {journal} {Nat.
  Nanotechnol.}\ }\textbf {\bibinfo {volume} {11}},\ \bibinfo {pages} {747}
  (\bibinfo {year} {2016})}\BibitemShut {NoStop}%
\bibitem [{\citenamefont {G{\"{u}}ttinger}\ \emph {et~al.}(2017)\citenamefont
  {G{\"{u}}ttinger}, \citenamefont {Noury}, \citenamefont {Weber},
  \citenamefont {Eriksson}, \citenamefont {Lagoin}, \citenamefont {Moser},
  \citenamefont {Eichler}, \citenamefont {Wallraff}, \citenamefont {Isacsson},\
  and\ \citenamefont {Bachtold}}]{Guttinger2017}%
  \BibitemOpen
  \bibfield  {author} {\bibinfo {author} {\bibfnamefont {J.}~\bibnamefont
  {G{\"{u}}ttinger}}, \bibinfo {author} {\bibfnamefont {A.}~\bibnamefont
  {Noury}}, \bibinfo {author} {\bibfnamefont {P.}~\bibnamefont {Weber}},
  \bibinfo {author} {\bibfnamefont {A.~M.}\ \bibnamefont {Eriksson}}, \bibinfo
  {author} {\bibfnamefont {C.}~\bibnamefont {Lagoin}}, \bibinfo {author}
  {\bibfnamefont {J.}~\bibnamefont {Moser}}, \bibinfo {author} {\bibfnamefont
  {C.}~\bibnamefont {Eichler}}, \bibinfo {author} {\bibfnamefont
  {A.}~\bibnamefont {Wallraff}}, \bibinfo {author} {\bibfnamefont
  {A.}~\bibnamefont {Isacsson}}, \ and\ \bibinfo {author} {\bibfnamefont
  {A.}~\bibnamefont {Bachtold}},\ }\href {\doibase 10.1038/nnano.2017.86}
  {\bibfield  {journal} {\bibinfo  {journal} {Nat. Nanotechnol.}\ }\textbf
  {\bibinfo {volume} {12}},\ \bibinfo {pages} {631} (\bibinfo {year}
  {2017})}\BibitemShut {NoStop}%
\bibitem [{\citenamefont {Singh}\ \emph {et~al.}(2014)\citenamefont {Singh},
  \citenamefont {Bosman}, \citenamefont {Schneider}, \citenamefont {Blanter},
  \citenamefont {Castellanos-Gomez},\ and\ \citenamefont {Steele}}]{Singh2014}%
  \BibitemOpen
  \bibfield  {author} {\bibinfo {author} {\bibfnamefont {V.}~\bibnamefont
  {Singh}}, \bibinfo {author} {\bibfnamefont {S.~J.}\ \bibnamefont {Bosman}},
  \bibinfo {author} {\bibfnamefont {B.~H.}\ \bibnamefont {Schneider}}, \bibinfo
  {author} {\bibfnamefont {Y.~M.}\ \bibnamefont {Blanter}}, \bibinfo {author}
  {\bibfnamefont {A.}~\bibnamefont {Castellanos-Gomez}}, \ and\ \bibinfo
  {author} {\bibfnamefont {G.~A.}\ \bibnamefont {Steele}},\ }\href {\doibase
  10.1038/nnano.2014.168} {\bibfield  {journal} {\bibinfo  {journal} {Nat.
  Nanotechnol.}\ }\textbf {\bibinfo {volume} {9}},\ \bibinfo {pages} {820}
  (\bibinfo {year} {2014})}\BibitemShut {NoStop}%
\bibitem [{\citenamefont {Song}\ \emph {et~al.}(2014)\citenamefont {Song},
  \citenamefont {Oksanen}, \citenamefont {Li}, \citenamefont {Hakonen},\ and\
  \citenamefont {Sillanp{\"{a}}{\"{a}}}}]{Song2014}%
  \BibitemOpen
  \bibfield  {author} {\bibinfo {author} {\bibfnamefont {X.}~\bibnamefont
  {Song}}, \bibinfo {author} {\bibfnamefont {M.}~\bibnamefont {Oksanen}},
  \bibinfo {author} {\bibfnamefont {J.}~\bibnamefont {Li}}, \bibinfo {author}
  {\bibfnamefont {P.~J.}\ \bibnamefont {Hakonen}}, \ and\ \bibinfo {author}
  {\bibfnamefont {M.~A.}\ \bibnamefont {Sillanp{\"{a}}{\"{a}}}},\ }\href
  {\doibase 10.1103/PhysRevLett.113.027404} {\bibfield  {journal} {\bibinfo
  {journal} {Phys. Rev. Lett.}\ }\textbf {\bibinfo {volume} {113}},\ \bibinfo
  {pages} {027404} (\bibinfo {year} {2014})}\BibitemShut {NoStop}%
\bibitem [{\citenamefont {Castellanos-Gomez}\ \emph {et~al.}(2013)\citenamefont
  {Castellanos-Gomez}, \citenamefont {van Leeuwen}, \citenamefont {Buscema},
  \citenamefont {van~der Zant}, \citenamefont {Steele},\ and\ \citenamefont
  {Venstra}}]{Castellanos-Gomez2013}%
  \BibitemOpen
  \bibfield  {author} {\bibinfo {author} {\bibfnamefont {A.}~\bibnamefont
  {Castellanos-Gomez}}, \bibinfo {author} {\bibfnamefont {R.}~\bibnamefont {van
  Leeuwen}}, \bibinfo {author} {\bibfnamefont {M.}~\bibnamefont {Buscema}},
  \bibinfo {author} {\bibfnamefont {H.~S.~J.}\ \bibnamefont {van~der Zant}},
  \bibinfo {author} {\bibfnamefont {G.~A.}\ \bibnamefont {Steele}}, \ and\
  \bibinfo {author} {\bibfnamefont {W.~J.}\ \bibnamefont {Venstra}},\ }\href
  {\doibase 10.1002/adma.201303569} {\bibfield  {journal} {\bibinfo  {journal}
  {Advanced Materials}\ }\textbf {\bibinfo {volume} {25}},\ \bibinfo {pages}
  {6719} (\bibinfo {year} {2013})}\BibitemShut {NoStop}%
\bibitem [{\citenamefont {Morell}\ \emph {et~al.}(2016)\citenamefont {Morell},
  \citenamefont {Reserbat-Plantey}, \citenamefont {Tsioutsios}, \citenamefont
  {Sch{\"{a}}dler}, \citenamefont {Dubin}, \citenamefont {Koppens},\ and\
  \citenamefont {Bachtold}}]{Morell2016}%
  \BibitemOpen
  \bibfield  {author} {\bibinfo {author} {\bibfnamefont {N.}~\bibnamefont
  {Morell}}, \bibinfo {author} {\bibfnamefont {A.}~\bibnamefont
  {Reserbat-Plantey}}, \bibinfo {author} {\bibfnamefont {I.}~\bibnamefont
  {Tsioutsios}}, \bibinfo {author} {\bibfnamefont {K.~G.}\ \bibnamefont
  {Sch{\"{a}}dler}}, \bibinfo {author} {\bibfnamefont {F.}~\bibnamefont
  {Dubin}}, \bibinfo {author} {\bibfnamefont {F.~H.~L.}\ \bibnamefont
  {Koppens}}, \ and\ \bibinfo {author} {\bibfnamefont {A.}~\bibnamefont
  {Bachtold}},\ }\href {\doibase 10.1021/acs.nanolett.6b02038} {\bibfield
  {journal} {\bibinfo  {journal} {Nano Lett.}\ }\textbf {\bibinfo {volume}
  {16}},\ \bibinfo {pages} {5102} (\bibinfo {year} {2016})}\BibitemShut
  {NoStop}%
\bibitem [{\citenamefont {Will}\ \emph {et~al.}(2017)\citenamefont {Will},
  \citenamefont {Hamer}, \citenamefont {Müller}, \citenamefont {Noury},
  \citenamefont {Weber}, \citenamefont {Bachtold}, \citenamefont {Gorbachev},
  \citenamefont {Stampfer},\ and\ \citenamefont {Güttinger}}]{Will2017}%
  \BibitemOpen
  \bibfield  {author} {\bibinfo {author} {\bibfnamefont {M.}~\bibnamefont
  {Will}}, \bibinfo {author} {\bibfnamefont {M.}~\bibnamefont {Hamer}},
  \bibinfo {author} {\bibfnamefont {M.}~\bibnamefont {Müller}}, \bibinfo
  {author} {\bibfnamefont {A.}~\bibnamefont {Noury}}, \bibinfo {author}
  {\bibfnamefont {P.}~\bibnamefont {Weber}}, \bibinfo {author} {\bibfnamefont
  {A.}~\bibnamefont {Bachtold}}, \bibinfo {author} {\bibfnamefont {R.~V.}\
  \bibnamefont {Gorbachev}}, \bibinfo {author} {\bibfnamefont {C.}~\bibnamefont
  {Stampfer}}, \ and\ \bibinfo {author} {\bibfnamefont {J.}~\bibnamefont
  {Güttinger}},\ }\href {\doibase 10.1021/acs.nanolett.7b01845} {\bibfield
  {journal} {\bibinfo  {journal} {Nano Lett.}\ }\textbf {\bibinfo {volume}
  {17}},\ \bibinfo {pages} {5950} (\bibinfo {year} {2017})}\BibitemShut
  {NoStop}%
\bibitem [{\citenamefont {Ye}\ \emph {et~al.}(2017)\citenamefont {Ye},
  \citenamefont {Lee},\ and\ \citenamefont {Feng}}]{Ye2017}%
  \BibitemOpen
  \bibfield  {author} {\bibinfo {author} {\bibfnamefont {F.}~\bibnamefont
  {Ye}}, \bibinfo {author} {\bibfnamefont {J.}~\bibnamefont {Lee}}, \ and\
  \bibinfo {author} {\bibfnamefont {P.~X.-L.}\ \bibnamefont {Feng}},\
  }\href@noop {} {\bibfield  {journal} {\bibinfo  {journal} {Nanoscale}\
  }\textbf {\bibinfo {volume} {9}},\ \bibinfo {pages} {18208} (\bibinfo {year}
  {2017})}\BibitemShut {NoStop}%
\bibitem [{\citenamefont {Kim}\ \emph {et~al.}(2018)\citenamefont {Kim},
  \citenamefont {Yu},\ and\ \citenamefont {van~der Zande}}]{Kim2018}%
  \BibitemOpen
  \bibfield  {author} {\bibinfo {author} {\bibfnamefont {S.}~\bibnamefont
  {Kim}}, \bibinfo {author} {\bibfnamefont {J.}~\bibnamefont {Yu}}, \ and\
  \bibinfo {author} {\bibfnamefont {A.~M.}\ \bibnamefont {van~der Zande}},\
  }\href {\doibase 10.1021/acs.nanolett.8b01926} {\bibfield  {journal}
  {\bibinfo  {journal} {Nano Lett.}\ }\textbf {\bibinfo {volume} {18}},\
  \bibinfo {pages} {6686} (\bibinfo {year} {2018})}\BibitemShut {NoStop}%
\bibitem [{\citenamefont {Koenig}\ \emph {et~al.}(2011)\citenamefont {Koenig},
  \citenamefont {Boddeti}, \citenamefont {Dunn},\ and\ \citenamefont
  {Bunch}}]{Koenig2011}%
  \BibitemOpen
  \bibfield  {author} {\bibinfo {author} {\bibfnamefont {S.~P.}\ \bibnamefont
  {Koenig}}, \bibinfo {author} {\bibfnamefont {N.~G.}\ \bibnamefont {Boddeti}},
  \bibinfo {author} {\bibfnamefont {M.~L.}\ \bibnamefont {Dunn}}, \ and\
  \bibinfo {author} {\bibfnamefont {J.~S.}\ \bibnamefont {Bunch}},\ }\href
  {\doibase 10.1038/nnano.2011.123} {\bibfield  {journal} {\bibinfo  {journal}
  {Nat. Nanotechnol.}\ }\textbf {\bibinfo {volume} {6}},\ \bibinfo {pages}
  {543} (\bibinfo {year} {2011})}\BibitemShut {NoStop}%
\bibitem [{\citenamefont {Lloyd}\ \emph {et~al.}(2017)\citenamefont {Lloyd},
  \citenamefont {Liu}, \citenamefont {Boddeti}, \citenamefont {Cantley},
  \citenamefont {Long}, \citenamefont {Dunn},\ and\ \citenamefont
  {Bunch}}]{Lloyd2017}%
  \BibitemOpen
  \bibfield  {author} {\bibinfo {author} {\bibfnamefont {D.}~\bibnamefont
  {Lloyd}}, \bibinfo {author} {\bibfnamefont {X.}~\bibnamefont {Liu}}, \bibinfo
  {author} {\bibfnamefont {N.}~\bibnamefont {Boddeti}}, \bibinfo {author}
  {\bibfnamefont {L.}~\bibnamefont {Cantley}}, \bibinfo {author} {\bibfnamefont
  {R.}~\bibnamefont {Long}}, \bibinfo {author} {\bibfnamefont {M.~L.}\
  \bibnamefont {Dunn}}, \ and\ \bibinfo {author} {\bibfnamefont {J.~S.}\
  \bibnamefont {Bunch}},\ }\href {\doibase 10.1021/acs.nanolett.7b01735}
  {\bibfield  {journal} {\bibinfo  {journal} {Nano Lett.}\ }\textbf {\bibinfo
  {volume} {17}},\ \bibinfo {pages} {5329} (\bibinfo {year}
  {2017})}\BibitemShut {NoStop}%
\bibitem [{\citenamefont {Dai}\ \emph {et~al.}(2019)\citenamefont {Dai},
  \citenamefont {Liu},\ and\ \citenamefont {Zhang}}]{Dai2019}%
  \BibitemOpen
  \bibfield  {author} {\bibinfo {author} {\bibfnamefont {Z.}~\bibnamefont
  {Dai}}, \bibinfo {author} {\bibfnamefont {L.}~\bibnamefont {Liu}}, \ and\
  \bibinfo {author} {\bibfnamefont {Z.}~\bibnamefont {Zhang}},\ }\href
  {\doibase 10.1002/adma.201805417} {\bibfield  {journal} {\bibinfo  {journal}
  {Advanced Materials}\ }\textbf {\bibinfo {volume} {31}},\ \bibinfo {pages}
  {1970322} (\bibinfo {year} {2019})}\BibitemShut {NoStop}%
\bibitem [{\citenamefont {Arcizet}\ \emph {et~al.}(2011)\citenamefont
  {Arcizet}, \citenamefont {Jacques}, \citenamefont {Siria}, \citenamefont
  {Poncharal}, \citenamefont {Vincent},\ and\ \citenamefont
  {Seidelin}}]{Arcizet2011}%
  \BibitemOpen
  \bibfield  {author} {\bibinfo {author} {\bibfnamefont {O.}~\bibnamefont
  {Arcizet}}, \bibinfo {author} {\bibfnamefont {V.}~\bibnamefont {Jacques}},
  \bibinfo {author} {\bibfnamefont {A.}~\bibnamefont {Siria}}, \bibinfo
  {author} {\bibfnamefont {P.}~\bibnamefont {Poncharal}}, \bibinfo {author}
  {\bibfnamefont {P.}~\bibnamefont {Vincent}}, \ and\ \bibinfo {author}
  {\bibfnamefont {S.}~\bibnamefont {Seidelin}},\ }\href {\doibase
  10.1038/nphys2070} {\bibfield  {journal} {\bibinfo  {journal} {Nature Phys.}\
  }\textbf {\bibinfo {volume} {7}},\ \bibinfo {pages} {879} (\bibinfo {year}
  {2011})}\BibitemShut {NoStop}%
\bibitem [{\citenamefont {Teissier}\ \emph {et~al.}(2014)\citenamefont
  {Teissier}, \citenamefont {Barfuss}, \citenamefont {Appel}, \citenamefont
  {Neu},\ and\ \citenamefont {Maletinsky}}]{Teissier2014}%
  \BibitemOpen
  \bibfield  {author} {\bibinfo {author} {\bibfnamefont {J.}~\bibnamefont
  {Teissier}}, \bibinfo {author} {\bibfnamefont {A.}~\bibnamefont {Barfuss}},
  \bibinfo {author} {\bibfnamefont {P.}~\bibnamefont {Appel}}, \bibinfo
  {author} {\bibfnamefont {E.}~\bibnamefont {Neu}}, \ and\ \bibinfo {author}
  {\bibfnamefont {P.}~\bibnamefont {Maletinsky}},\ }\href {\doibase
  10.1103/PhysRevLett.113.020503} {\bibfield  {journal} {\bibinfo  {journal}
  {Phys. Rev. Lett.}\ }\textbf {\bibinfo {volume} {113}},\ \bibinfo {pages}
  {020503} (\bibinfo {year} {2014})}\BibitemShut {NoStop}%
\bibitem [{\citenamefont {Ovartchaiyapong}\ \emph {et~al.}(2014)\citenamefont
  {Ovartchaiyapong}, \citenamefont {Lee}, \citenamefont {Myers},\ and\
  \citenamefont {Jayich}}]{Ovartchaiyapong2014}%
  \BibitemOpen
  \bibfield  {author} {\bibinfo {author} {\bibfnamefont {P.}~\bibnamefont
  {Ovartchaiyapong}}, \bibinfo {author} {\bibfnamefont {K.~W.}\ \bibnamefont
  {Lee}}, \bibinfo {author} {\bibfnamefont {B.~A.}\ \bibnamefont {Myers}}, \
  and\ \bibinfo {author} {\bibfnamefont {A.~C.~B.}\ \bibnamefont {Jayich}},\
  }\href {\doibase 10.1038/ncomms5429} {\bibfield  {journal} {\bibinfo
  {journal} {Nat. Commun.}\ }\textbf {\bibinfo {volume} {5}},\ \bibinfo {pages}
  {4429} (\bibinfo {year} {2014})}\BibitemShut {NoStop}%
\bibitem [{\citenamefont {Yeo}\ \emph {et~al.}(2014)\citenamefont {Yeo},
  \citenamefont {de~Assis}, \citenamefont {Gloppe}, \citenamefont
  {Dupont-Ferrier}, \citenamefont {Verlot}, \citenamefont {Malik},
  \citenamefont {Dupuy}, \citenamefont {Claudon}, \citenamefont {G{\'{e}}rard},
  \citenamefont {Auff{\`{e}}ves}, \citenamefont {Nogues}, \citenamefont
  {Seidelin}, \citenamefont {Poizat}, \citenamefont {Arcizet},\ and\
  \citenamefont {Richard}}]{Yeo2014}%
  \BibitemOpen
  \bibfield  {author} {\bibinfo {author} {\bibfnamefont {I.}~\bibnamefont
  {Yeo}}, \bibinfo {author} {\bibfnamefont {P.-L.}\ \bibnamefont {de~Assis}},
  \bibinfo {author} {\bibfnamefont {a.}~\bibnamefont {Gloppe}}, \bibinfo
  {author} {\bibfnamefont {E.}~\bibnamefont {Dupont-Ferrier}}, \bibinfo
  {author} {\bibfnamefont {P.}~\bibnamefont {Verlot}}, \bibinfo {author}
  {\bibfnamefont {N.~S.}\ \bibnamefont {Malik}}, \bibinfo {author}
  {\bibfnamefont {E.}~\bibnamefont {Dupuy}}, \bibinfo {author} {\bibfnamefont
  {J.}~\bibnamefont {Claudon}}, \bibinfo {author} {\bibfnamefont {J.-M.}\
  \bibnamefont {G{\'{e}}rard}}, \bibinfo {author} {\bibfnamefont
  {a.}~\bibnamefont {Auff{\`{e}}ves}}, \bibinfo {author} {\bibfnamefont
  {G.}~\bibnamefont {Nogues}}, \bibinfo {author} {\bibfnamefont
  {S.}~\bibnamefont {Seidelin}}, \bibinfo {author} {\bibfnamefont {J.-p.}\
  \bibnamefont {Poizat}}, \bibinfo {author} {\bibfnamefont {O.}~\bibnamefont
  {Arcizet}}, \ and\ \bibinfo {author} {\bibfnamefont {M.}~\bibnamefont
  {Richard}},\ }\href {\doibase 10.1038/nnano.2013.274} {\bibfield  {journal}
  {\bibinfo  {journal} {Nat. Nanotechnol.}\ }\textbf {\bibinfo {volume} {9}},\
  \bibinfo {pages} {106} (\bibinfo {year} {2014})}\BibitemShut {NoStop}%
\bibitem [{\citenamefont {Ferrari}\ and\ \citenamefont
  {Basko}(2013)}]{Ferrari2013}%
  \BibitemOpen
  \bibfield  {author} {\bibinfo {author} {\bibfnamefont {A.~C.}\ \bibnamefont
  {Ferrari}}\ and\ \bibinfo {author} {\bibfnamefont {D.~M.}\ \bibnamefont
  {Basko}},\ }\href {\doibase 10.1038/nnano.2013.46} {\bibfield  {journal}
  {\bibinfo  {journal} {Nat. Nanotechnol.}\ }\textbf {\bibinfo {volume} {8}},\
  \bibinfo {pages} {235} (\bibinfo {year} {2013})}\BibitemShut {NoStop}%
\bibitem [{\citenamefont {Mohiuddin}\ \emph {et~al.}(2009)\citenamefont
  {Mohiuddin}, \citenamefont {Lombardo}, \citenamefont {Nair}, \citenamefont
  {Bonetti}, \citenamefont {Savini}, \citenamefont {Jalil}, \citenamefont
  {Bonini}, \citenamefont {Basko}, \citenamefont {Galiotis}, \citenamefont
  {Marzari}, \citenamefont {Novoselov}, \citenamefont {Geim},\ and\
  \citenamefont {Ferrari}}]{Mohiuddin2009}%
  \BibitemOpen
  \bibfield  {author} {\bibinfo {author} {\bibfnamefont {T.~M.~G.}\
  \bibnamefont {Mohiuddin}}, \bibinfo {author} {\bibfnamefont {A.}~\bibnamefont
  {Lombardo}}, \bibinfo {author} {\bibfnamefont {R.~R.}\ \bibnamefont {Nair}},
  \bibinfo {author} {\bibfnamefont {A.}~\bibnamefont {Bonetti}}, \bibinfo
  {author} {\bibfnamefont {G.}~\bibnamefont {Savini}}, \bibinfo {author}
  {\bibfnamefont {R.}~\bibnamefont {Jalil}}, \bibinfo {author} {\bibfnamefont
  {N.}~\bibnamefont {Bonini}}, \bibinfo {author} {\bibfnamefont {D.~M.}\
  \bibnamefont {Basko}}, \bibinfo {author} {\bibfnamefont {C.}~\bibnamefont
  {Galiotis}}, \bibinfo {author} {\bibfnamefont {N.}~\bibnamefont {Marzari}},
  \bibinfo {author} {\bibfnamefont {K.~S.}\ \bibnamefont {Novoselov}}, \bibinfo
  {author} {\bibfnamefont {A.~K.}\ \bibnamefont {Geim}}, \ and\ \bibinfo
  {author} {\bibfnamefont {A.~C.}\ \bibnamefont {Ferrari}},\ }\href {\doibase
  10.1103/PhysRevB.79.205433} {\bibfield  {journal} {\bibinfo  {journal} {Phys.
  Rev. B}\ }\textbf {\bibinfo {volume} {79}},\ \bibinfo {pages} {205433}
  (\bibinfo {year} {2009})}\BibitemShut {NoStop}%
\bibitem [{\citenamefont {Metten}\ \emph {et~al.}(2014)\citenamefont {Metten},
  \citenamefont {Federspiel}, \citenamefont {Romeo},\ and\ \citenamefont
  {Berciaud}}]{Metten2014}%
  \BibitemOpen
  \bibfield  {author} {\bibinfo {author} {\bibfnamefont {D.}~\bibnamefont
  {Metten}}, \bibinfo {author} {\bibfnamefont {F.}~\bibnamefont {Federspiel}},
  \bibinfo {author} {\bibfnamefont {M.}~\bibnamefont {Romeo}}, \ and\ \bibinfo
  {author} {\bibfnamefont {S.}~\bibnamefont {Berciaud}},\ }\href {\doibase
  10.1103/PhysRevApplied.2.054008} {\bibfield  {journal} {\bibinfo  {journal}
  {Phys. Rev. Applied}\ }\textbf {\bibinfo {volume} {2}},\ \bibinfo {pages}
  {054008} (\bibinfo {year} {2014})}\BibitemShut {NoStop}%
\bibitem [{\citenamefont {Androulidakis}\ \emph {et~al.}()\citenamefont
  {Androulidakis}, \citenamefont {Koukaras}, \citenamefont {Parthenios},
  \citenamefont {Kalosakas}, \citenamefont {Papagelis},\ and\ \citenamefont
  {Galiotis}}]{Androulidakis2015}%
  \BibitemOpen
  \bibfield  {author} {\bibinfo {author} {\bibfnamefont {C.}~\bibnamefont
  {Androulidakis}}, \bibinfo {author} {\bibfnamefont {E.~N.}\ \bibnamefont
  {Koukaras}}, \bibinfo {author} {\bibfnamefont {J.}~\bibnamefont
  {Parthenios}}, \bibinfo {author} {\bibfnamefont {G.}~\bibnamefont
  {Kalosakas}}, \bibinfo {author} {\bibfnamefont {K.}~\bibnamefont
  {Papagelis}}, \ and\ \bibinfo {author} {\bibfnamefont {C.}~\bibnamefont
  {Galiotis}},\ }\href@noop {} {\bibfield  {journal} {\bibinfo  {journal}
  {Scientific Reports}\ }\textbf {\bibinfo {volume} {5}},\ \bibinfo {pages}
  {18219}}\BibitemShut {NoStop}%
\bibitem [{\citenamefont {Zhang}\ \emph {et~al.}(2015)\citenamefont {Zhang},
  \citenamefont {Qiao}, \citenamefont {Shi}, \citenamefont {Wu}, \citenamefont
  {Jiang},\ and\ \citenamefont {Tan}}]{Zhangx2015}%
  \BibitemOpen
  \bibfield  {author} {\bibinfo {author} {\bibfnamefont {X.}~\bibnamefont
  {Zhang}}, \bibinfo {author} {\bibfnamefont {X.-F.}\ \bibnamefont {Qiao}},
  \bibinfo {author} {\bibfnamefont {W.}~\bibnamefont {Shi}}, \bibinfo {author}
  {\bibfnamefont {J.-B.}\ \bibnamefont {Wu}}, \bibinfo {author} {\bibfnamefont
  {D.-S.}\ \bibnamefont {Jiang}}, \ and\ \bibinfo {author} {\bibfnamefont
  {P.-H.}\ \bibnamefont {Tan}},\ }\href {\doibase 10.1039/C4CS00282B}
  {\bibfield  {journal} {\bibinfo  {journal} {Chem. Soc. Rev.}\ }\textbf
  {\bibinfo {volume} {44}},\ \bibinfo {pages} {2757} (\bibinfo {year}
  {2015})}\BibitemShut {NoStop}%
\bibitem [{\citenamefont {Metten}\ \emph {et~al.}(2016)\citenamefont {Metten},
  \citenamefont {Froehlicher},\ and\ \citenamefont {Berciaud}}]{Metten2016}%
  \BibitemOpen
  \bibfield  {author} {\bibinfo {author} {\bibfnamefont {D.}~\bibnamefont
  {Metten}}, \bibinfo {author} {\bibfnamefont {G.}~\bibnamefont {Froehlicher}},
  \ and\ \bibinfo {author} {\bibfnamefont {S.}~\bibnamefont {Berciaud}},\
  }\href {\doibase 10.1088/2053-1583/4/1/014004} {\bibfield  {journal}
  {\bibinfo  {journal} {2D Mater.}\ }\textbf {\bibinfo {volume} {4}},\ \bibinfo
  {pages} {014004} (\bibinfo {year} {2016})}\BibitemShut {NoStop}%
\bibitem [{\citenamefont {Pomeroy}\ \emph {et~al.}(2008)\citenamefont
  {Pomeroy}, \citenamefont {Gkotsis}, \citenamefont {Zhu}, \citenamefont
  {Leighton}, \citenamefont {Kirby},\ and\ \citenamefont
  {Kuball}}]{Pomeroy2008}%
  \BibitemOpen
  \bibfield  {author} {\bibinfo {author} {\bibfnamefont {J.~W.}\ \bibnamefont
  {Pomeroy}}, \bibinfo {author} {\bibfnamefont {P.}~\bibnamefont {Gkotsis}},
  \bibinfo {author} {\bibfnamefont {M.}~\bibnamefont {Zhu}}, \bibinfo {author}
  {\bibfnamefont {G.}~\bibnamefont {Leighton}}, \bibinfo {author}
  {\bibfnamefont {P.}~\bibnamefont {Kirby}}, \ and\ \bibinfo {author}
  {\bibfnamefont {M.}~\bibnamefont {Kuball}},\ }\href {\doibase
  10.1109/JMEMS.2008.2004849} {\bibfield  {journal} {\bibinfo  {journal}
  {Journal of Microelectromechanical Systems}\ }\textbf {\bibinfo {volume}
  {17}},\ \bibinfo {pages} {1315} (\bibinfo {year} {2008})}\BibitemShut
  {NoStop}%
\bibitem [{\citenamefont {Xue}\ \emph {et~al.}(2007)\citenamefont {Xue},
  \citenamefont {Zheng}, \citenamefont {Zhang}, \citenamefont {Zhang},\ and\
  \citenamefont {Jian}}]{Xue2007}%
  \BibitemOpen
  \bibfield  {author} {\bibinfo {author} {\bibfnamefont {C.}~\bibnamefont
  {Xue}}, \bibinfo {author} {\bibfnamefont {L.}~\bibnamefont {Zheng}}, \bibinfo
  {author} {\bibfnamefont {W.}~\bibnamefont {Zhang}}, \bibinfo {author}
  {\bibfnamefont {B.}~\bibnamefont {Zhang}}, \ and\ \bibinfo {author}
  {\bibfnamefont {A.}~\bibnamefont {Jian}},\ }\href {\doibase 10.1002/jrs.1673}
  {\bibfield  {journal} {\bibinfo  {journal} {Journal of Raman Spectroscopy}\
  }\textbf {\bibinfo {volume} {38}},\ \bibinfo {pages} {467} (\bibinfo {year}
  {2007})}\BibitemShut {NoStop}%
\bibitem [{\citenamefont {Reserbat-Plantey}\ \emph {et~al.}(2012)\citenamefont
  {Reserbat-Plantey}, \citenamefont {Marty}, \citenamefont {Arcizet},
  \citenamefont {Bendiab},\ and\ \citenamefont
  {Bouchiat}}]{Reserbat-Plantey2012}%
  \BibitemOpen
  \bibfield  {author} {\bibinfo {author} {\bibfnamefont {A.}~\bibnamefont
  {Reserbat-Plantey}}, \bibinfo {author} {\bibfnamefont {L.}~\bibnamefont
  {Marty}}, \bibinfo {author} {\bibfnamefont {O.}~\bibnamefont {Arcizet}},
  \bibinfo {author} {\bibfnamefont {N.}~\bibnamefont {Bendiab}}, \ and\
  \bibinfo {author} {\bibfnamefont {V.}~\bibnamefont {Bouchiat}},\ }\href
  {\doibase 10.1038/nnano.2011.250} {\bibfield  {journal} {\bibinfo  {journal}
  {Nat. Nanotechnol.}\ }\textbf {\bibinfo {volume} {7}},\ \bibinfo {pages}
  {151} (\bibinfo {year} {2012})}\BibitemShut {NoStop}%
\bibitem [{\citenamefont {Midolo}\ \emph {et~al.}(2018)\citenamefont {Midolo},
  \citenamefont {Schliesser},\ and\ \citenamefont {Fiore}}]{Midolo2018}%
  \BibitemOpen
  \bibfield  {author} {\bibinfo {author} {\bibfnamefont {L.}~\bibnamefont
  {Midolo}}, \bibinfo {author} {\bibfnamefont {A.}~\bibnamefont {Schliesser}},
  \ and\ \bibinfo {author} {\bibfnamefont {A.}~\bibnamefont {Fiore}},\ }\href
  {\doibase 10.1038/s41565-017-0039-1} {\bibfield  {journal} {\bibinfo
  {journal} {Nat. Nanotechnol.}\ }\textbf {\bibinfo {volume} {13}},\ \bibinfo
  {pages} {11} (\bibinfo {year} {2018})}\BibitemShut {NoStop}%
\bibitem [{\citenamefont {Sampathkumar}\ \emph {et~al.}(2006)\citenamefont
  {Sampathkumar}, \citenamefont {Murray},\ and\ \citenamefont
  {Ekinci}}]{Sampathkumar2006}%
  \BibitemOpen
  \bibfield  {author} {\bibinfo {author} {\bibfnamefont {A.}~\bibnamefont
  {Sampathkumar}}, \bibinfo {author} {\bibfnamefont {T.~W.}\ \bibnamefont
  {Murray}}, \ and\ \bibinfo {author} {\bibfnamefont {K.~L.}\ \bibnamefont
  {Ekinci}},\ }\href@noop {} {\bibfield  {journal} {\bibinfo  {journal}
  {Applied Physics Letters}\ }\textbf {\bibinfo {volume} {88}},\ \bibinfo
  {pages} {223104} (\bibinfo {year} {2006})}\BibitemShut {NoStop}%
\bibitem [{\citenamefont {Pisana}\ \emph {et~al.}(2007)\citenamefont {Pisana},
  \citenamefont {Lazzeri}, \citenamefont {Casiraghi}, \citenamefont
  {Novoselov}, \citenamefont {Geim}, \citenamefont {Ferrari},\ and\
  \citenamefont {Mauri}}]{Pisana2007}%
  \BibitemOpen
  \bibfield  {author} {\bibinfo {author} {\bibfnamefont {S.}~\bibnamefont
  {Pisana}}, \bibinfo {author} {\bibfnamefont {M.}~\bibnamefont {Lazzeri}},
  \bibinfo {author} {\bibfnamefont {C.}~\bibnamefont {Casiraghi}}, \bibinfo
  {author} {\bibfnamefont {K.~S.}\ \bibnamefont {Novoselov}}, \bibinfo {author}
  {\bibfnamefont {A.~K.}\ \bibnamefont {Geim}}, \bibinfo {author}
  {\bibfnamefont {A.~C.}\ \bibnamefont {Ferrari}}, \ and\ \bibinfo {author}
  {\bibfnamefont {F.}~\bibnamefont {Mauri}},\ }\href {\doibase
  10.1038/nmat1846} {\bibfield  {journal} {\bibinfo  {journal} {Nat. Mater.}\
  }\textbf {\bibinfo {volume} {6}},\ \bibinfo {pages} {198} (\bibinfo {year}
  {2007})}\BibitemShut {NoStop}%
\bibitem [{\citenamefont {Lee}\ \emph {et~al.}(2012{\natexlab{a}})\citenamefont
  {Lee}, \citenamefont {Ahn}, \citenamefont {Shim}, \citenamefont {Lee},\ and\
  \citenamefont {Ryu}}]{Lee2012a}%
  \BibitemOpen
  \bibfield  {author} {\bibinfo {author} {\bibfnamefont {J.~E.}\ \bibnamefont
  {Lee}}, \bibinfo {author} {\bibfnamefont {G.}~\bibnamefont {Ahn}}, \bibinfo
  {author} {\bibfnamefont {J.}~\bibnamefont {Shim}}, \bibinfo {author}
  {\bibfnamefont {Y.~S.}\ \bibnamefont {Lee}}, \ and\ \bibinfo {author}
  {\bibfnamefont {S.}~\bibnamefont {Ryu}},\ }\href {\doibase
  10.1038/ncomms2022} {\bibfield  {journal} {\bibinfo  {journal} {Nat.
  Commun.}\ }\textbf {\bibinfo {volume} {3}},\ \bibinfo {pages} {1024}
  (\bibinfo {year} {2012}{\natexlab{a}})}\BibitemShut {NoStop}%
\bibitem [{\citenamefont {Froehlicher}\ and\ \citenamefont
  {Berciaud}(2015)}]{Froehlicher2015}%
  \BibitemOpen
  \bibfield  {author} {\bibinfo {author} {\bibfnamefont {G.}~\bibnamefont
  {Froehlicher}}\ and\ \bibinfo {author} {\bibfnamefont {S.}~\bibnamefont
  {Berciaud}},\ }\href {\doibase 10.1103/PhysRevB.91.205413} {\bibfield
  {journal} {\bibinfo  {journal} {Phys. Rev. B}\ }\textbf {\bibinfo {volume}
  {91}},\ \bibinfo {pages} {205413} (\bibinfo {year} {2015})}\BibitemShut
  {NoStop}%
\bibitem [{\citenamefont {Singh}\ \emph {et~al.}(2010)\citenamefont {Singh},
  \citenamefont {Sengupta}, \citenamefont {Solanki}, \citenamefont {Dhall},
  \citenamefont {Allain}, \citenamefont {Dhara}, \citenamefont {Pant},\ and\
  \citenamefont {Deshmukh}}]{Singh2010}%
  \BibitemOpen
  \bibfield  {author} {\bibinfo {author} {\bibfnamefont {V.}~\bibnamefont
  {Singh}}, \bibinfo {author} {\bibfnamefont {S.}~\bibnamefont {Sengupta}},
  \bibinfo {author} {\bibfnamefont {H.~S.}\ \bibnamefont {Solanki}}, \bibinfo
  {author} {\bibfnamefont {R.}~\bibnamefont {Dhall}}, \bibinfo {author}
  {\bibfnamefont {A.}~\bibnamefont {Allain}}, \bibinfo {author} {\bibfnamefont
  {S.}~\bibnamefont {Dhara}}, \bibinfo {author} {\bibfnamefont
  {P.}~\bibnamefont {Pant}}, \ and\ \bibinfo {author} {\bibfnamefont {M.~M.}\
  \bibnamefont {Deshmukh}},\ }\href {\doibase 10.1088/0957-4484/21/16/165204}
  {\bibfield  {journal} {\bibinfo  {journal} {Nanotechnology}\ }\textbf
  {\bibinfo {volume} {21}},\ \bibinfo {pages} {165204} (\bibinfo {year}
  {2010})}\BibitemShut {NoStop}%
\bibitem [{\citenamefont {Lee}\ \emph {et~al.}(2008)\citenamefont {Lee},
  \citenamefont {Wei}, \citenamefont {Kysar},\ and\ \citenamefont
  {Hone}}]{Lee2008}%
  \BibitemOpen
  \bibfield  {author} {\bibinfo {author} {\bibfnamefont {C.}~\bibnamefont
  {Lee}}, \bibinfo {author} {\bibfnamefont {X.}~\bibnamefont {Wei}}, \bibinfo
  {author} {\bibfnamefont {J.~W.}\ \bibnamefont {Kysar}}, \ and\ \bibinfo
  {author} {\bibfnamefont {J.}~\bibnamefont {Hone}},\ }\href {\doibase
  10.1126/science.1157996} {\bibfield  {journal} {\bibinfo  {journal}
  {Science}\ }\textbf {\bibinfo {volume} {321}},\ \bibinfo {pages} {385}
  (\bibinfo {year} {2008})}\BibitemShut {NoStop}%
\bibitem [{\citenamefont {Nayfeh}\ and\ \citenamefont {Mook}(2007)}]{NO_book}%
  \BibitemOpen
  \bibfield  {author} {\bibinfo {author} {\bibfnamefont {A.~H.}\ \bibnamefont
  {Nayfeh}}\ and\ \bibinfo {author} {\bibfnamefont {D.~T.}\ \bibnamefont
  {Mook}},\ }\href@noop {} {\emph {\bibinfo {title} {Nonlinear Oscillations}}}\
  (\bibinfo  {publisher} {Wiley},\ \bibinfo {year} {2007})\BibitemShut
  {NoStop}%
\bibitem [{\citenamefont {Hauer}\ \emph {et~al.}(2013)\citenamefont {Hauer},
  \citenamefont {Doolin}, \citenamefont {Beach},\ and\ \citenamefont
  {Davis}}]{Hauer2013}%
  \BibitemOpen
  \bibfield  {author} {\bibinfo {author} {\bibfnamefont {B.~D.}\ \bibnamefont
  {Hauer}}, \bibinfo {author} {\bibfnamefont {C.}~\bibnamefont {Doolin}},
  \bibinfo {author} {\bibfnamefont {K.~S.}\ \bibnamefont {Beach}}, \ and\
  \bibinfo {author} {\bibfnamefont {J.~P.}\ \bibnamefont {Davis}},\ }\href
  {\doibase 10.1016/j.aop.2013.08.003} {\bibfield  {journal} {\bibinfo
  {journal} {Annals of Physics}\ }\textbf {\bibinfo {volume} {339}},\ \bibinfo
  {pages} {181} (\bibinfo {year} {2013})}\BibitemShut {NoStop}%
\bibitem [{\citenamefont {Eichler}\ \emph {et~al.}(2013)\citenamefont
  {Eichler}, \citenamefont {Moser}, \citenamefont {Dykman},\ and\ \citenamefont
  {Bachtold}}]{Eichler2013}%
  \BibitemOpen
  \bibfield  {author} {\bibinfo {author} {\bibfnamefont {A.}~\bibnamefont
  {Eichler}}, \bibinfo {author} {\bibfnamefont {J.}~\bibnamefont {Moser}},
  \bibinfo {author} {\bibfnamefont {M.~I.}\ \bibnamefont {Dykman}}, \ and\
  \bibinfo {author} {\bibfnamefont {A.}~\bibnamefont {Bachtold}},\ }\href
  {\doibase 10.1038/ncomms3843} {\bibfield  {journal} {\bibinfo  {journal}
  {Nat. Commun.}\ }\textbf {\bibinfo {volume} {4}},\ \bibinfo {pages} {2843}
  (\bibinfo {year} {2013})}\BibitemShut {NoStop}%
\bibitem [{\citenamefont {Fandan}\ \emph {et~al.}(2020)\citenamefont {Fandan},
  \citenamefont {Pedrós}, \citenamefont {Hernández-Mínguez}, \citenamefont
  {Iikawa}, \citenamefont {Santos}, \citenamefont {Boscá},\ and\ \citenamefont
  {Calle}}]{Fandan2020}%
  \BibitemOpen
  \bibfield  {author} {\bibinfo {author} {\bibfnamefont {R.}~\bibnamefont
  {Fandan}}, \bibinfo {author} {\bibfnamefont {J.}~\bibnamefont {Pedrós}},
  \bibinfo {author} {\bibfnamefont {A.}~\bibnamefont {Hernández-Mínguez}},
  \bibinfo {author} {\bibfnamefont {F.}~\bibnamefont {Iikawa}}, \bibinfo
  {author} {\bibfnamefont {P.~V.}\ \bibnamefont {Santos}}, \bibinfo {author}
  {\bibfnamefont {A.}~\bibnamefont {Boscá}}, \ and\ \bibinfo {author}
  {\bibfnamefont {F.}~\bibnamefont {Calle}},\ }\href {\doibase
  10.1021/acs.nanolett.9b04085} {\bibfield  {journal} {\bibinfo  {journal}
  {Nano Letters}\ }\textbf {\bibinfo {volume} {20}},\ \bibinfo {pages} {402}
  (\bibinfo {year} {2020})}\BibitemShut {NoStop}%
\bibitem [{\citenamefont {Lee}\ \emph {et~al.}(2012{\natexlab{b}})\citenamefont
  {Lee}, \citenamefont {Yoon},\ and\ \citenamefont {Cheong}}]{Lee2012NL}%
  \BibitemOpen
  \bibfield  {author} {\bibinfo {author} {\bibfnamefont {J.-U.}\ \bibnamefont
  {Lee}}, \bibinfo {author} {\bibfnamefont {D.}~\bibnamefont {Yoon}}, \ and\
  \bibinfo {author} {\bibfnamefont {H.}~\bibnamefont {Cheong}},\ }\href
  {\doibase 10.1021/nl301073q} {\bibfield  {journal} {\bibinfo  {journal} {Nano
  Lett.}\ }\textbf {\bibinfo {volume} {12}},\ \bibinfo {pages} {4444} (\bibinfo
  {year} {2012}{\natexlab{b}})}\BibitemShut {NoStop}%
\bibitem [{\citenamefont {Schmid}\ \emph {et~al.}(2016)\citenamefont {Schmid},
  \citenamefont {Villanueva},\ and\ \citenamefont {Roukes}}]{Schmid2016}%
  \BibitemOpen
  \bibfield  {author} {\bibinfo {author} {\bibfnamefont {S.}~\bibnamefont
  {Schmid}}, \bibinfo {author} {\bibfnamefont {L.~G.}\ \bibnamefont
  {Villanueva}}, \ and\ \bibinfo {author} {\bibfnamefont {M.~L.}\ \bibnamefont
  {Roukes}},\ }\href@noop {} {\emph {\bibinfo {title} {Fundamentals of
  nanomechanical resonators}}}\ (\bibinfo  {publisher} {Springer},\ \bibinfo
  {year} {2016})\BibitemShut {NoStop}%
\bibitem [{\citenamefont {Yang}\ \emph {et~al.}(2019)\citenamefont {Yang},
  \citenamefont {Rochau}, \citenamefont {Huber}, \citenamefont {Brieussel},
  \citenamefont {Rastelli}, \citenamefont {Weig},\ and\ \citenamefont
  {Scheer}}]{Yang2019}%
  \BibitemOpen
  \bibfield  {author} {\bibinfo {author} {\bibfnamefont {F.}~\bibnamefont
  {Yang}}, \bibinfo {author} {\bibfnamefont {F.}~\bibnamefont {Rochau}},
  \bibinfo {author} {\bibfnamefont {J.~S.}\ \bibnamefont {Huber}}, \bibinfo
  {author} {\bibfnamefont {A.}~\bibnamefont {Brieussel}}, \bibinfo {author}
  {\bibfnamefont {G.}~\bibnamefont {Rastelli}}, \bibinfo {author}
  {\bibfnamefont {E.~M.}\ \bibnamefont {Weig}}, \ and\ \bibinfo {author}
  {\bibfnamefont {E.}~\bibnamefont {Scheer}},\ }\href {\doibase
  10.1103/PhysRevLett.122.154301} {\bibfield  {journal} {\bibinfo  {journal}
  {Phys. Rev. Lett.}\ }\textbf {\bibinfo {volume} {122}},\ \bibinfo {pages}
  {154301} (\bibinfo {year} {2019})}\BibitemShut {NoStop}%
\bibitem [{\citenamefont {Cattiaux}\ \emph {et~al.}(2019)\citenamefont
  {Cattiaux}, \citenamefont {Kumar}, \citenamefont {Zhou}, \citenamefont
  {Fefferman},\ and\ \citenamefont {Collin}}]{Cattiaux2019}%
  \BibitemOpen
  \bibfield  {author} {\bibinfo {author} {\bibfnamefont {D.}~\bibnamefont
  {Cattiaux}}, \bibinfo {author} {\bibfnamefont {S.}~\bibnamefont {Kumar}},
  \bibinfo {author} {\bibfnamefont {X.}~\bibnamefont {Zhou}}, \bibinfo {author}
  {\bibfnamefont {A.}~\bibnamefont {Fefferman}}, \ and\ \bibinfo {author}
  {\bibfnamefont {E.}~\bibnamefont {Collin}},\ }\href@noop {} {\bibfield
  {journal} {\bibinfo  {journal} {arXiv preprint arXiv:1910.02852}\ } (\bibinfo
  {year} {2019})}\BibitemShut {NoStop}%
\bibitem [{\citenamefont {Sajadi}\ \emph {et~al.}(2017)\citenamefont {Sajadi},
  \citenamefont {Alijani}, \citenamefont {Davidovikj}, \citenamefont {Goosen},
  \citenamefont {Steeneken},\ and\ \citenamefont {van Keulen}}]{Banafsheh2017}%
  \BibitemOpen
  \bibfield  {author} {\bibinfo {author} {\bibfnamefont {B.}~\bibnamefont
  {Sajadi}}, \bibinfo {author} {\bibfnamefont {F.}~\bibnamefont {Alijani}},
  \bibinfo {author} {\bibfnamefont {D.}~\bibnamefont {Davidovikj}}, \bibinfo
  {author} {\bibfnamefont {J.~H.}\ \bibnamefont {Goosen}}, \bibinfo {author}
  {\bibfnamefont {P.~G.}\ \bibnamefont {Steeneken}}, \ and\ \bibinfo {author}
  {\bibfnamefont {F.}~\bibnamefont {van Keulen}},\ }\href@noop {} {\bibfield
  {journal} {\bibinfo  {journal} {Journal of Applied Physics}\ }\textbf
  {\bibinfo {volume} {122}},\ \bibinfo {pages} {234302} (\bibinfo {year}
  {2017})}\BibitemShut {NoStop}%
\bibitem [{\citenamefont {Atalaya}\ \emph {et~al.}(2008)\citenamefont
  {Atalaya}, \citenamefont {Isacsson},\ and\ \citenamefont
  {Kinaret}}]{Atalaya2008}%
  \BibitemOpen
  \bibfield  {author} {\bibinfo {author} {\bibfnamefont {J.}~\bibnamefont
  {Atalaya}}, \bibinfo {author} {\bibfnamefont {A.}~\bibnamefont {Isacsson}}, \
  and\ \bibinfo {author} {\bibfnamefont {J.~M.}\ \bibnamefont {Kinaret}},\
  }\href {\doibase 10.1021/nl801733d} {\bibfield  {journal} {\bibinfo
  {journal} {Nano Lett.}\ }\textbf {\bibinfo {volume} {8}},\ \bibinfo {pages}
  {4196} (\bibinfo {year} {2008})}\BibitemShut {NoStop}%
\bibitem [{\citenamefont {Ackerman}\ \emph {et~al.}(2016)\citenamefont
  {Ackerman}, \citenamefont {Kumar}, \citenamefont {Neek-Amal}, \citenamefont
  {Thibado}, \citenamefont {Peeters},\ and\ \citenamefont
  {Singh}}]{Ackerman2016}%
  \BibitemOpen
  \bibfield  {author} {\bibinfo {author} {\bibfnamefont {M.~L.}\ \bibnamefont
  {Ackerman}}, \bibinfo {author} {\bibfnamefont {P.}~\bibnamefont {Kumar}},
  \bibinfo {author} {\bibfnamefont {M.}~\bibnamefont {Neek-Amal}}, \bibinfo
  {author} {\bibfnamefont {P.~M.}\ \bibnamefont {Thibado}}, \bibinfo {author}
  {\bibfnamefont {F.~M.}\ \bibnamefont {Peeters}}, \ and\ \bibinfo {author}
  {\bibfnamefont {S.}~\bibnamefont {Singh}},\ }\href {\doibase
  10.1103/PhysRevLett.117.126801} {\bibfield  {journal} {\bibinfo  {journal}
  {Phys. Rev. Lett.}\ }\textbf {\bibinfo {volume} {117}},\ \bibinfo {pages}
  {126801} (\bibinfo {year} {2016})}\BibitemShut {NoStop}%
\bibitem [{\citenamefont {Kang}\ \emph {et~al.}(2013)\citenamefont {Kang},
  \citenamefont {Kim}, \citenamefont {Kim},\ and\ \citenamefont
  {Lee}}]{Kang2013}%
  \BibitemOpen
  \bibfield  {author} {\bibinfo {author} {\bibfnamefont {J.~W.}\ \bibnamefont
  {Kang}}, \bibinfo {author} {\bibfnamefont {H.-W.}\ \bibnamefont {Kim}},
  \bibinfo {author} {\bibfnamefont {K.-S.}\ \bibnamefont {Kim}}, \ and\
  \bibinfo {author} {\bibfnamefont {J.~H.}\ \bibnamefont {Lee}},\ }\href
  {\doibase https://doi.org/10.1016/j.cap.2012.12.007} {\bibfield  {journal}
  {\bibinfo  {journal} {Current Applied Physics}\ }\textbf {\bibinfo {volume}
  {13}},\ \bibinfo {pages} {789 } (\bibinfo {year} {2013})}\BibitemShut
  {NoStop}%
\bibitem [{\citenamefont {Wang}\ \emph {et~al.}(2018)\citenamefont {Wang},
  \citenamefont {Chernikov}, \citenamefont {Glazov}, \citenamefont {Heinz},
  \citenamefont {Marie}, \citenamefont {Amand},\ and\ \citenamefont
  {Urbaszek}}]{Wang2018}%
  \BibitemOpen
  \bibfield  {author} {\bibinfo {author} {\bibfnamefont {G.}~\bibnamefont
  {Wang}}, \bibinfo {author} {\bibfnamefont {A.}~\bibnamefont {Chernikov}},
  \bibinfo {author} {\bibfnamefont {M.~M.}\ \bibnamefont {Glazov}}, \bibinfo
  {author} {\bibfnamefont {T.~F.}\ \bibnamefont {Heinz}}, \bibinfo {author}
  {\bibfnamefont {X.}~\bibnamefont {Marie}}, \bibinfo {author} {\bibfnamefont
  {T.}~\bibnamefont {Amand}}, \ and\ \bibinfo {author} {\bibfnamefont
  {B.}~\bibnamefont {Urbaszek}},\ }\href {\doibase
  10.1103/RevModPhys.90.021001} {\bibfield  {journal} {\bibinfo  {journal}
  {Rev. Mod. Phys.}\ }\textbf {\bibinfo {volume} {90}},\ \bibinfo {pages}
  {021001} (\bibinfo {year} {2018})}\BibitemShut {NoStop}%
\bibitem [{\citenamefont {Palacios-Berraquero}\ \emph
  {et~al.}(2017)\citenamefont {Palacios-Berraquero}, \citenamefont {Kara},
  \citenamefont {Montblanch}, \citenamefont {Barbone}, \citenamefont
  {Latawiec}, \citenamefont {Yoon}, \citenamefont {Ott}, \citenamefont
  {Loncar}, \citenamefont {Ferrari},\ and\ \citenamefont
  {Atatüre}}]{Palacios2017}%
  \BibitemOpen
  \bibfield  {author} {\bibinfo {author} {\bibfnamefont {C.}~\bibnamefont
  {Palacios-Berraquero}}, \bibinfo {author} {\bibfnamefont {D.~M.}\
  \bibnamefont {Kara}}, \bibinfo {author} {\bibfnamefont {A.~R.-P.}\
  \bibnamefont {Montblanch}}, \bibinfo {author} {\bibfnamefont
  {M.}~\bibnamefont {Barbone}}, \bibinfo {author} {\bibfnamefont
  {P.}~\bibnamefont {Latawiec}}, \bibinfo {author} {\bibfnamefont
  {D.}~\bibnamefont {Yoon}}, \bibinfo {author} {\bibfnamefont {A.~K.}\
  \bibnamefont {Ott}}, \bibinfo {author} {\bibfnamefont {M.}~\bibnamefont
  {Loncar}}, \bibinfo {author} {\bibfnamefont {A.~C.}\ \bibnamefont {Ferrari}},
  \ and\ \bibinfo {author} {\bibfnamefont {M.}~\bibnamefont {Atatüre}},\
  }\href@noop {} {\bibfield  {journal} {\bibinfo  {journal} {Nat. Commun.}\
  }\textbf {\bibinfo {volume} {8}},\ \bibinfo {pages} {15093} (\bibinfo {year}
  {2017})}\BibitemShut {NoStop}%
\bibitem [{\citenamefont {Branny}\ \emph {et~al.}(2017)\citenamefont {Branny},
  \citenamefont {Kumar}, \citenamefont {Proux},\ and\ \citenamefont
  {Gerardot}}]{Branny2017}%
  \BibitemOpen
  \bibfield  {author} {\bibinfo {author} {\bibfnamefont {A.}~\bibnamefont
  {Branny}}, \bibinfo {author} {\bibfnamefont {S.}~\bibnamefont {Kumar}},
  \bibinfo {author} {\bibfnamefont {R.}~\bibnamefont {Proux}}, \ and\ \bibinfo
  {author} {\bibfnamefont {B.~D.}\ \bibnamefont {Gerardot}},\ }\href@noop {}
  {\bibfield  {journal} {\bibinfo  {journal} {Nat. Commun.}\ }\textbf {\bibinfo
  {volume} {8}},\ \bibinfo {pages} {15053} (\bibinfo {year}
  {2017})}\BibitemShut {NoStop}%
\bibitem [{\citenamefont {Zhou}\ \emph {et~al.}(2020)\citenamefont {Zhou},
  \citenamefont {Scuri}, \citenamefont {Sung}, \citenamefont {Gelly},
  \citenamefont {Wild}, \citenamefont {De~Greve}, \citenamefont {Joe},
  \citenamefont {Taniguchi}, \citenamefont {Watanabe}, \citenamefont {Kim},
  \citenamefont {Lukin},\ and\ \citenamefont {Park}}]{Zhou2019}%
  \BibitemOpen
  \bibfield  {author} {\bibinfo {author} {\bibfnamefont {Y.}~\bibnamefont
  {Zhou}}, \bibinfo {author} {\bibfnamefont {G.}~\bibnamefont {Scuri}},
  \bibinfo {author} {\bibfnamefont {J.}~\bibnamefont {Sung}}, \bibinfo {author}
  {\bibfnamefont {R.~J.}\ \bibnamefont {Gelly}}, \bibinfo {author}
  {\bibfnamefont {D.~S.}\ \bibnamefont {Wild}}, \bibinfo {author}
  {\bibfnamefont {K.}~\bibnamefont {De~Greve}}, \bibinfo {author}
  {\bibfnamefont {A.~Y.}\ \bibnamefont {Joe}}, \bibinfo {author} {\bibfnamefont
  {T.}~\bibnamefont {Taniguchi}}, \bibinfo {author} {\bibfnamefont
  {K.}~\bibnamefont {Watanabe}}, \bibinfo {author} {\bibfnamefont
  {P.}~\bibnamefont {Kim}}, \bibinfo {author} {\bibfnamefont {M.~D.}\
  \bibnamefont {Lukin}}, \ and\ \bibinfo {author} {\bibfnamefont
  {H.}~\bibnamefont {Park}},\ }\href {\doibase 10.1103/PhysRevLett.124.027401}
  {\bibfield  {journal} {\bibinfo  {journal} {Phys. Rev. Lett.}\ }\textbf
  {\bibinfo {volume} {124}},\ \bibinfo {pages} {027401} (\bibinfo {year}
  {2020})}\BibitemShut {NoStop}%
\bibitem [{\citenamefont {Gibertini}\ \emph {et~al.}(2019)\citenamefont
  {Gibertini}, \citenamefont {Koperski}, \citenamefont {Morpurgo},\ and\
  \citenamefont {Novoselov}}]{Gibertini2019}%
  \BibitemOpen
  \bibfield  {author} {\bibinfo {author} {\bibfnamefont {M.}~\bibnamefont
  {Gibertini}}, \bibinfo {author} {\bibfnamefont {M.}~\bibnamefont {Koperski}},
  \bibinfo {author} {\bibfnamefont {A.~F.}\ \bibnamefont {Morpurgo}}, \ and\
  \bibinfo {author} {\bibfnamefont {K.~S.}\ \bibnamefont {Novoselov}},\ }\href
  {\doibase 10.1038/s41565-019-0438-6} {\bibfield  {journal} {\bibinfo
  {journal} {Nat. Nanotechnol.}\ }\textbf {\bibinfo {volume} {14}},\ \bibinfo
  {pages} {408} (\bibinfo {year} {2019})}\BibitemShut {NoStop}%
\bibitem [{\citenamefont {{\v{S}}i{\v{s}}kins}\ \emph
  {et~al.}(2020)\citenamefont {{\v{S}}i{\v{s}}kins}, \citenamefont {Lee},
  \citenamefont {Ma{\~{n}}as-Valero}, \citenamefont {Coronado}, \citenamefont
  {Blanter}, \citenamefont {van~der Zant},\ and\ \citenamefont
  {Steeneken}}]{Siskins2020}%
  \BibitemOpen
  \bibfield  {author} {\bibinfo {author} {\bibfnamefont {M.}~\bibnamefont
  {{\v{S}}i{\v{s}}kins}}, \bibinfo {author} {\bibfnamefont {M.}~\bibnamefont
  {Lee}}, \bibinfo {author} {\bibfnamefont {S.}~\bibnamefont
  {Ma{\~{n}}as-Valero}}, \bibinfo {author} {\bibfnamefont {E.}~\bibnamefont
  {Coronado}}, \bibinfo {author} {\bibfnamefont {Y.~M.}\ \bibnamefont
  {Blanter}}, \bibinfo {author} {\bibfnamefont {H.~S.~J.}\ \bibnamefont
  {van~der Zant}}, \ and\ \bibinfo {author} {\bibfnamefont {P.~G.}\
  \bibnamefont {Steeneken}},\ }\href {\doibase 10.1038/s41467-020-16430-2}
  {\bibfield  {journal} {\bibinfo  {journal} {Nat. Communs}\ }\textbf {\bibinfo
  {volume} {11}},\ \bibinfo {pages} {2698} (\bibinfo {year}
  {2020})}\BibitemShut {NoStop}%
\bibitem [{\citenamefont {Jiang}\ \emph {et~al.}(2020)\citenamefont {Jiang},
  \citenamefont {Xie}, \citenamefont {Shan},\ and\ \citenamefont
  {Mak}}]{Jiang2020}%
  \BibitemOpen
  \bibfield  {author} {\bibinfo {author} {\bibfnamefont {S.}~\bibnamefont
  {Jiang}}, \bibinfo {author} {\bibfnamefont {H.}~\bibnamefont {Xie}}, \bibinfo
  {author} {\bibfnamefont {J.}~\bibnamefont {Shan}}, \ and\ \bibinfo {author}
  {\bibfnamefont {K.~F.}\ \bibnamefont {Mak}},\ }\href {\doibase
  10.1038/s41563-020-0712-x} {\bibfield  {journal} {\bibinfo  {journal} {Nat.
  Mater.}\ } (\bibinfo {year} {2020}),\ 10.1038/s41563-020-0712-x}\BibitemShut
  {NoStop}%
\bibitem [{\citenamefont {Huang}\ \emph {et~al.}(2018)\citenamefont {Huang},
  \citenamefont {Wang}, \citenamefont {Zhang}, \citenamefont {Chen},
  \citenamefont {Li}, \citenamefont {Wang}, \citenamefont {Huang},
  \citenamefont {Zhu}, \citenamefont {Zhang}, \citenamefont {Bacsa},
  \citenamefont {Ding},\ and\ \citenamefont {Ruoff}}]{Huang2018}%
  \BibitemOpen
  \bibfield  {author} {\bibinfo {author} {\bibfnamefont {Y.}~\bibnamefont
  {Huang}}, \bibinfo {author} {\bibfnamefont {X.}~\bibnamefont {Wang}},
  \bibinfo {author} {\bibfnamefont {X.}~\bibnamefont {Zhang}}, \bibinfo
  {author} {\bibfnamefont {X.}~\bibnamefont {Chen}}, \bibinfo {author}
  {\bibfnamefont {B.}~\bibnamefont {Li}}, \bibinfo {author} {\bibfnamefont
  {B.}~\bibnamefont {Wang}}, \bibinfo {author} {\bibfnamefont {M.}~\bibnamefont
  {Huang}}, \bibinfo {author} {\bibfnamefont {C.}~\bibnamefont {Zhu}}, \bibinfo
  {author} {\bibfnamefont {X.}~\bibnamefont {Zhang}}, \bibinfo {author}
  {\bibfnamefont {W.~S.}\ \bibnamefont {Bacsa}}, \bibinfo {author}
  {\bibfnamefont {F.}~\bibnamefont {Ding}}, \ and\ \bibinfo {author}
  {\bibfnamefont {R.~S.}\ \bibnamefont {Ruoff}},\ }\href {\doibase
  10.1103/PhysRevLett.120.186104} {\bibfield  {journal} {\bibinfo  {journal}
  {Phys. Rev. Lett.}\ }\textbf {\bibinfo {volume} {120}},\ \bibinfo {pages}
  {186104} (\bibinfo {year} {2018})}\BibitemShut {NoStop}%
\bibitem [{\citenamefont {Nicholl}\ \emph {et~al.}(2017)\citenamefont
  {Nicholl}, \citenamefont {Lavrik}, \citenamefont {Vlassiouk}, \citenamefont
  {Srijanto},\ and\ \citenamefont {Bolotin}}]{Nicholl2017}%
  \BibitemOpen
  \bibfield  {author} {\bibinfo {author} {\bibfnamefont {R.~J.~T.}\
  \bibnamefont {Nicholl}}, \bibinfo {author} {\bibfnamefont {N.~V.}\
  \bibnamefont {Lavrik}}, \bibinfo {author} {\bibfnamefont {I.}~\bibnamefont
  {Vlassiouk}}, \bibinfo {author} {\bibfnamefont {B.~R.}\ \bibnamefont
  {Srijanto}}, \ and\ \bibinfo {author} {\bibfnamefont {K.~I.}\ \bibnamefont
  {Bolotin}},\ }\href {\doibase 10.1103/PhysRevLett.118.266101} {\bibfield
  {journal} {\bibinfo  {journal} {Phys. Rev. Lett.}\ }\textbf {\bibinfo
  {volume} {118}},\ \bibinfo {pages} {266101} (\bibinfo {year}
  {2017})}\BibitemShut {NoStop}%
\bibitem [{\citenamefont {Berciaud}\ \emph {et~al.}(2009)\citenamefont
  {Berciaud}, \citenamefont {Ryu}, \citenamefont {Brus},\ and\ \citenamefont
  {Heinz}}]{Berciaud2009}%
  \BibitemOpen
  \bibfield  {author} {\bibinfo {author} {\bibfnamefont {S.}~\bibnamefont
  {Berciaud}}, \bibinfo {author} {\bibfnamefont {S.}~\bibnamefont {Ryu}},
  \bibinfo {author} {\bibfnamefont {L.~E.}\ \bibnamefont {Brus}}, \ and\
  \bibinfo {author} {\bibfnamefont {T.~F.}\ \bibnamefont {Heinz}},\ }\href
  {\doibase 10.1021/nl8031444} {\bibfield  {journal} {\bibinfo  {journal} {Nano
  Lett.}\ }\textbf {\bibinfo {volume} {9}},\ \bibinfo {pages} {346} (\bibinfo
  {year} {2009})}\BibitemShut {NoStop}%
\bibitem [{\citenamefont {Berciaud}\ \emph {et~al.}(2013)\citenamefont
  {Berciaud}, \citenamefont {Li}, \citenamefont {Htoon}, \citenamefont {Brus},
  \citenamefont {Doorn},\ and\ \citenamefont {Heinz}}]{Berciaud2013}%
  \BibitemOpen
  \bibfield  {author} {\bibinfo {author} {\bibfnamefont {S.}~\bibnamefont
  {Berciaud}}, \bibinfo {author} {\bibfnamefont {X.}~\bibnamefont {Li}},
  \bibinfo {author} {\bibfnamefont {H.}~\bibnamefont {Htoon}}, \bibinfo
  {author} {\bibfnamefont {L.~E.}\ \bibnamefont {Brus}}, \bibinfo {author}
  {\bibfnamefont {S.~K.}\ \bibnamefont {Doorn}}, \ and\ \bibinfo {author}
  {\bibfnamefont {T.~F.}\ \bibnamefont {Heinz}},\ }\href {\doibase
  10.1021/nl400917e} {\bibfield  {journal} {\bibinfo  {journal} {Nano Lett.}\
  }\textbf {\bibinfo {volume} {13}},\ \bibinfo {pages} {3517} (\bibinfo {year}
  {2013})}\BibitemShut {NoStop}%
\bibitem [{\citenamefont {Bonini}\ \emph {et~al.}(2007)\citenamefont {Bonini},
  \citenamefont {Lazzeri}, \citenamefont {Marzari},\ and\ \citenamefont
  {Mauri}}]{Bonini2007}%
  \BibitemOpen
  \bibfield  {author} {\bibinfo {author} {\bibfnamefont {N.}~\bibnamefont
  {Bonini}}, \bibinfo {author} {\bibfnamefont {M.}~\bibnamefont {Lazzeri}},
  \bibinfo {author} {\bibfnamefont {N.}~\bibnamefont {Marzari}}, \ and\
  \bibinfo {author} {\bibfnamefont {F.}~\bibnamefont {Mauri}},\ }\href
  {\doibase 10.1103/PhysRevLett.99.176802} {\bibfield  {journal} {\bibinfo
  {journal} {Phys. Rev. Lett.}\ }\textbf {\bibinfo {volume} {99}},\ \bibinfo
  {pages} {176802} (\bibinfo {year} {2007})}\BibitemShut {NoStop}%
\bibitem [{\citenamefont {Zabel}\ \emph {et~al.}(2012)\citenamefont {Zabel},
  \citenamefont {Nair}, \citenamefont {Ott}, \citenamefont {Georgiou},
  \citenamefont {Geim}, \citenamefont {Novoselov},\ and\ \citenamefont
  {Casiraghi}}]{Zabel2012a}%
  \BibitemOpen
  \bibfield  {author} {\bibinfo {author} {\bibfnamefont {J.}~\bibnamefont
  {Zabel}}, \bibinfo {author} {\bibfnamefont {R.~R.}\ \bibnamefont {Nair}},
  \bibinfo {author} {\bibfnamefont {A.}~\bibnamefont {Ott}}, \bibinfo {author}
  {\bibfnamefont {T.}~\bibnamefont {Georgiou}}, \bibinfo {author}
  {\bibfnamefont {A.~K.}\ \bibnamefont {Geim}}, \bibinfo {author}
  {\bibfnamefont {K.~S.}\ \bibnamefont {Novoselov}}, \ and\ \bibinfo {author}
  {\bibfnamefont {C.}~\bibnamefont {Casiraghi}},\ }\href {\doibase
  10.1021/nl203359n} {\bibfield  {journal} {\bibinfo  {journal} {Nano Lett.}\
  }\textbf {\bibinfo {volume} {12}},\ \bibinfo {pages} {617} (\bibinfo {year}
  {2012})}\BibitemShut {NoStop}%
\bibitem [{\citenamefont {Eichler}\ \emph {et~al.}(2011)\citenamefont
  {Eichler}, \citenamefont {Moser}, \citenamefont {Chaste}, \citenamefont
  {Zdrojek}, \citenamefont {Wilson-Rae},\ and\ \citenamefont
  {Bachtold}}]{Eichler2011a}%
  \BibitemOpen
  \bibfield  {author} {\bibinfo {author} {\bibfnamefont {A.}~\bibnamefont
  {Eichler}}, \bibinfo {author} {\bibfnamefont {J.}~\bibnamefont {Moser}},
  \bibinfo {author} {\bibfnamefont {J.}~\bibnamefont {Chaste}}, \bibinfo
  {author} {\bibfnamefont {M.}~\bibnamefont {Zdrojek}}, \bibinfo {author}
  {\bibfnamefont {I.}~\bibnamefont {Wilson-Rae}}, \ and\ \bibinfo {author}
  {\bibfnamefont {A.}~\bibnamefont {Bachtold}},\ }\href {\doibase
  10.1038/nnano.2011.71} {\bibfield  {journal} {\bibinfo  {journal} {Nat.
  Nanotechnol.}\ }\textbf {\bibinfo {volume} {6}},\ \bibinfo {pages} {339}
  (\bibinfo {year} {2011})}\BibitemShut {NoStop}%
\bibitem [{\citenamefont {Imboden}\ \emph {et~al.}(2013)\citenamefont
  {Imboden}, \citenamefont {Williams},\ and\ \citenamefont
  {Mohanty}}]{Imboden2013}%
  \BibitemOpen
  \bibfield  {author} {\bibinfo {author} {\bibfnamefont {M.}~\bibnamefont
  {Imboden}}, \bibinfo {author} {\bibfnamefont {O.}~\bibnamefont {Williams}}, \
  and\ \bibinfo {author} {\bibfnamefont {P.}~\bibnamefont {Mohanty}},\ }\href
  {\doibase 10.1063/1.4794907} {\bibfield  {journal} {\bibinfo  {journal}
  {Appl. Phys. Lett.}\ }\textbf {\bibinfo {volume} {102}},\ \bibinfo {pages}
  {103502} (\bibinfo {year} {2013})}\BibitemShut {NoStop}%
\bibitem [{\citenamefont {Maultzsch}\ \emph {et~al.}(2004)\citenamefont
  {Maultzsch}, \citenamefont {Reich},\ and\ \citenamefont
  {Thomsen}}]{Maultzsch2004}%
  \BibitemOpen
  \bibfield  {author} {\bibinfo {author} {\bibfnamefont {J.}~\bibnamefont
  {Maultzsch}}, \bibinfo {author} {\bibfnamefont {S.}~\bibnamefont {Reich}}, \
  and\ \bibinfo {author} {\bibfnamefont {C.}~\bibnamefont {Thomsen}},\ }\href
  {\doibase 10.1103/PhysRevB.70.155403} {\bibfield  {journal} {\bibinfo
  {journal} {Phys. Rev. B}\ }\textbf {\bibinfo {volume} {70}},\ \bibinfo
  {pages} {155403} (\bibinfo {year} {2004})}\BibitemShut {NoStop}%
\bibitem [{\citenamefont {Basko}(2008)}]{Basko2008}%
  \BibitemOpen
  \bibfield  {author} {\bibinfo {author} {\bibfnamefont {D.~M.}\ \bibnamefont
  {Basko}},\ }\href {\doibase 10.1103/PhysRevB.78.125418} {\bibfield  {journal}
  {\bibinfo  {journal} {Phys. Rev. B}\ }\textbf {\bibinfo {volume} {78}},\
  \bibinfo {pages} {125418} (\bibinfo {year} {2008})}\BibitemShut {NoStop}%
\bibitem [{\citenamefont {Venezuela}\ \emph {et~al.}(2011)\citenamefont
  {Venezuela}, \citenamefont {Lazzeri},\ and\ \citenamefont
  {Mauri}}]{Venezuela2011}%
  \BibitemOpen
  \bibfield  {author} {\bibinfo {author} {\bibfnamefont {P.}~\bibnamefont
  {Venezuela}}, \bibinfo {author} {\bibfnamefont {M.}~\bibnamefont {Lazzeri}},
  \ and\ \bibinfo {author} {\bibfnamefont {F.}~\bibnamefont {Mauri}},\ }\href
  {\doibase 10.1103/PhysRevB.84.035433} {\bibfield  {journal} {\bibinfo
  {journal} {Phys. Rev. B}\ }\textbf {\bibinfo {volume} {84}},\ \bibinfo
  {pages} {1} (\bibinfo {year} {2011})}\BibitemShut {NoStop}%
\bibitem [{\citenamefont {Chen}\ \emph {et~al.}(2011)\citenamefont {Chen},
  \citenamefont {Park}, \citenamefont {Boudouris}, \citenamefont {Horng},
  \citenamefont {Geng}, \citenamefont {Girit}, \citenamefont {Zettl},
  \citenamefont {Crommie}, \citenamefont {Segalman}, \citenamefont {Louie},\
  and\ \citenamefont {Wang}}]{Chen2011}%
  \BibitemOpen
  \bibfield  {author} {\bibinfo {author} {\bibfnamefont {C.-F.}\ \bibnamefont
  {Chen}}, \bibinfo {author} {\bibfnamefont {C.-H.}\ \bibnamefont {Park}},
  \bibinfo {author} {\bibfnamefont {B.~W.}\ \bibnamefont {Boudouris}}, \bibinfo
  {author} {\bibfnamefont {J.}~\bibnamefont {Horng}}, \bibinfo {author}
  {\bibfnamefont {B.}~\bibnamefont {Geng}}, \bibinfo {author} {\bibfnamefont
  {C.}~\bibnamefont {Girit}}, \bibinfo {author} {\bibfnamefont
  {A.}~\bibnamefont {Zettl}}, \bibinfo {author} {\bibfnamefont {M.~F.}\
  \bibnamefont {Crommie}}, \bibinfo {author} {\bibfnamefont {R.~A.}\
  \bibnamefont {Segalman}}, \bibinfo {author} {\bibfnamefont {S.~G.}\
  \bibnamefont {Louie}}, \ and\ \bibinfo {author} {\bibfnamefont
  {F.}~\bibnamefont {Wang}},\ }\href {\doibase 10.1038/nature09866} {\bibfield
  {journal} {\bibinfo  {journal} {Nature}\ }\textbf {\bibinfo {volume} {471}},\
  \bibinfo {pages} {617} (\bibinfo {year} {2011})}\BibitemShut {NoStop}%
\bibitem [{\citenamefont {Blake}\ \emph {et~al.}(2007)\citenamefont {Blake},
  \citenamefont {Hill}, \citenamefont {{Castro Neto}}, \citenamefont
  {Novoselov}, \citenamefont {Jiang}, \citenamefont {Yang}, \citenamefont
  {Booth},\ and\ \citenamefont {Geim}}]{Blake2007}%
  \BibitemOpen
  \bibfield  {author} {\bibinfo {author} {\bibfnamefont {P.}~\bibnamefont
  {Blake}}, \bibinfo {author} {\bibfnamefont {E.~W.}\ \bibnamefont {Hill}},
  \bibinfo {author} {\bibfnamefont {A.~H.}\ \bibnamefont {{Castro Neto}}},
  \bibinfo {author} {\bibfnamefont {K.~S.}\ \bibnamefont {Novoselov}}, \bibinfo
  {author} {\bibfnamefont {D.}~\bibnamefont {Jiang}}, \bibinfo {author}
  {\bibfnamefont {R.}~\bibnamefont {Yang}}, \bibinfo {author} {\bibfnamefont
  {T.~J.}\ \bibnamefont {Booth}}, \ and\ \bibinfo {author} {\bibfnamefont
  {A.~K.}\ \bibnamefont {Geim}},\ }\href {\doibase 10.1063/1.2768624}
  {\bibfield  {journal} {\bibinfo  {journal} {Applied Physics Letters}\
  }\textbf {\bibinfo {volume} {91}},\ \bibinfo {pages} {2007} (\bibinfo {year}
  {2007})}\BibitemShut {NoStop}%
\bibitem [{\citenamefont {Yoon}\ \emph {et~al.}(2009)\citenamefont {Yoon},
  \citenamefont {Moon}, \citenamefont {Son}, \citenamefont {Choi},
  \citenamefont {Park}, \citenamefont {Cha}, \citenamefont {Kim},\ and\
  \citenamefont {Cheong}}]{Yoon2009}%
  \BibitemOpen
  \bibfield  {author} {\bibinfo {author} {\bibfnamefont {D.}~\bibnamefont
  {Yoon}}, \bibinfo {author} {\bibfnamefont {H.}~\bibnamefont {Moon}}, \bibinfo
  {author} {\bibfnamefont {Y.-W.}\ \bibnamefont {Son}}, \bibinfo {author}
  {\bibfnamefont {J.~S.}\ \bibnamefont {Choi}}, \bibinfo {author}
  {\bibfnamefont {B.~H.}\ \bibnamefont {Park}}, \bibinfo {author}
  {\bibfnamefont {Y.~H.}\ \bibnamefont {Cha}}, \bibinfo {author} {\bibfnamefont
  {Y.~D.}\ \bibnamefont {Kim}}, \ and\ \bibinfo {author} {\bibfnamefont
  {H.}~\bibnamefont {Cheong}},\ }\href {\doibase 10.1103/PhysRevB.80.125422}
  {\bibfield  {journal} {\bibinfo  {journal} {Phys. Rev. B}\ }\textbf {\bibinfo
  {volume} {80}},\ \bibinfo {pages} {125422} (\bibinfo {year}
  {2009})}\BibitemShut {NoStop}%
\bibitem [{\citenamefont {Metten}\ \emph {et~al.}(2015)\citenamefont {Metten},
  \citenamefont {Froehlicher},\ and\ \citenamefont {Berciaud}}]{Metten2015}%
  \BibitemOpen
  \bibfield  {author} {\bibinfo {author} {\bibfnamefont {D.}~\bibnamefont
  {Metten}}, \bibinfo {author} {\bibfnamefont {G.}~\bibnamefont {Froehlicher}},
  \ and\ \bibinfo {author} {\bibfnamefont {S.}~\bibnamefont {Berciaud}},\
  }\href@noop {} {\bibfield  {journal} {\bibinfo  {journal} {Phys. Status
  Solidi B}\ }\textbf {\bibinfo {volume} {252}},\ \bibinfo {pages} {2390}
  (\bibinfo {year} {2015})}\BibitemShut {NoStop}%
\bibitem [{\citenamefont {Nicholl}\ \emph {et~al.}(2015)\citenamefont
  {Nicholl}, \citenamefont {Conley}, \citenamefont {Lavrik}, \citenamefont
  {Vlassiouk}, \citenamefont {Puzyrev}, \citenamefont {Sreenivas},
  \citenamefont {Pantelides},\ and\ \citenamefont {Bolotin}}]{Nicholl2015}%
  \BibitemOpen
  \bibfield  {author} {\bibinfo {author} {\bibfnamefont {R.~J.}\ \bibnamefont
  {Nicholl}}, \bibinfo {author} {\bibfnamefont {H.~J.}\ \bibnamefont {Conley}},
  \bibinfo {author} {\bibfnamefont {N.~V.}\ \bibnamefont {Lavrik}}, \bibinfo
  {author} {\bibfnamefont {I.}~\bibnamefont {Vlassiouk}}, \bibinfo {author}
  {\bibfnamefont {Y.~S.}\ \bibnamefont {Puzyrev}}, \bibinfo {author}
  {\bibfnamefont {V.~P.}\ \bibnamefont {Sreenivas}}, \bibinfo {author}
  {\bibfnamefont {S.~T.}\ \bibnamefont {Pantelides}}, \ and\ \bibinfo {author}
  {\bibfnamefont {K.~I.}\ \bibnamefont {Bolotin}},\ }\href {\doibase
  10.1038/ncomms9789} {\bibfield  {journal} {\bibinfo  {journal} {Nat.
  Commun.}\ }\textbf {\bibinfo {volume} {6}},\ \bibinfo {pages} {8789}
  (\bibinfo {year} {2015})}\BibitemShut {NoStop}%
\bibitem [{\citenamefont {Bolotin}\ \emph {et~al.}(2008)\citenamefont
  {Bolotin}, \citenamefont {Sikes}, \citenamefont {Jiang}, \citenamefont
  {Klima}, \citenamefont {Fudenberg}, \citenamefont {Hone}, \citenamefont
  {Kim},\ and\ \citenamefont {Stormer}}]{Bolotin2008}%
  \BibitemOpen
  \bibfield  {author} {\bibinfo {author} {\bibfnamefont {K.~I.}\ \bibnamefont
  {Bolotin}}, \bibinfo {author} {\bibfnamefont {K.~J.}\ \bibnamefont {Sikes}},
  \bibinfo {author} {\bibfnamefont {Z.}~\bibnamefont {Jiang}}, \bibinfo
  {author} {\bibfnamefont {M.}~\bibnamefont {Klima}}, \bibinfo {author}
  {\bibfnamefont {G.}~\bibnamefont {Fudenberg}}, \bibinfo {author}
  {\bibfnamefont {J.}~\bibnamefont {Hone}}, \bibinfo {author} {\bibfnamefont
  {P.}~\bibnamefont {Kim}}, \ and\ \bibinfo {author} {\bibfnamefont {H.~L.}\
  \bibnamefont {Stormer}},\ }\href@noop {} {\bibfield  {journal} {\bibinfo
  {journal} {Solid State Commun.}\ }\textbf {\bibinfo {volume} {146}},\
  \bibinfo {pages} {351} (\bibinfo {year} {2008})}\BibitemShut {NoStop}%
\bibitem [{\citenamefont {Berciaud}\ \emph {et~al.}(2014)\citenamefont
  {Berciaud}, \citenamefont {Potemski},\ and\ \citenamefont
  {Faugeras}}]{Berciaud2014}%
  \BibitemOpen
  \bibfield  {author} {\bibinfo {author} {\bibfnamefont {S.}~\bibnamefont
  {Berciaud}}, \bibinfo {author} {\bibfnamefont {M.}~\bibnamefont {Potemski}},
  \ and\ \bibinfo {author} {\bibfnamefont {C.}~\bibnamefont {Faugeras}},\
  }\href@noop {} {\bibfield  {journal} {\bibinfo  {journal} {Nano Lett.}\
  }\textbf {\bibinfo {volume} {14}},\ \bibinfo {pages} {4548} (\bibinfo {year}
  {2014})}\BibitemShut {NoStop}%
\bibitem [{Note1()}]{Note1}%
  \BibitemOpen
  \bibinfo {note} {In principle, Eq.~(S9) could include other non-linear
  contributions, and in particular a non-linear damping term $(\propto \protect
  \dot {z}z^2)$~\cite {Eichler2011a,Imboden2013}. Non-linear damping may
  broaden the frequency-dependent mechanical susceptibility of our drums,
  reduce its resonant amplitude and may thus act against the enhancement of
  $\varepsilon _{\protect \mathrm {d}}$. As a result, more pronounced
  dynamically-induced strain enhancement could be achieved provided non-linear
  damping in minimized.}\BibitemShut {Stop}%
\bibitem [{\citenamefont {Schwarz}(2016)}]{Schwarz-thesis}%
  \BibitemOpen
  \bibfield  {author} {\bibinfo {author} {\bibfnamefont {C.}~\bibnamefont
  {Schwarz}},\ }\href@noop {} {\bibinfo {type} {{PhD} dissertation}},\ \bibinfo
   {school} {Institut Neel, Grenoble} (\bibinfo {year} {2016})\BibitemShut
  {NoStop}%
\bibitem [{\citenamefont {Calizo}\ \emph {et~al.}(2007)\citenamefont {Calizo},
  \citenamefont {Balandin}, \citenamefont {Bao}, \citenamefont {Miao},\ and\
  \citenamefont {Lau}}]{Calizo2007}%
  \BibitemOpen
  \bibfield  {author} {\bibinfo {author} {\bibfnamefont {I.}~\bibnamefont
  {Calizo}}, \bibinfo {author} {\bibfnamefont {A.~A.}\ \bibnamefont
  {Balandin}}, \bibinfo {author} {\bibfnamefont {W.}~\bibnamefont {Bao}},
  \bibinfo {author} {\bibfnamefont {F.}~\bibnamefont {Miao}}, \ and\ \bibinfo
  {author} {\bibfnamefont {C.~N.}\ \bibnamefont {Lau}},\ }\href {\doibase
  10.1021/nl071033g} {\bibfield  {journal} {\bibinfo  {journal} {Nano Lett.}\
  }\textbf {\bibinfo {volume} {7}},\ \bibinfo {pages} {2645} (\bibinfo {year}
  {2007})}\BibitemShut {NoStop}%
\bibitem [{\citenamefont {Yoon}\ \emph {et~al.}(2011)\citenamefont {Yoon},
  \citenamefont {Son},\ and\ \citenamefont {Cheong}}]{Yoon2011}%
  \BibitemOpen
  \bibfield  {author} {\bibinfo {author} {\bibfnamefont {D.}~\bibnamefont
  {Yoon}}, \bibinfo {author} {\bibfnamefont {Y.-W.}\ \bibnamefont {Son}}, \
  and\ \bibinfo {author} {\bibfnamefont {H.}~\bibnamefont {Cheong}},\ }\href
  {\doibase 10.1021/nl201488g} {\bibfield  {journal} {\bibinfo  {journal} {Nano
  Lett.}\ }\textbf {\bibinfo {volume} {11}},\ \bibinfo {pages} {3227} (\bibinfo
  {year} {2011})}\BibitemShut {NoStop}%
\end{thebibliography}
\end{document}